\documentclass[aps,pra,twocolumn,amsmath,amssymb,superscriptaddress,longbibliography]{revtex4-2}
\usepackage[english]{babel}

\usepackage{amssymb}
\usepackage{dcolumn}
\usepackage{bm}
\usepackage{amsmath}
\usepackage{tabularx}
\usepackage{graphicx}        
\usepackage{makecell}                             
\graphicspath{{pict/}{}}
\newcommand{\tabincell}[2]{\begin{tabular}{@{}#1@{}}#2\end{tabular}}
\usepackage{xcolor}
\usepackage{multirow}
\usepackage{dcolumn}
\usepackage{bm}
\def\red{\textcolor{red}}
\def\blue{\textcolor{blue}}

\usepackage[normalem]{ulem}
\usepackage{comment}

\usepackage{hhline}

\begin{document}

\title{Topological classification of Higher-order topological phases with nested band inversion surfaces}

\author{Zhoutao Lei}
\affiliation{Guangdong Provincial Key Laboratory of Quantum Metrology and Sensing $\&$ School of Physics and Astronomy, Sun Yat-Sen University (Zhuhai Campus), Zhuhai 519082, China}
\author{Yuangang Deng}\email{dengyg3@mail.sysu.edu.cn}
\affiliation{Guangdong Provincial Key Laboratory of Quantum Metrology and Sensing $\&$ School of Physics and Astronomy, Sun Yat-Sen University (Zhuhai Campus), Zhuhai 519082, China}
\author{Linhu Li}\email{lilh56@mail.sysu.edu.cn}
\affiliation{Guangdong Provincial Key Laboratory of Quantum Metrology and Sensing $\&$ School of Physics and Astronomy, Sun Yat-Sen University (Zhuhai Campus), Zhuhai 519082, China}

\date{\today}

\begin{abstract}
Higher-order topological phases (HOTPs) hold gapped bulk bands
and topological boundary states localized in boundaries with codimension higher than one.
In this paper, we provide a unified construction and topological characterization of HOTPs for the full Altland-Zirnbauer tenfold symmetry classes, based on a method known as nested band inversion surfaces (BISs).
%
Specifically, HOTPs built on this method are decomposed into a series of subsystems, and higher-order topological boundary states emerges from the interplay of their first-order topology.
Our analysis begins with a general discussion of HOTPs in continuous Hamiltonians for each symmetry class,
then moves on to several lattice examples illustrating the topological characterization based on the nested-BIS method.
Despite the example minimal models possessing several spatial symmetries,
our method does not rely on any spatial symmetry, and can be easily extended into arbitrary orders of topology in dimensions.
Furthermore, we extend our discussion to systems with asymmetric boundary states induced by two different mechanisms, namely, crossed BISs that break a $\mathcal{C}_4$ rotation symmetry, and non-Clifford operators that break certain chiral-mirror symmetries of the minimal models.
\end{abstract}

\maketitle

\section{introduction\label{Sec1}}
Topological quantum matters, which are characterized by topological indices and host in-gap boundary states, have attracted
much attention over the past decade~\cite{RevModPhys.82.3045,RevModPhys.83.1057}.
In the last few years, higher-order topological phases (HOTPs) beyond the conventional
bulk-boundary correspondence principle have been introduced, i.e., a $d$-dimensional ($d$D) $n$th-order topological phase supports topologically protected boundary states in their ($d-n$)D boundaries~\cite{Benalcazar61,PhysRevB.96.245115,PhysRevLett.119.246401,PhysRevLett.119.246402}.
In contrast to conventional (first-order) boundary states that have one dimension lower than the bulk,
the emergence of higher-order boundary states (e.g., corner states in 2D or higher) may depend not only on the bulk band topology, but also on properties of the system's boundaries.
Accordingly, HOTPs have been classified into two categories: (i) ``intrinsic" HOTPs associated with spatial-symmetry-protected bulk topology and (ii) ``extrinsic" HOTPs, 
whose topological properties rely on boundary termination of the system, instead of protection by a bulk crystalline symmetry \cite{PhysRevB.97.205135,PhysRevX.9.011012}.
In particular, a concept closely related to the latter is the boundary-obstructed HOTP, 
which hosts robust higher-order boundary states in association with bulk quantities even their emergence and disappearance do not involve bulk gap closing under the periodic boundary condition (PBC)
 \cite{ezawa2020edge,asaga2020boundary,wu2020boundary,tiwari2020chiral,khalaf2021boundary}.


Owing to the rich and sophisticated topological origins rooted in the bulk and/or boundary of the systems,
constructing and topologically characterizing HOTPs has been of great interest since their discovery~
\cite{Benalcazar61,PhysRevB.96.245115,PhysRevLett.119.246401,PhysRevLett.119.246402,PhysRevLett.120.026801,
ezawa2020edge,asaga2020boundary,wu2020boundary,tiwari2020chiral,khalaf2021boundary,
PhysRevB.97.205135,PhysRevB.97.205136,PhysRevB.97.241405,PhysRevB.98.205147,PhysRevB.98.205422,PhysRevB.99.020304,Serra-Garcia2018,Schindler2018,Peterson2018,PhysRevLett.122.076801,PhysRevLett.123.016805,PhysRevLett.123.016806,
PhysRevLett.123.060402,PhysRevLett.123.073601,PhysRevX.9.011012,PhysRevLett.123.167001,PhysRevLett.123.186401,PhysRevLett.123.256402,
PhysRevLett.125.166801,PhysRevB.99.041301,PhysRevLett.124.156601,WANG2022788,
PhysRevB.104.224303,PhysRevLett.128.127601,LI2021,2209.14811}.
Recently, boundary-obstructed HOTPs with nontrivial bulk topological invariants have been unveiled for the ${\rm A}$ (without chiral symmetry) and ${\rm AIII}$ (with chiral symmetry) classes~\cite{LI2021,PhysRevLett.128.127601} of the Altland-Zirnbauer (AZ) tenfold classes \cite{doi:10.1063/1.531675,PhysRevB.55.1142,Ryu_2010,RevModPhys.88.035005}, broadening the scope of topological matters for these two complex symmetry classes.
On the other hand, complete classification involves another two non-spatial symmetries, namely, time-reversal symmetry $\mathcal{T}$ and particle-hole symmetry $\mathcal{C}$.
These two anti-unitary symmetries not only give rise to the eight real symmetry classes of the AZ classification,
but are also tightly related to some intriguing properties of the systems.
For instance, a time-reversal symmetric system with half-integer spin ensures Kramers degeneracy of its eigenvalues,
and particle-hole symmetry naturally arises in the Bogoliubov-de Gennes Hamiltonian describing superconductors and superfluids, which allows for the passibility of inducing zero-energy Majorana modes~\cite{Kitaev_2001,RevModPhys.87.137}.
For intrinsic HOTPs, symmetry classification has been established with consideration of both the AZ classes and extra crystalline symmetries~\cite{PhysRevB.97.205135,PhysRevB.97.205136,PhysRevX.9.011012,PhysRevB.99.085127,PhysRevB.101.085137}.
Nonetheless, a universal method for constructing and characterizing boundary-obstructed HOTPs with nontrivial bulk topological invariants for the full AZ classes is still absent in contemporary literature.

The starting point of this paper is a method known as nested band inversion surfaces (BISs), which has been employed to explore HOTPs in A and AIII classes of the AZ classes \cite{LI2021}.
%
This method is built on the concept of BISs, which offers a powerful and innovative approach to probe first-order topological invariants through quantum dynamics~\cite{ZHANG20181385,PRXQuantum.2.020320,LI20211817}.
%
%
Inhabiting this advantage, the nested-BIS method allows us to construct higher-order topology from the first-order one of different parts of the system's Hamiltonian.
To give an overview of HOTPs for the full AZ classes in arbitrary dimensions, we start from a systematic analysis of minimal continuous models supporting HOTPs in each symmetry class.
The characterization of these HOTPs based on the nested-BIS method is then given by considering several 2D and 3D lattice examples converted from the continuous models.
%
Specifically, $n$th-order topological phases with $Z$-type topology can be categorized as $Z^n$ and $2Z^n$ classes, both characterized by $n$-fold $Z$ topological invariants, while the number of higher-order boundary states will be double for the latter one.
On the other hand, HOTPs with $Z_2$ topology are characterized by a $Z_2$ invariant and a set of $Z$ invariants, denoted as $Z_2\times Z^{n-1}$ classes accordingly.
These topological indexes are shifted between different rows or columns in the symmetry classification table for different $n$ (orders of topology), meaning that many HOTPs would be identified as trivial phases, or phases with different types of topology, in the conventional (first-order) topological band theory.
%
%
Finally, we extend our discussion to two scenarios beyond the standard nested-BIS method, namely, asymmetric parameters along different directions may lead to crossed BISs, and a nested relation can only be recovered by taking into account high-order BISs \cite{PRXQuantum.2.020320}. Second, the standard nested-BIS method requires the Hamiltonian to be formed by Clifford operators \cite{LI2021}, and a generalization of this method for systems with certain non-Clifford operators is established with extra effective surface BISs.
Interestingly, in both cases, certain spatial symmetries are broken by the extra modulations,
resulting in asymmetric properties of higher-order boundary states in our systems.
%

The rest of this paper is organized as follows. In Sec.~\ref{Sec2},
we review the nested-BIS method and establish a general Hamiltonian for HOTPs.
Section~\ref{Sec3} presents the symmetry classification and the continuous Hamiltonians of HOTPs characterized by integer topological invariants.
The corresponding minimal lattice Hamiltonians for several examples are provided in Sec~\ref{Sec4}.
Next we derive HOTPs with a $Z_2$ topological invariant in Sec~\ref{Sec5}, which complete the five topologically nontrivial classes in every spatial dimension of the AZ classification table.
In Sec~\ref{Sec6}, we explore asymmetric properties of HOTPs, and extend the nested-BIS to Hamiltonians beyond the Clifford algebra.
Lastly, a brief summary and discussion of our results are given in Sec.~\ref{Sec7}.

\section{The nested band inversion surfaces\label{Sec2}}

We begin with an introduction of the nested-BIS method for constructing and characterizing HOTPs with arbitrary orders of topology in arbitrary dimensions
~\cite{LI2021}.
This method is built on the BISs and high-order BISs used to dynamically characterize the first-order topological phases, where $m$th-order BISs are given by a special region in the Brillouin zone with $m$ vanishing pseudo-spin components of the Hamiltonian, denoted as $m$-BISs~\cite{ZHANG20181385,PRXQuantum.2.020320}.
Following the nested-BIS method,
a $d$D Hamiltonian hosting $n$th-order topology can be constructed as
\begin{eqnarray}\label{ReHa0}
H=\mathbf{h}(\mathbf{k})\cdot\bm{\gamma}=\sum_j^Jh_j(\mathbf{k})\gamma_j,
\end{eqnarray}
where $\gamma_j$ are the operators from the Clifford algebra satisfying $\{\gamma_i,\gamma_j\}=2\delta_{i,j}$, $\mathbf{k}=(k_1,k_2,...,k_d)$ is the $d$D momentum, and $J=d+n$.
We assume that $k_i$ is only contained in $(h_1,h_2,...,h_{2i})$ for $\forall i<n$
(in principle, this condition can always be satisfied through some rotations and deformations for systems with only nearest neighbor hoppings),
e.g. $k_1$ is contained only in $h_1$ and $h_2$,
and dividing the Hamiltonian of Eq.~(\ref{ReHa0}) into $n-1$ two-component Hamiltonian terms and one ($d-n+2$)-component Hamiltonian term:
\begin{eqnarray}\label{ReHadi}
H_1(\mathbf{k})=&&h_1(\mathbf{k})\gamma_1+h_2(\mathbf{k})\gamma_2, \nonumber \\
H_2(\mathbf{k}_{1,\parallel})=&&h_3(\mathbf{k}_{1,\parallel})\gamma_3+h_4(\mathbf{k}_{1,\parallel})\gamma_4, \nonumber \\
&&...\nonumber \\
H_{n-1}(\mathbf{k}_{n-2,\parallel})=&&h_{2n-3}(\mathbf{k}_{n-2,\parallel})\gamma_{2n-3}\nonumber \\
&&+h_{2n-2}(\mathbf{k}_{n-2,\parallel})\gamma_{2n-2}, \nonumber \\
H_{n}(\mathbf{k}_{n-1,\parallel})=&&\sum_{j=2n-1}^{J}h_{j}(\mathbf{k}_{n-1,\parallel})\gamma_{j},
\end{eqnarray}
where $\mathbf{k}_{i,\parallel}=(k_{i+1},k_{i+2},....,k_d)$.
We note that every sub-Hamiltonian supports first-order topology. 
More specifically, 
we choose each $H_j$ with $j=1,..,n-1$ to have two components so it can support surface states along $k_j$-direction.
On the other hand, the last sub-Hamiltonian $H_n$ with $(d-n+2)$ components can support first order topology in a $(d-n+1)$D subsystem. In total we need $J=2(n-1)+(d-n+2)=d+n$ anticommuting terms to assure $n$th-order topological boundary states.
With this construction, we can now define a BIS and a topological invariant for each $H_i$, and the overall $n$th-order topology can be characterized by the nested relation of these BISs and the collection of these topological invariants, as elaborated below.

%
For each two-component Hamiltonian $H_i$ with $i<n$, a winding number $v_i$ can be defined as a function of the $(d-i)$D momentum $\mathbf{k}_{i,\parallel}$:
\begin{eqnarray}\label{winding}
v_i(\mathbf{k}_{i,\parallel})=\frac{1}{2\pi}\oint_{k_i}\frac{h_{2i-1}dh_{2i}-h_{2i}dh_{2i-1}}{h^2_{2i-1}+h^2_{2i}}.
\end{eqnarray}
The ($d-i$)D surface Brillouin zone (BZ) expanded by $\mathbf{k}_{i,\parallel}$ is then divided into distinct regimes with different values of $v_i$.
The boundaries between these regimes are given by $h_{2i-1}=h_{2i}=0$, which by definition are a second-order BIS (2-BIS) of the system~\cite{PRXQuantum.2.020320}, denoted as $S_i^{m=2}$ with $m$ specifying the order of the BIS.
%
Finally, another topological invariant $v_n$ is defined for the $(d-n + 1)D$ Hamiltonian $H_n$ and can be extracted from the pseudospin textures at its BIS $S_n^m$~\cite{ZHANG20181385}.
Note that in contrast to the rest of the 2-BIS $S_i^{m=2}$ with $i<n$, the order of $S_n^m$ is not specified, for reasons that will become clear later.

\begin{figure}
\includegraphics[width=1\columnwidth]{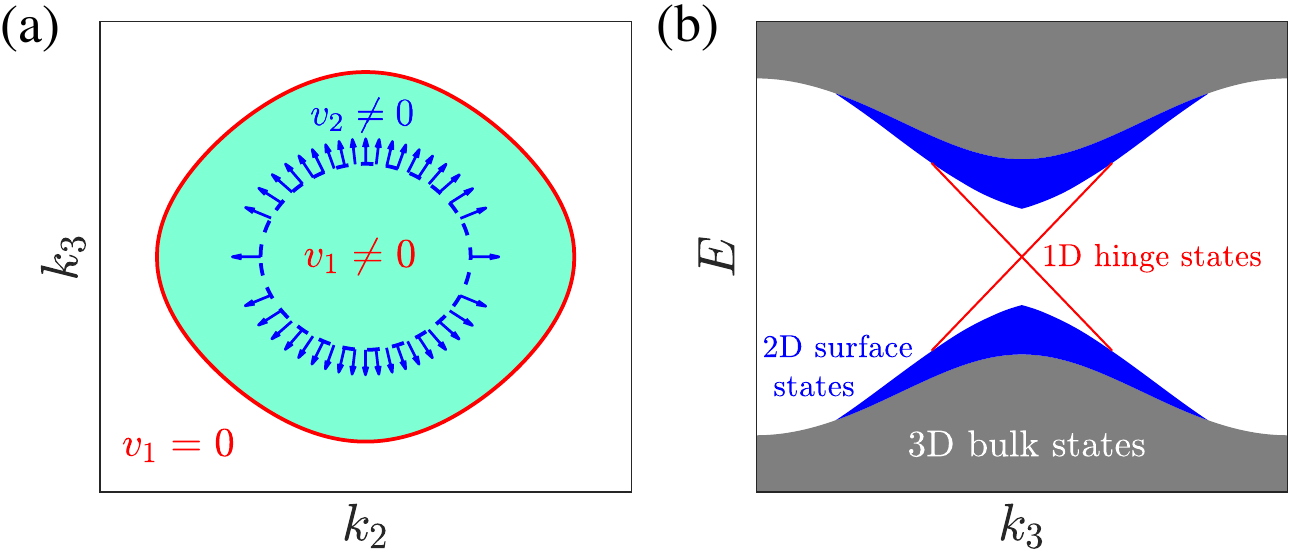}
\caption{\label{illu}(color online) (a) An illustration of nested BISs for a $3$D second-order topological phase. Here the red solid (blue dashed) loop indicates the BIS $S_1^{m=2}$ ($S_2^{m=1}$), and the blue arrows are the normalized pseudospin texture along $S_2^{m=1}$ for $H_2$ [as in Eq. \eqref{ReHadi} with $n=2$]. (b) An illustration of corresponding energy spectrum, with OBCs taken along $r_1$- and $r_2$ directions of the system.
}
\end{figure}

With these preparations, we are now able to examine the topological properties of the system order by order.
We first take $n=2$ as an example with a sketched illustration in Fig. \ref{illu}.
Since $k_1$ is contained only in $H_1$, and the rest of the Hamiltonian anti-commutes with $H_1$,
a nontrivial $v_1(\mathbf{k}_{1,\parallel})$ indicates that the system supports first-order boundary states at momentum $\mathbf{k}_{1,\parallel}$ under the open boundary condition (OBC) along  $r_1$-direction, whose eigenenergies are determined by $H_2$~\cite{PhysRevB.83.125109,PhysRevB.96.235424}.
Furthermore, the bulk topological properties of $H_2$ can also be inherited by these first-order boundary states of $H_1$, provided a BIS of $H_2$ (i.e., $S^m_2$) falls in the regime with a nontrivial $v_1$~\cite{LI2021}.
In this way, first-order topology of $H_2$ is manifested as second-order topology of $H=H_1+H_2$, and the system supports second-order topological boundary states characterized by both $v_1$ and $v_2$.
In the example in Fig. \ref{illu}(a), $S^{m=1}_2$ (blue dashed loop) is ``nested" in the regime with a nontrivial  $\nu_1$ (green area) bounded by $S^{m=2}_1$ (red loop), and the pseudo-spin texture exhibits a nontrivial winding along $S^{m=1}_2$ (blue arrows), described by a winding number $v_2$. Consequently, 1D chiral-like hinge states will appear in this 3D system under the OBC, as illustrated in Fig. \ref{illu}(b).

In the above procedure, introducing the 2-BIS $S^{m=2}_1$ reduces both the dimension and the order of topology by $1$ for the effective Hamiltonian $H-H_1=H_2$. In other words, the first-order topology of $H_2$ is captured by the first-order surface states of $H_1$, thus giving rise to second-order topological boundary states of the overall system.
For $n>2$,
this procedure can be repeatedly applied to the Hamiltonian until a first-order topological Hamiltonian $H_n$ is obtained.
For example, for $n=3$, the first-order topology of $H_3$ is captured by the first-order surface states of $H_2$, and the resultant second-order topology of this subsystem is further captured by the first-order surface states of $H_1$, leading to third-order topological boundary states of the overall system.
The inheritance of the topology only requires that each $S^m_i$ falls within the nontrivial regime associated with $S^m_{i-1}$.
Finally, since the order of topology cannot be further reduced, the last BIS $S_n^m$ is defined only for topologically characterized $H_n$ and is not restricted to a second-order one.

%
%
%
%

%
%
%

\section{classification of higher-order topological phases with $Z$ invariants\label{Sec3}}
Based on the time-reversal, particle-hole, and chiral symmetries,
the tenfold AZ classification provides a systematic scheme for analyzing various topological properties.
%
%
In this section, we focus on AZ classes with Z-type topological invariants, and construct corresponding continuous Hamiltonians for HOTPs in the form of Eq. \eqref{ReHadi}.
We will first give a brief overview of the two complex classes (AIII and A) discussed in Ref. \cite{LI2021}, then analyze the other eight real classes in more detail.
This analysis allows us to further generate lattice models of HOTPs characterized by the nested BISs in different classes, with examples given in Sec~\ref{Sec4}.

\subsection{Higher-order topological phases for the complex symmetry classes\label{Sec31}}
The nested-BIS method requires the Hamiltonian to satisfy Clifford algebra, which naturally leads to the presence or absence of the Chiral symmetry,
\begin{eqnarray}
\mathcal{S} H(\mathbf{k}) \mathcal{S}^{-1}=-H(\mathbf{k}),
\end{eqnarray}
with $\mathcal{S}$ an unitary operator,
and the system falls into the two complex symmetry classes AIII and A, respectively.
%
%
To see this, consider the Hamiltonian of Eqs.~\eqref{ReHa0} and ~\eqref{ReHadi}, which requires $d+n$ operators $\gamma_j$ from Clifford algebra to generate $d$D $n$-th HOTPs.
These operators can be given by the Kronecker product of $p$ sets of Pauli matrices and the $2\times2$ identity matrix, which generates a set of $2^p\times 2^p$ matrices, where at most $2p+1$ of them anti-commute with each other.
Therefore, for $d+n=2p$, there is one extra term absent in the Hamiltonian, acting as the chiral symmetry operator $\mathcal{S}$ for the system, and for $d+n=2p+1$, no such a symmetry operator can be defined, unless some extra degrees of freedom are introduced to the system (which increases the dimension of the Hamiltonian matrix).
Note that in these two classes, there is no further restriction of the exact coefficients of $\gamma_j$, since the chiral symmetry does not involve different momenta $\mathbf{k}$.
%

%
%
In our construction of the Hamiltonian in Eq.~\eqref{ReHadi}, we can calculate the winding number $v_i$ with $i=1,2,...,n-1$ for every two-component subsystem $H_i$ through Eq.~\eqref{winding}.
The last effective $(d-n+1)$D Hamiltonian $H_n$ contains ($d-n+2$) anti-communting terms, meaning it describes either an odd-dimensional system of class AIII with a chiral symmetry, or an even-dimensional system of class A without a chiral symmetry,
both support $Z$-type first-order topology~\cite{doi:10.1063/1.531675,PhysRevB.55.1142,RevModPhys.88.035005,Ryu_2010}.
Therefore, another $Z$-class topological invariant $v_n$ can be defined through the spin texture at a BIS of $H_n$.
As discussed in Sec.~\ref{Sec2}, nontrivial HOTPs are induced when the BISs of these Hamiltonian form the nesting relation,
thus these HOTPs can be characterized by these $n$ $Z$-class topological invariants, indexed as $Z^n$, which also corresponds to the number of boundary states at an $n$-th boundary.
We emphasize that for even $n$, the AIII (A) class can hold HOTPs in even (odd) dimensions, which is topologically trivial in the conventional
topological band theory.
The classification and topological invariants of second-order topological phases for the two complex classes are shown in the first two rows of Table~\ref{Classification}.
%

\begin{table*}
\centering
\caption{Classification of second-order topological phases based on the nested-BIS method, according to the presence or absence of time-reversal,
particle-hole, and chiral symmetries, labeled by $\mathcal{T}$, $\mathcal{C}$, and $\mathcal{S}$ repsectively in the table. ``$\pm$" represents different types of anti-unitary symmetries square to $\pm1$.
The continuous Hamiltonians for topological phases characterized by two $Z$-invariants are provided with $H^{({I})}$ in Eq.~\eqref{H_1}, $H^{(II)}$ in Eq.~\eqref{H_2}, $H^{(III)}$ in Eq.~\eqref{H_3}, and $H^{(IV)}$ in Eq.~\eqref{H_4}.
Note that in order to support second-order topology and corresponding boundary states in a lattice model, the spatial dimension must be $d\geqslant2$.}
\label{Classification}
\begin{tabular}{|c||c|c|c||c|c|c|c|c|c|c|c|}
   \hline
    \multirow{2}{*}{Class}&\multicolumn{3}{c||}{Symmetry}& \multicolumn{8}{c|}{Dimension $d$ (mod $8$)}\\ \cline{2-12}
    & ~$\mathcal{T}$~ & ~$\mathcal{C}$~ & ~$\mathcal{S}$~ & 0 & 1 & 2 & 3 & 4 & 5& 6 & 7  \\
\hhline{|============|}
${\rm A}$   & 0 & 0 & 0 & 0 & $Z\times Z$ & 0 & $Z\times Z$ & 0&  $Z\times Z$ & 0 & $Z\times Z$  \\
\hline
${\rm AIII}$   & 0 & 0 & $1$           & $Z\times Z$ & 0 & $Z\times Z$ & 0 & $Z\times Z$ & 0 &$Z\times Z$ & 0  \\
\hline
${\rm AI}$ & $+$ & 0 & 0 & $Z\times Z_2$ & \tabincell{c}{$Z\times Z$\\$H^{(\uppercase\expandafter{\romannumeral1})}$} & 0 & 0 & 0& \tabincell{c}{$2Z\times Z$\\$H^{(\uppercase\expandafter{\romannumeral3})}$} & 0 & $Z\times Z_2$  \\
\hline
${\rm BDI}$   & $+$ & $+$ & $1$                                  & $Z\times Z_2$ & $Z\times Z_2$ & \tabincell{c}{$Z\times Z$\\$H^{(\uppercase\expandafter{\romannumeral2})}$} & 0 & 0 & 0& \tabincell{c}{$2Z\times Z$\\$H^{(\uppercase\expandafter{\romannumeral4})}$} & 0  \\
\hline
{\rm D}                           & 0 & $+$ & 0         & 0 & $Z\times Z_2$ & $Z\times Z_2$ & \tabincell{c}{$Z\times Z$\\$H^{(\uppercase\expandafter{\romannumeral1})}$} & 0 & 0& 0 & \tabincell{c}{$2Z\times Z$\\$H^{(\uppercase\expandafter{\romannumeral3})}$}  \\
\hline
${\rm DIII}$      & $-$ & $+$ & $1$     & \tabincell{c}{$2Z\times Z$\\$H^{(\uppercase\expandafter{\romannumeral4})}$} & 0 & $Z\times Z_2$ & $Z\times Z_2$ & \tabincell{c}{$Z\times Z$\\$H^{(\uppercase\expandafter{\romannumeral2})}$} & 0& 0 & 0  \\
\hline
${\rm AII}$      & $-$ & 0 & 0        & 0 &\tabincell{c}{$2Z\times Z$\\$H^{(\uppercase\expandafter{\romannumeral3})}$} & 0 & $Z\times Z_2$ & $Z\times Z_2$ & \tabincell{c}{$Z\times Z$\\$H^{(\uppercase\expandafter{\romannumeral1})}$}& 0 & 0  \\
\hline
${\rm CII}$      & $-$ & $-$ & $1$      & 0 & 0 & \tabincell{c}{$2Z\times Z$\\$H^{(\uppercase\expandafter{\romannumeral4})}$} & 0 & $Z\times Z_2$ & $Z\times Z_2$& \tabincell{c}{$Z\times Z$\\$H^{(\uppercase\expandafter{\romannumeral2})}$} & 0  \\
\hline
${\rm C}$                               & 0 & $-$ & 0        & 0 & 0 & 0 & \tabincell{c}{$2Z\times Z$\\$H^{(\uppercase\expandafter{\romannumeral3})}$} & 0 & $Z\times Z_2$& $Z\times Z_2$ & \tabincell{c}{$Z\times Z$\\$H^{(\uppercase\expandafter{\romannumeral1})}$}  \\
\hline
${\rm CI}$      & $+$ & $-$ & $1$            & \tabincell{c}{$Z\times Z$\\$H^{(\uppercase\expandafter{\romannumeral2})}$}& 0 & 0 & 0& \tabincell{c}{$2Z\times Z$\\$H^{(\uppercase\expandafter{\romannumeral4})}$} & 0& $Z\times Z_2$ & $Z\times Z_2$  \\
\hline
\end{tabular}
\end{table*}

\subsection{Second-order topological phases for the real classes\label{Sec32}}
We now consider the HOTPs in real AZ symmetry classes described by integer topological invariants, and HOTPs containing $Z_2$ invariants will be discussed in Sec~\ref{Sec5}.
A Hamiltonian in these classes holds time-reversal symmetry $\mathcal{T}$ and/or particle-hole symmetry $\mathcal{C}$, with
\begin{eqnarray}
\mathcal{T} H(\mathbf{ k})\mathcal{T}^{-1}=H(-\mathbf{k}),~~
\mathcal{C} H(\mathbf{ k})\mathcal{C}^{-1}=-H(-\mathbf{k}).\label{TRS_PHS}
\end{eqnarray}
These anti-unitary symmetries have symmetry operators square to $+1$ or $-1$, giving rise to $3^2-1=8$ real symmetry classes in total.
In this subsection, we explore the classification of second-order topological phases and the corresponding continuous Hamiltonian.

To study the Hamiltonian with anti-unitary symmetry conveniently, we define the $2p+1$ anti-commuting $2^p\times2^p$ matrices from Clifford algebra as
\begin{eqnarray}\label{Clifford}
\Gamma_{(2p+1)}^{2a-1}&&=\underbrace{\sigma_z\bigotimes...\sigma_z}_{a-1}\bigotimes\sigma_x\bigotimes\underbrace{\sigma_0\bigotimes...\sigma_0}_{p-a}, \nonumber \\
\Gamma_{(2p+1)}^{2a}&&=\underbrace{\sigma_z\bigotimes...\sigma_z}_{a-1}\bigotimes\sigma_y\bigotimes\underbrace{\sigma_0\bigotimes...\sigma_0}_{p-a},  \nonumber \\
\Gamma_{(2p+1)}^{2p+1}&&=\sigma_z\bigotimes\sigma_z...\bigotimes\sigma_z\bigotimes\sigma_z\bigotimes...\sigma_z
\end{eqnarray}
with $a=1,2,...,p$.
In this representation, $\Gamma_{(2p+1)}^{\alpha}$ is purely real (imaginary) when $\alpha$ is odd (even), similar to the one used in Ref.~\cite{Ryu_2010}.
Thus, an anti-unitary operator can be defined as
\begin{eqnarray}\label{antiuA0}
\hat{A}_0:=\Pi_{\alpha=0}^{p}\Gamma_{(2p+1)}^{2\alpha+1}\mathcal{K}
\end{eqnarray}
where $\mathcal{K}$ is the complex conjugate operator.
For a Hamiltonian formed by the matrices in Eq. \eqref{Clifford},
$\hat{A}_0$ represents different symmetries ($\mathcal{T}$ or $\mathcal{C}$ with $\pm$ in Table \ref{Classification}) for different values of $p$,
as
it satisfies
\begin{eqnarray}\label{squareA0}
\hat{A}_0^2=&&(-1)^{p(p+1)/2}, \nonumber \\
\hat{A}_0\Gamma_{(2p+1)}^{\alpha}\hat{A}_0^{-1}=&&(-1)^{p}\Gamma_{(2p+1)}^{\alpha}.
\end{eqnarray}

\subsubsection{Gapless Hamiltonian as a starting point}
We begin by writing a $(2p+1)$D gapless Dirac Hamiltonian,
\begin{eqnarray}\label{gapless}
H^{(0)}(\mathbf{k})=\sum_{\alpha=1}^{2p+1}k_{\alpha}\Gamma_{(2p+1)}^{\alpha}.
\end{eqnarray}
It satisfies an anti-unitary symmetry,
\begin{eqnarray}\label{symgapless}
\hat{A}_0H^{(0)}(\mathbf{k})\hat{A}_0^{-1}=(-1)^{p+1}H^{(0)}(-\mathbf{k}),
\end{eqnarray}
which represents the time-reversal (particle-hole) symmetry when $p$ is odd (even).
%
The symmetry class of $H^{(0)}(\mathbf{k})$ is determined by further considering the square of $\hat{A}_0$ calculated through Eq.~(\ref{squareA0}). That is, when $p$ changes from $0$ to $3$, we get a series of odd-dimensional Hamiltonians with symmetry classes ${\rm D}\rightarrow{\rm AII}\rightarrow{\rm C}\rightarrow{\rm AI}$,
with an eightfold periodicity of the spatial dimension $d=2p+1$.
%
%

\subsubsection{Second-order $Z$-class Hamiltonian without chiral symmetry}
Starting from the ($2p+1$)D gapless Dirac Hamiltonian in~Eq. \eqref{gapless}, a ($2p$)D gapped phase can be  obtained by replacing one momentum component with a mass term $m_1$.
Nontrivial first-order topology may arise if $m_1$ takes different signs at different high-symmetric points and generate nontrivial winding of the Hamiltonian vector throughout the ($2p$)D Brillouin zone.
On the other hand,
a Hamiltonian in the form of Eq.~\eqref{ReHadi} supporting second-order topological phases can be divided into two subsystems with nontrivial first-order topology.
Therefore we convert two terms with real matrices 
from Eq.~\eqref{gapless} to mass terms ($k_1\rightarrow m_1$ and $k_3\rightarrow m_2$) and rewrite the resultant $(2p-1)$D Hamiltonian as $H^{({\uppercase\expandafter{\romannumeral1}})}(\mathbf{k})=H^{({\uppercase\expandafter{\romannumeral1}})}_1(\mathbf{k})+H^{({\uppercase\expandafter{\romannumeral1}})}_2(\mathbf{k})$
\begin{eqnarray}\label{H_1}
H^{({\uppercase\expandafter{\romannumeral1}})}_1(\mathbf{k})=&&m_1\Gamma_{(2p+1)}^{1}+k_{1}\Gamma_{(2p+1)}^{2}, \nonumber \\
H^{({\uppercase\expandafter{\romannumeral1}})}_2(\mathbf{k}_{1,\parallel})=&&m_2\Gamma_{(2p+1)}^{3}+\sum_{\alpha=2}^{2p-1}k_{\alpha}\Gamma_{(2p+1)}^{\alpha+2}.
\end{eqnarray}
Note the momentum components are reindexed to have $\alpha$ ranging from $1$ to $2p-1$ for $k_\alpha$.
The anti-unitary symmetry of Eq.~\eqref{antiuA0} is broken by these mass terms, but another one emerges for the system, given by
\begin{eqnarray}\label{squareA1}
\hat{A}_1H^{({\uppercase\expandafter{\romannumeral1}})}(\mathbf{k})\hat{A}_1^{-1}=&&(-1)^{p-1}H^{({\uppercase\expandafter{\romannumeral1}})}(-\mathbf{k}), \nonumber \\
\hat{A}_1^2=&&(-1)^{(p-1)(p-2)/2},
\end{eqnarray}
where
$\hat{A}_1=\Gamma_{(2p+1)}^{3}\Gamma_{(2p+1)}^{1}\hat{A}_0$.
It is straightforward to see that $\hat{A}_1$ also represents different symmetries for different $p$,
and the corresponding symmetry class changes periodically as ${\rm AI}\rightarrow{\rm D}\rightarrow{\rm AII}\rightarrow{\rm C}$
when $p$ increases from $1$ to $4$, as shown in Table \ref{Classification}.

To unveil the topological characterization of $H^{(I)}(\mathbf{k})$, note that $H_1^{(I)}(\mathbf{k})$ can be viewed as a 1D Hamiltonian of $k_1$, whose lattice counterpart can be characterized by a winding number defined as in Eq. \eqref{winding}.
Meanwhile, $H_2^{(I)}(\mathbf{k}_{1,\parallel})$ is a ($2p-2$)D Hamiltonian with ($2p-1$) anti-commuting terms, belonging to the same symmetry class as that of $H^{(I)}(\mathbf{k})$ since they share the same symmetry conditions \footnote{A chiral symmetry seem to emerge for $H_2^{(I)}(\mathbf{k}_{1,\parallel})$ due to the absence of $\Gamma^{1,2}_{(2p+1)}$, yet the Hilbert space of this effective Hamiltonian can be reduced to $2^{p-1}$ dimension, where the chiral symmetry is ruled out.}.
According to the standard AZ class, this $H_2^{(I)}(\mathbf{k}_{1,\parallel})$ is also characterized by a $Z$ invariant.
Thus the total system $H^{(I)}(\mathbf{k})$ is characterized two $Z$ invariants, indexed as $Z\times Z$ in Table \ref{Classification}.

%

\subsubsection{ Second-order $Z$-class Hamiltonian with chiral symmetry}
The Hamiltonian in Eq.~\eqref{H_1} includes a full set of $2^p\times 2^p$ anti-commuting matrices defined in Eq. \eqref{Clifford},
which excludes chiral symmetry for the system.
To constructed a chiral-symmetric Hamiltonian, we remove another term $k_{2p-1}\Gamma_{(2p+1)}^{2p+1}$ from Eq.~(\ref{H_1}), and obtain a $(2p-2)$D Hamiltonian $H^{({\uppercase\expandafter{\romannumeral2}})}(\mathbf{k})=H^{({\uppercase\expandafter{\romannumeral2}})}_1(\mathbf{k})
+H^{({\uppercase\expandafter{\romannumeral2}})}_2(\mathbf{k}_{1,\parallel})$, with
\begin{eqnarray}\label{H_2}
H^{({\uppercase\expandafter{\romannumeral2}})}_1(\mathbf{k})=&&m_1\Gamma_{(2p+1)}^{1}+k_{1}\Gamma_{(2p+1)}^{2}, \nonumber \\
H^{({\uppercase\expandafter{\romannumeral2}})}_2(\mathbf{k}_{1,\parallel})=&&m_2\Gamma_{(2p+1)}^{3}+\sum_{\alpha=2}^{2p-2}k_{\alpha}\Gamma_{(2p+1)}^{\alpha+2}.
\end{eqnarray}
This Hamiltonian exhibits the same anti-unitary symmetry of $\hat{A}_1$ in Eq.~\eqref{squareA1}, and also a chiral symmetry as the $\Gamma_{(2p+1)}^{2p+1}$ term is removed:
\begin{eqnarray}\label{chiral}
\Gamma_{(2p+1)}^{2p+1}H^{({\uppercase\expandafter{\romannumeral2}})}(\mathbf{k})\Gamma_{(2p+1)}^{2p+1}
=-H^{({\uppercase\expandafter{\romannumeral2}})}(\mathbf{k}).
\end{eqnarray}
Combining these two symmetries, another anti-unitary symmetry arises for the system,
\begin{eqnarray}\label{squareA2}
\hat{A}_2H^{({\uppercase\expandafter{\romannumeral2}})}(\mathbf{k})\hat{A}_2^{-1}=&&(-1)^{p}H^{({\uppercase\expandafter{\romannumeral2}})}(-\mathbf{k}), \nonumber \\
\hat{A}_2^2=&&(-1)^{(p-2)(p-3)/2},
\end{eqnarray}
with
\begin{eqnarray}\label{antiuA2}
\hat{A}_2=\Gamma_{(2p+1)}^{2p+1}\hat{A}_1.
\end{eqnarray}
When $p$ increases from $1$ to $4$, the symmetry class for this $(2p-2)$D Hamiltonian will change in the sequence of
${\rm CI}\rightarrow{\rm BDI}\rightarrow{\rm DIII}\rightarrow{\rm CII}$, as shown in Table~\ref{Classification}.
Similarly to the previous case, here $H_1^{(II)}(\mathbf{k})$ is associated with a 1D winding topology, and $H_2^{(II)}(\mathbf{k}_{2,\parallel})$ corresponds to a ($2p-3$)D system of the same symmetry class as $H^{(II)}(\mathbf{k})$, which is characterized by a $Z$ invariant.
Therefore the overall Hamiltonian also possesses $Z\times Z$-type second-order topology.

\subsubsection{Second-order $2Z$-class Hamiltonian without chiral symmetry}\label{sec:2Z_non_chiral}
So far, we have constructed Hamiltonians for two classes (one real and one complex) in each spatial dimension, where the second-order topology stems from the hybridization of a 1D winding topology and a first-order $Z$-type topology of the effective Hamiltonian $H_2(\mathbf{k}_{1,\parallel})$ in the same symmetry class but with one spatial dimension less than the total system.
Based on the standard AZ classification, there is another class with 1st-order topology characterized by a $2Z$ invariant in every spatial dimension \cite{Ryu_2010},
which can also be used to build HOTPs.

To construct such a second-order $2Z$-class Hamiltonian in the absence of chiral symmetry,
we replace four terms of the gapless Hamiltonian of Eq.~\eqref{gapless},
$k_{\alpha}\Gamma_{2n+1}^{\alpha}$ with $\alpha=1,2,3,5$, with two mass terms
$-im_1\Pi_{\alpha=1}^{3}\Gamma_{2n+1}^{\alpha}$ and $m_{2}\Gamma_{2n+1}^{5}$,
and obtain a gapped $(2p-3)$D Hamiltonian $H^{({\uppercase\expandafter{\romannumeral3}})}(\mathbf{k})=H^{({\uppercase\expandafter{\romannumeral3}})}_1(\mathbf{k})
+H^{({\uppercase\expandafter{\romannumeral3}})}_2(\mathbf{k}_{1,\parallel})$, with
\begin{eqnarray}\label{H_3}
H^{({\uppercase\expandafter{\romannumeral3}})}_1(\mathbf{k})=&&-im_1\Pi_{\alpha=1}^{3}\Gamma_{(2p+1)}^{\alpha}+k_{1}\Gamma_{(2p+1)}^{4}, \nonumber \\
H^{({\uppercase\expandafter{\romannumeral3}})}_2(\mathbf{k}_{1,\parallel})=&&m_2\Gamma_{(2p+1)}^{5}+\sum_{\alpha=2}^{2p-3}k_{\alpha}\Gamma_{(2p+1)}^{\alpha+4}
\end{eqnarray}
with $k_\alpha$ reindexed from $k_1$ to $k_{2p-3}$.
This Hamiltonian holds an anti-unitary symmetry
\begin{eqnarray*}\label{antiuA3}
\hat{A}_3=\Gamma_{(2p+1)}^{5}\hat{A}_0,
\end{eqnarray*}
with
\begin{eqnarray*}\label{squareA3}
\hat{A}_3H^{({\uppercase\expandafter{\romannumeral3}})}(\mathbf{k})\hat{A}_3^{-1}=&&(-1)^{p}H^{({\uppercase\expandafter{\romannumeral3}})}(-\mathbf{k}), \nonumber \\
\hat{A}_3^2=&&(-1)^{p(p-1)/2}.
\end{eqnarray*}
When $p$ changes from $2$ to $5$, this $(2p-3)$D Hamiltonian's symmetry class changes as ${\rm AII}\rightarrow{\rm C}\rightarrow{\rm AI}\rightarrow{\rm D}$, as shown in Table~\ref{Classification}.

To unveil the topological properties of this Hamiltonian, notice that the extra unitary matrix introduced for the first mass term satisfies
$-i\Pi_{\alpha=1}^{3}\Gamma_{2n+1}^{\alpha}=\sigma_0\otimes\Gamma_{(2p-1)}^{1}$.
Thus we rewrite the Hamiltonian of Eq. \eqref{H_3} as
\begin{eqnarray*}\label{H3dia1}
H^{({\uppercase\expandafter{\romannumeral3}})}_1(\mathbf{k})=&&m_1\sigma_0\otimes\Gamma_{(2p-1)}^{1}+k_{1}\sigma_z\otimes\Gamma_{(2p-1)}^{2}, \nonumber \\
H^{({\uppercase\expandafter{\romannumeral3}})}_2(\mathbf{k}_{1,\parallel})=&&m_2\sigma_z\otimes\Gamma_{(2p-1)}^{3}+\sum_{\alpha=2}^{2p-3}k_{\alpha}\sigma_z\otimes\Gamma_{(2p-1)}^{\alpha+2}, \nonumber \\
\end{eqnarray*}
which can be reduced to the direct sum of two $H^{(I)}$-type Hamiltonians in a $2^{p-1}\times 2^{p-1}$ Hillbert space:
\begin{eqnarray}
H^{({\uppercase\expandafter{\romannumeral3}})}_{2^p\times2^p}(\mathbf{k})&=&H^{({\uppercase\expandafter{\romannumeral1}})}_{2^{p-1}\times2^{p-1}}(\mathbf{k},m_1,m_2)\nonumber\\
&&\oplus H^{({\uppercase\expandafter{\romannumeral1}})}_{2^{p-1}\times2^{p-1}}(-\mathbf{k},m_1,-m_2).\nonumber
\end{eqnarray}
 In addition, the two $H^{(I)}$-type Hamiltonians can be mapped to each other through a unitary transformation,
 $\Gamma_{(2p-1)}^{1}H^{(I)}_{2^{p-1}\times2^{p-1}}(\mathbf{k},m_1,m_2)\Gamma_{(2p-1)}^{1}=H^{(I)}_{2^{p-1}\times2^{p-1}}(-\mathbf{k},m_1,-m_2)$.
Consequently, the original Hamiltonian $H^{({\uppercase\expandafter{\romannumeral3}})}(\mathbf{k})$ possesses two copies of the $Z\times Z$-type second-order topology of $H^{(I)}(\mathbf{k})$, indexed as $2Z\times Z$ in Table \ref{Classification}.

\subsubsection{Second-order $2Z$-class Hamiltonian with chiral symmetry}
Finally, by removing $k_{2p-3}\Gamma^{2p+1}_{(2p+1)}$ from Hamiltonian Eq.~(\ref{H_3}), we can obtain a $(2p-3)$D second-order $2Z$-class Hamiltonian
$H^{(IV)}(\mathbf{k})=H^{(IV)}_1(\mathbf{k})
+H^{(IV)}_2(\mathbf{k}_{1,\parallel})$
with chiral symmetry:
\begin{eqnarray}\label{H_4}
H^{({\uppercase\expandafter{\romannumeral4}})}_1(\mathbf{k})=&&-im_1\Pi_{\alpha=1}^{3}\Gamma_{(2p+1)}^{\alpha}+k_{1}\Gamma_{(2p+1)}^{4}, \nonumber \\
H^{({\uppercase\expandafter{\romannumeral4}})}_2(\mathbf{k}_{1,\parallel})=&&m_2\Gamma_{(2p+1)}^{5}+\sum_{\alpha=2}^{2p-4}k_{\alpha}\Gamma_{(2p+1)}^{\alpha+4}.
\end{eqnarray}

Obviously, this Hamiltonian keeps the anti-unitary symmetry of $\hat{A}_3$ in Eq.~\eqref{antiuA3} and the same chiral symmetry as in Eq. \eqref{chiral}, and their combination gives rise to another anti-unitary symmetry:
\begin{eqnarray*}\label{squareC}
\hat{A}_4H^{({\uppercase\expandafter{\romannumeral4}})}(\mathbf{k})\hat{A}_4^{-1}=&&(-1)^{p-1}H^{({\uppercase\expandafter{\romannumeral4}})}(-\mathbf{k}), \nonumber \\
\hat{A}_4^2=&&(-1)^{(p-1)(p-2)/2}
\end{eqnarray*}
with \begin{eqnarray*}\label{antiuC}
\hat{A}_4=\Gamma_{(2p+1)}^{2p+1}\hat{A}_3.
\end{eqnarray*}
When $p$ changes from $2$ to $5$, the symmetry class of this $(2p-4)$D Hamiltonian changes in the sequence of ${\rm DIII}\rightarrow {\rm CII}
\rightarrow {\rm CI}\rightarrow {\rm BDI}$, as shown in Table~\ref{Classification}.
Similar to the previous discussion of $H^{(III)}(\mathbf{k})$, $H^{(IV)}(\mathbf{k})$ can be reduced to two copies of $H^{(II)}(\mathbf{k})$ in a ($2^{p-1}$)D Hillbert space, and thus possesses $2Z\times Z$-type second-order topology.

%

\subsection{General higher-order topological phases with $Z$ invariants for the real classes\label{Sec33}}
Similarly, HOTPs with an arbitrary order of topology can be obtained from the gapless Hamiltonian Eq.~(\ref{gapless}) through converting different terms with real matrices into mass terms.
Here we list the general forms of Hamiltonians with HOTPs:

(1). $(2p-n+1)$D higher-order $Z$-class Hamiltonian without chiral symmetry:
\begin{eqnarray}\label{thH_1}
H_1^{(I)}(\mathbf{k})=&&m_1\Gamma_{(2p+1)}^{1}+k_{1}\Gamma_{(2p+1)}^{2},\nonumber \\
H_2^{(I)}(\mathbf{k}_{1,\parallel})=&&m_2\Gamma_{(2p+1)}^{3}+k_{2}\Gamma_{(2p+1)}^{4},\nonumber \\
....\nonumber \\
H_{n-1}^{(I)}(\mathbf{k}_{n-2,\parallel})=&&m_{n-1}\Gamma_{(2p+1)}^{2n-3}+k_{n-1}\Gamma_{(2p+1)}^{2n-2},\nonumber \\
H_{n}^{(I)}(\mathbf{k}_{n-1,\parallel})=&&m_{n}\Gamma_{(2p+1)}^{2n-1}+\sum^{2p-n+1}_{\alpha=n}k_\alpha\Gamma_{(2p+1)}^{n+\alpha};
\end{eqnarray}

(2). $(2p-n)$D higher-order $Z$-class Hamiltonian with chiral symmetry:
\begin{eqnarray}\label{thH_2}
H_1^{(II)}(\mathbf{k})=&&m_1\Gamma_{(2p+1)}^{1}+k_{1}\Gamma_{(2p+1)}^{2},\nonumber \\
H_2^{(II)}(\mathbf{k}_{1,\parallel})=&&m_2\Gamma_{(2p+1)}^{3}+k_{2}\Gamma_{(2p+1)}^{4},\nonumber \\
....\nonumber \\
H_{n-1}^{(II)}(\mathbf{k}_{n-2,\parallel})=&&m_{n-1}\Gamma_{(2p+1)}^{2n-3}+k_{n-1}\Gamma_{(2p+1)}^{2n-2},\nonumber \\
H_{n}^{(II)}(\mathbf{k}_{n-1,\parallel})=&&m_{n}\Gamma_{(2p+1)}^{2n-1}+\sum^{2p-n}_{\alpha=n}k_\alpha\Gamma_{(2p+1)}^{n+\alpha};
\end{eqnarray}

(3). $(2p-n-1)$D higher-order $2Z$-class Hamiltonian without chiral symmetry:
\begin{eqnarray}\label{thH_3}
H_1^{(III)}(\mathbf{k})=&&-im_1\Pi_{\alpha=1}^{3}\Gamma_{(2p+1)}^{\alpha}+k_{1}\Gamma_{(2p+1)}^{4},\nonumber \\
H_2^{(III)}(\mathbf{k}_{1,\parallel})=&&m_2\Gamma_{(2p+1)}^{5}+k_{2}\Gamma_{(2p+1)}^{6},\nonumber \\
....\nonumber \\
H_{n-1}^{(III)}(\mathbf{k}_{n-2,\parallel})=&&m_{n-1}\Gamma_{(2p+1)}^{2n-1}+k_{n-1}\Gamma_{(2p+1)}^{2n},\nonumber \\
H_{n}^{(III)}(\mathbf{k}_{n-1,\parallel})=&&m_{n}\Gamma_{(2p+1)}^{2n+1}+\sum^{2p-n-1}_{\alpha=n}k_\alpha\Gamma_{(2p+1)}^{n+\alpha+2};
\end{eqnarray}

(4). $(2p-n-2)$D higher-order $2Z$-class Hamiltonian with chiral symmetry:
\begin{eqnarray}\label{thH_4}
H_1^{(IV)}(\mathbf{k})=&&-im_1\Pi_{\alpha=1}^{3}\Gamma_{(2p+1)}^{\alpha}+k_{1}\Gamma_{(2p+1)}^{4},\nonumber \\
H_2^{(IV)}(\mathbf{k}_{1,\parallel})=&&m_2\Gamma_{(2p+1)}^{5}+k_{2}\Gamma_{(2p+1)}^{6},\nonumber \\
....\nonumber \\
H_{n-1}^{(IV)}(\mathbf{k}_{n-2,\parallel})=&&m_{n-1}\Gamma_{(2p+1)}^{2n-1}+k_{n-1}\Gamma_{(2p+1)}^{2n},\nonumber \\
H_{n}^{(IV)}(\mathbf{k}_{n-1,\parallel})=&&m_{n}\Gamma_{(2p+1)}^{2n+1}+\sum^{2p-n-2}_{\alpha=n}k_\alpha\Gamma_{(2p+1)}^{n+\alpha+2}.
\end{eqnarray}

Among these Hamiltonians, those in Eqs.~\eqref{thH_1} and~\eqref{thH_2} can be characterized by $n$ $Z$-class invariant and indexed by $Z^n$.
Each of the remaining two Hamiltonians in Eqs.~\eqref{thH_3} and~\eqref{thH_4} can be reduced to two $Z$-class Hamiltonians and indexed by $2Z^n$.
Obviously, these results with $n=1$ recovers conventional (first-order) topological phases,
and previous results of second-order topological phases can be recovered by setting $n=2$.
%
%
Note that when the order of topology $n$ increases by one and $p$ remains the same, the spatial dimension $d$ decreases by one for the above Hamiltonians, shifting all topological indexes one column to the left in the symmetry classification table.
Meanwhile, as detailed in Appendix  \ref{appA}, particle-hole symmetry with $\mathcal{C}^2=\pm1$ will convert to time-reversal symmetry with $\mathcal{T}^2=\pm1$,
and time-reversal symmetry with $\mathcal{T}^2=\pm1$ becomes particle-hole symmetry with $\mathcal{C}^2=\mp1$, shifting all topological indexes up two rows in the symmetry classification table.
Putting these results together, the AZ classification table for ($n+1$)th-order topological phases can be obtained from that for $n$th-order topological phases by shifting all topological indexes up one row. A general classification table for topological phases with arbitrary orders of topology can be obtained accordingly, as given in Appendix \ref{appA}.
%

%

\section{Lattice Hamiltonian and examples\label{Sec4}}
After obtaining the continuous Hamiltonian,
we now provide some specific lattice models to illustrate the HOTPs with nested BISs in 2D and 3D.
Note that a continuous Hamiltonian can be considered as the low-energy expansion near high symmetric points of a lattice Hamiltonian.
%
%
Explicitly, we apply the transformation:
\begin{eqnarray}\label{trans}
k_{\alpha}\rightarrow h_\alpha(\mathbf{k}):=&&\sin k_\alpha,\nonumber \\
m_\alpha\rightarrow M_\alpha(\mathbf{k}):=&&m_\alpha+\sum_{i=\alpha}^d(1-\cos k_i).
\end{eqnarray}
Then a minimum Hamiltonian with nearest-neighbor hoppings and unitary topological charges belonging to a certain symmetry class is obtained.
For simplicity but without loss of generality, we consider the regime with $|m_{i}|<2$, where the Hamiltonian has band inversion at $\mathbf{k}=0$.
The low-energy expansion of lattice Hamiltonian near $\mathbf{k}=0$ is the corresponding continuous Hamiltonian.
We will provide several examples and display their topological properties in this section.
%

\subsection{Examples: $3$D and $2$D $Z$-class second-order topological phases\label{Sec41}}
To begin with, we illustrate two examples of second-order topological phases, namely a D class $3$D Hamiltonian and a BDI class $2$D Hamiltonian,
obtained by applying the transformation of Eq.~(\ref{trans}) to Hamiltonian $H^{(I)}(\mathbf{k})$ of~Eq. (\ref{H_1}) and Hamiltonian $H^{(II)}(\mathbf{k})$ of~Eq. (\ref{H_2}) respectively.
The $3$D Hamiltonian with second-order topology in D class reads $H^{({\rm D})}(\mathbf{k})=H^{({\rm D})}_1(\mathbf{k})+
H^{({\rm D})}_2(\mathbf{k}_{1,\parallel})$, with
\begin{eqnarray}\label{3dDsec}
H^{({\rm D})}_1(\mathbf{k})=&&M_1(\mathbf{k})\sigma_x\tau_0+h_{1}(\mathbf{k})\sigma_y\tau_0, \nonumber \\
H^{({\rm D})}_2(\mathbf{k}_{1,\parallel})=&&M_2(\mathbf{k}_{1,\parallel})\sigma_z\tau_x+h_{2}(\mathbf{k}_{1,\parallel})\sigma_z\tau_y \nonumber \\
&&+h_{3}(\mathbf{k}_{1,\parallel})\sigma_z\tau_z,
\end{eqnarray}
where $M_1=m_1+3-\cos k_1-\cos k_2-\cos k_3$, $M_2=m_2+2-\cos k_2-\cos k_3$, $h_1=\sin k_1$, $h_2=\sin k_2$ and $h_3=\sin k_3$.
Here $\sigma_{\beta=x,y,z}$ and $\tau_{\beta=x,y,z}$ are two sets of Pauli matrices, and $\tau_0$ is the two-by-two identity matrix.
Following previous discussions, this Hamiltonian supports a particle-hole symmetry, $\mathcal{C}H^{(D)}(\mathbf{k})\mathcal{C}^{-1}=-H^{(D)}(-\mathbf{k})$ with $\mathcal{C}=\sigma_z\tau_z\mathcal{K}$ and $\mathcal{C}^2=1$.

\begin{figure}[!htp]
\includegraphics[width=1\columnwidth]{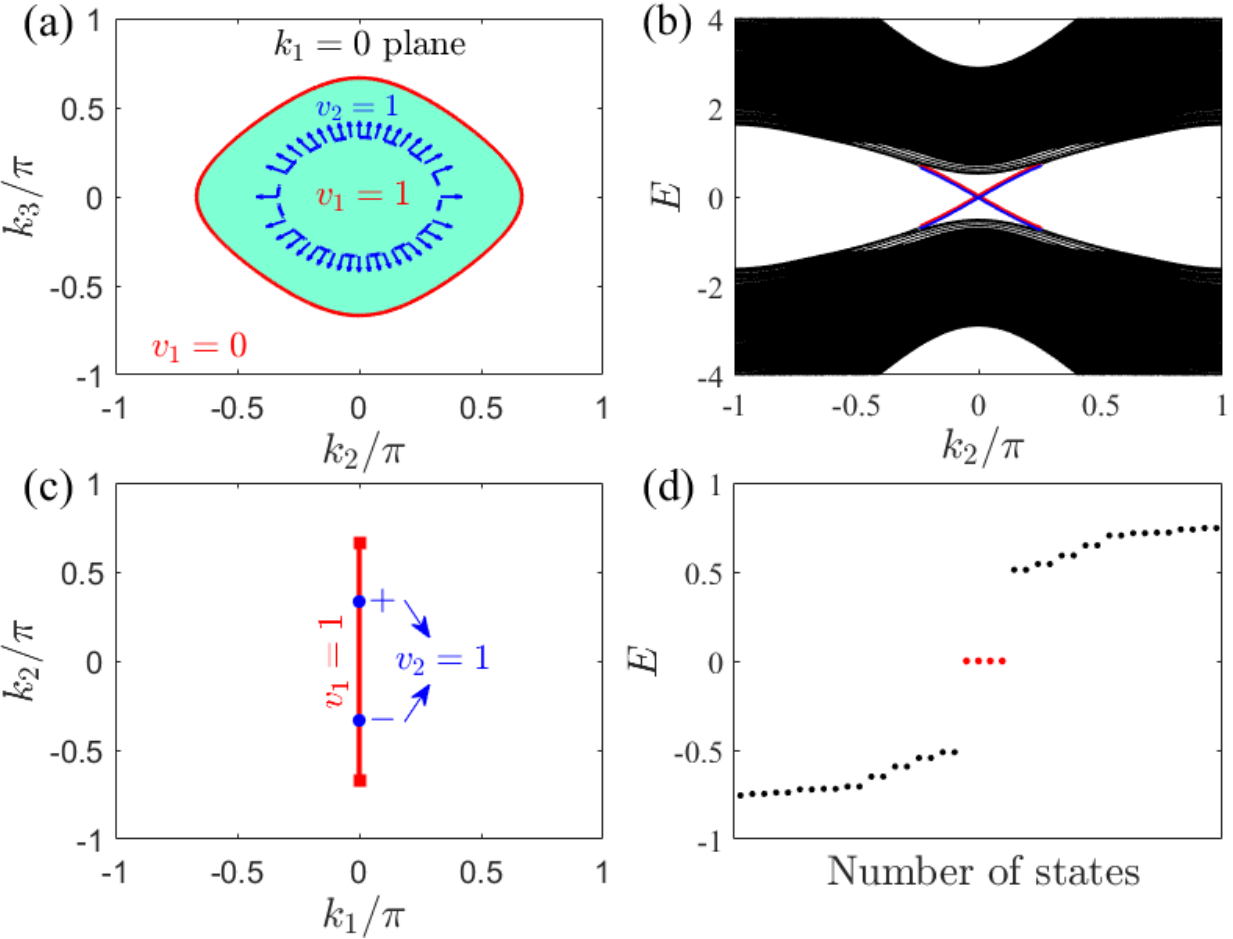}
\caption{\label{secondEX}(color online) (a) Nested BISs for Hamiltonian $H^{(D)}(\mathbf{k})$ of Eq.~\eqref{3dDsec}, where ${S}_1^{m=2}$ and ${S}_2^{m=1}$ are indicated by the red solid loop and blue dashed loop, respectively. Blue arrows indicate the normalized vector of $(h_2,h_3)$. (b) Energy spectrum for Hamiltonian in Eq.~\eqref{3dDsec} with the PBC along $r_3$-direction and OBCs for the other two directions. The hinge states on $(100)$ and $(\bar{1}00)$ are indicated by red and blue lines, respectively. (c) Nested BISs for Hamiltonian of Eq.~\eqref{2dBD1sec}. The red squares and blue dots indicate the BISs ${S}_1^{m=2}$ and ${S}_2^{m=1}$, respectively.
Red line indicates the region of $k_2$ with $v_1=1$.
(d) Eigenenergies for Hamiltonian of Eq.~\eqref{2dBD1sec} with zero-energy corner states (red dots). In all panels, $m_1=-1.5$ and $m_2=-0.5$ have been set.}
\end{figure}

Based on the nested-BIS method, we can define a nontrivial winding number $v_1(\mathbf{k}_{\parallel})$ along the $r_1$ direction for $H^{(D)}_1(\mathbf{k})$:
\begin{eqnarray}
\label{windingDse}
v_1(\mathbf{k}_{1,\parallel})=\frac{1}{2\pi}\oint_{k_1}\frac{h_{1}dM_{1}-M_{1}dh_{1}}{M^2_{1}+h^2_{1}}.
\end{eqnarray}
The second-BIS $S_1^{m=2}$ given by $H_1^{(D)}(\mathbf{k})=0$ acts as a boundary between a nontrivial region with $v_1\neq0$ and a trivial region with $v_1=0$, as shown in Fig.~\ref{secondEX}(a).
According to the bulk-boundary correspondence, the nontrivial invariant $v_1\neq0$ corresponds to the appearance of surface states in the $(100)$ and $(\bar{1}00)$ surface.
We then define a first-BIS $S_2^{m=1}$ of $H^{(D)}_2(\mathbf{k})$  in the 2D BZ of $\mathbf{k}_{1,\parallel}$ as the area with $M_2(\mathbf{k}_{1,\parallel})=0$.
When $S_2^{m=1}$ falls within the region with nontrivial $v_1\neq0$, the topological properties of the Hamiltonian $H^{(D)}_2(\mathbf{k}_{1,\parallel})$ can be captured by these surface states \cite{LI2021}.
Then we obtain an effective $2$D Hamiltonian in the subspace associated with each surface state~\cite{PhysRevB.97.205135,PhysRevX.9.011012}:
\begin{eqnarray}\label{effH}
H^{({\rm D})}_{{\rm eff},\pm}=&&P_{1,\pm}H^{({\rm D})}_2P_{1,\pm} \nonumber \\
=&&P_{1,\pm}(M_2\sigma_z\tau_x+h_{2}\sigma_z\tau_y
+h_{3}\sigma_z\tau_z)P_{1,\pm},\nonumber \\
P_{1,\pm}=&&[1\pm i(\sigma_y\tau_0)(\sigma_x\tau_0)]/2=(1\pm\sigma_z\tau_0)/2,
\end{eqnarray}
where $P_{1,+}$ and $P_{1,-}$ indicate the projection operators onto the subspace of low-energy surface states on $(100)$ and $(\bar{1}00)$ surfaces, respectively.
Therefore, the surface states share the same topological properties with $H_2^{({\rm D})}$, whose nontrivial topology gives rise to second-order surface states of the parent 3D system.

%
To topologically characterize $H^{({\rm D})}_2(\mathbf{k}_{1,\parallel})$, a winding number can be defined for the pseudo-spin texture of $(h_2, h_3)$ along the first-BIS $S_2^{m=1}$,
\begin{eqnarray}\label{chernH2}
v_2=\frac{1}{2\pi}\int_{S_2^{m=1}}d\theta,
\end{eqnarray}
with $\theta=\arctan(h_3/h_2)$,
which is equivalent to the Chern number defined in the 2D BZ \cite{PhysRevB.94.165117,ZHANG20181385}.
Chiral-like hinge states are seen when the two BISs are nested in the proper order, i.e., when $S_2^{m=1}$ falls within the region with nontrivial $v_1\neq0$ bounded by $S_1^{m=2}$ in the 2D BZ of $(k_2,k_3)$,
as displayed in Fig.~\ref{secondEX}(b).
%
We note that the Hamiltonian $H^{(D)}(\mathbf{k})$ of Eq.~\eqref{3dDsec} can describe the 3D topological insulators materials with noncollinear antiferromagnetic order~\cite{doi:10.1126/sciadv.aat0346}, as discussed in Appendix \ref{appadd}.

By removing terms associated with $k_3$ from the Hamiltonian~(\ref{3dDsec}) (or by transforming $H^{(II)}(\mathbf{k})$ in 2D into a lattice model), we can get a $2$D Hamiltonian in  BDI class $H^{({\rm BDI})}(\mathbf{k})=H^{({\rm BDI})}_1(\mathbf{k})+
H^{({\rm BDI})}_2(\mathbf{k}_{1,\parallel})$ with
\begin{eqnarray}\label{2dBD1sec}
H^{({\rm BDI})}_1(\mathbf{k})=&&M_1(\mathbf{k})\sigma_x\tau_0+h_{1}(\mathbf{k})\sigma_y\tau_0, \nonumber \\
H^{({\rm BDI})}_2(\mathbf{k}_{1,\parallel})=&&M_2(\mathbf{k}_{1,\parallel})\sigma_z\tau_x+h_{2}(\mathbf{k}_{1,\parallel})\sigma_z\tau_y,
\end{eqnarray}
where $M_1=m_1+2-\cos k_1-\cos k_2$, $M_2=m_2+1-\cos k_2$, $h_1=\sin k_1$ and $h_2=\sin k_2$.
In addition to the particle-hole symmetry $\mathcal{C}=\sigma_z\tau_z\mathcal{K}$, this Hamiltonian also holds a time-reversal symmetry: $\mathcal{T}H^{({\rm BDI})}(\mathbf{k})\mathcal{T}^{-1}=H^{({\rm BDI})}(-\mathbf{k})$, with $\mathcal{T}=\mathcal{K}$ and $\mathcal{T}^2=1$, and a chiral symmetry
$\mathcal{S}H^{({\rm BDI})}(\mathbf{k})\mathcal{S}^{-1}=-H^{({\rm BDI})}(\mathbf{k})$, with $\mathcal{S}=\mathcal{C}\mathcal{T}=\sigma_z\tau_z$.
%

Similarly, the second-BIS $S_1^{m=2}$ separate the $k_2$ axis into nontrivial region and trivial region, as shown in Fig.~\ref{secondEX}(c).
The topological invariant of $H^{({\rm BDI})}_2(\mathbf{k})$ can be obtained through \cite{PRXQuantum.2.020320}
\begin{eqnarray}\label{windingH2}
v_2=\frac{1}{2}\sum_{S_2^{m=1}}{\rm Sgn}[\frac{\partial{M_2}}{\partial{k_2}}h_2].
\end{eqnarray}
where $S_2^{m=1}$ indicates the discrete points satisfying $M_2=0$. 
When $S_2^{m=1}$ falls within the region with $v_1\neq0$ bounded by $S_1^{m=2}$ in the 1D BZ of $k_2$, the $0$D corner states appear, as shown in Fig.~\ref{secondEX}(d).

For these two Hamiltonians, the topological invariants can be indexed as $Z\times Z$, i.e., the two parts of the Hamiltonian can be indexed by $Z$-class topological invariants and second-order topological phases appear when their BISs satisfy the nested relation.
It also establishes for all Hamiltonians $H^{(\uppercase\expandafter{\romannumeral1})}$ of Eq.~(\ref{H_1}) and $H^{(\uppercase\expandafter{\romannumeral2})}$ of Eq.~(\ref{H_2}), indexed as $Z\times Z$ in Table.~\ref{Classification}.
The other two cases, namely $H^{(\uppercase\expandafter{\romannumeral3})}$ of Eq.~(\ref{H_3}) and $H^{(\uppercase\expandafter{\romannumeral4})}$ of Eq.~(\ref{H_4}), are indexed as $2Z\times Z$ in Table~\ref{Classification}, as their BISs and pseudospin textures are similar to the first two classes but the number of topological boundary states will double, and an example will be given in the following subsection.

\subsection{Examples: $3$D $Z$-class and $2Z$-class third-order topological phases\label{Sec42}}
Next we give two 3D examples of third-order topological phases with $Z$ and $2Z$ invariants, by applying the transformation of Eq.~(\ref{trans}) to Hamiltonian $H^{(I)}(\mathbf{k})$ of~Eq. (\ref{thH_1}) and Hamiltonian $H^{(III)}(\mathbf{k})$ of~Eq. (\ref{thH_3}) with $n=3$ for both cases, respectively.
The first one falls in the BDI class, described by the Hamiltonian
$H^{({\rm BDI})}(\mathbf{k})=H^{({\rm BDI})}_1(\mathbf{k})+
H^{({\rm BDI})}_2(\mathbf{k}_{1,\parallel})+
H^{({\rm BDI})}_3(k_3)$ with
\begin{eqnarray}\label{3dBD1th}
H^{({\rm BDI})}_1(\mathbf{k})=&&M_1(\mathbf{k})\sigma_x\tau_0s_0+h_{1}(\mathbf{k})\sigma_y\tau_0s_0, \nonumber \\
H^{({\rm BDI})}_2(\mathbf{k}_{1,\parallel})=&&M_2(\mathbf{k}_{1,\parallel})\sigma_z\tau_xs_0+h_{2}(\mathbf{k}_{1,\parallel})\sigma_z\tau_ys_0, \nonumber \\
H^{({\rm BDI})}_3(k_3)=&&M_3(k_3)\sigma_z\tau_zs_x+h_{3}(k_3)\sigma_z\tau_zs_y,
\end{eqnarray}
where $M_1=m_1+3-\cos k_1-\cos k_2-\cos k_3$, $M_2=m_2+2-\cos k_2-\cos k_3$, $M_3=m_3+1-\cos k_3$, $h_1=\sin k_1$, $h_2=\sin k_2$, and $h_3=\sin k_3$.
Here we have introduced another set of Pauli matrices indexed by $s_{x,y,z}$.
This Hamiltonian satisfies a particle-hole symmetry $\mathcal{C}=\sigma_z\tau_zs_z\mathcal{K}$ with $\mathcal{C}^2=1$, a time-reversal symmetry $\mathcal{T}=\mathcal{K}$ with $\mathcal{T}^2=1$, and a chiral symmetry $\mathcal{S}=\mathcal{C}\mathcal{T}=\sigma_z\tau_zs_z$.

\begin{figure}[!htp]
\includegraphics[width=1\columnwidth]{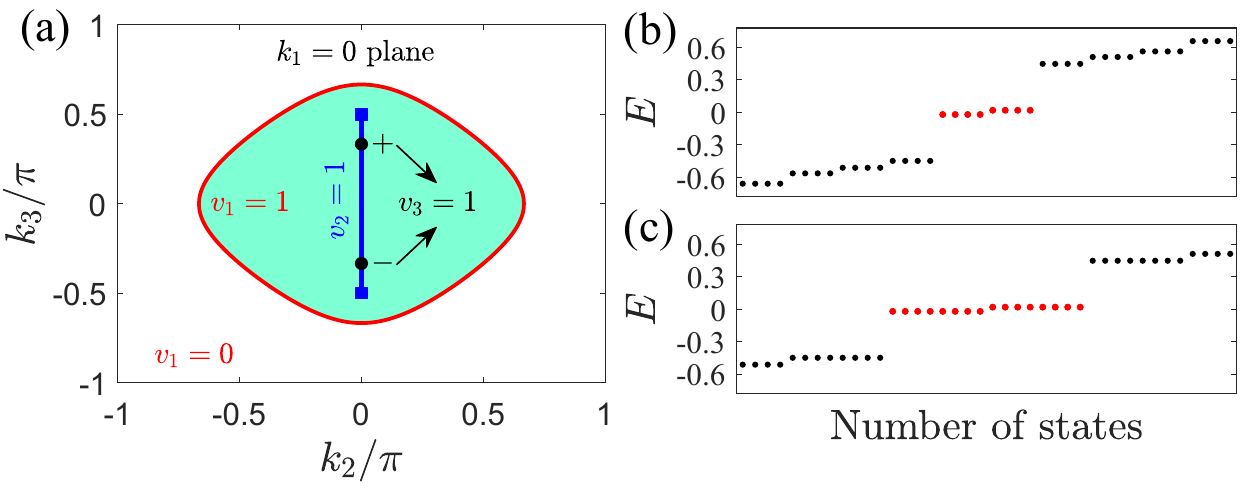}
\caption{\label{thirdEX}(color online) (a) Nested BISs for Hamiltonian $ H^{({\rm BDI})}(\mathbf{k})$ of Eq.~\eqref{3dBD1th} and $ H^{({\rm CII})}(\mathbf{k})$ of Eq.~\eqref{3dC2th}.
Red loop, blue squares, and black dots indicate ${S}_1^{m=2}$, ${S}_2^{m=2}$, and ${S}_3^{m=1}$, respectively.
(b), (c) Energy spectra for $H^{({\rm BDI})}(\mathbf{k})$ of Eq.~\eqref{3dBD1th} and $ H^{({\rm CII})}(\mathbf{k})$ of Eq.~\eqref{3dC2th},
respectively, where all three directions are under OBCs.  Other parameters are $m_1=-1.5$, $m_2=-1$ and $m_3=-0.5$ in all panels.
Red dots represent topological corner states in these models, which acquire small nonzero energies due to finite-size effect ($8\times8\times8$ unit cells are considered in our calculation).}
\end{figure}

To apply the nested-BIS method, we can define three BISs as $S_1^{m=2}: H^{({\rm BDI})}_1=0$, $S_2^{m=2}: H^{({\rm BDI})}_2=0$ and $S_3^{m=1}: M_3=0$.
Then, when these BISs correspond to nontrivial topology and form nested relations [e.g., as shown in Fig.~\ref{thirdEX}(a)], $Z\times Z\times Z$ corner states shall emerge at each corner, give rise to eight in-gap topological states for a 3D system, as displayed in Fig.~\ref{thirdEX}(b).
%

Similarly, substituting  Eq.~(\ref{trans}) to Hamiltonian of Eq.~(\ref{thH_3}),
we can get a $3$D third-order topological phase in CII class
$H^{({\rm CII})}(\mathbf{k})=H^{({\rm CII})}_1(\mathbf{k})+
H^{({\rm CII})}_2(\mathbf{k}_{1,\parallel})+
H^{({\rm CII})}_3(k_3)$ with
\begin{eqnarray}\label{3dC2th}
H^{({\rm CII})}_1(\mathbf{k})=&&M_1(\mathbf{k})\sigma_0\tau_xs_0\varsigma_0+h_{1}(\mathbf{k})\sigma_z\tau_ys_0\varsigma_0, \nonumber \\
H^{({\rm CII})}_2(\mathbf{k}_{1,\parallel})=&&M_2(\mathbf{k}_{1,\parallel})\sigma_z\tau_zs_x\varsigma_0 +h_{2}(\mathbf{k}_{1,\parallel})\sigma_z\tau_zs_y\varsigma_0,  \nonumber \\
H^{({\rm CII})}_3(k_3)=&&M_3(k_3)\sigma_z\tau_zs_z\varsigma_x+h_{3}(k_3)\sigma_z\tau_zs_z\varsigma_y, \nonumber \\
\end{eqnarray}
where the coefficients $M_{1,2,3}$ and $h_{1,2,3}$ are the same as those of Hamiltonian $H^{({\rm BDI})}(\mathbf{k})$ in Eq.~(\ref{3dBD1th}), but another set of Pauli matrices $\varsigma_{x,y,z}$ has been introduced.
This Hamiltonian holds a particle-hole symmetry $\mathcal{C}=i\sigma_x\tau_ys_z\varsigma_z\mathcal{K}$ with $\mathcal{C}^2=-1$, a time-reversal symmetry $\mathcal{T}=i\sigma_y\tau_xs_0\varsigma_0\mathcal{K}$ with $\mathcal{T}^2=-1$, and a chiral symmetry $\mathcal{S}=\mathcal{C}\mathcal{T}=\sigma_z\tau_zs_z\varsigma_z$.
By definition, the BISs and their corresponding topological invariants of $H^{({\rm CII})}(\mathbf{k})$ are the same as those of $H^{({\rm BDI})}(\mathbf{k})$. The only difference is that the pseudospin space is doubled by the extra set of Pauli matrices, leading to double corner states at each corner, namely, $16$ in-gap topological states in the topologically nontrivial regime,  as shown in Fig.~\ref{thirdEX}(c).
Thus the topological properties of this model are indexed as $2Z\times Z\times Z$.
%
%
%
%
%

\section{Higher-order topological phases with $Z_2$ topology \label{Sec5}}
In this section, we derive HOTPs with $Z_2$ properties from two parent Hamiltonians indexed by $Z^n$, i.e., $H^{(\uppercase\expandafter{\romannumeral1})}$ in Eq.~(\ref{thH_1}) and $H^{(\uppercase\expandafter{\romannumeral2})}$ in Eq.~(\ref{thH_2}).
That is, we convert the last subsystem $H_n$ to a $Z_2$-class Hamiltonian and keep the rest of the two-component parts in Eq.~\eqref{ReHadi} unchanged.
Then their topological invariants can be indexed as $Z^{n-1}\times Z_2$.

In particular, for the symmetry class without chiral symmetry, the HOTPs with $Z_2$ properties can be obtained from $H^{(\uppercase\expandafter{\romannumeral1})}$ in Eq.~\eqref{thH_1} through the conversion~\cite{PhysRevB.78.195424}:
\begin{eqnarray}\label{Z2withoutChiral}
H^{({\uppercase\expandafter{\romannumeral1}})}_n(\mathbf{k}_{n-1,\parallel})=&&m_n\Gamma_{(2p+1)}^{2n-1}+\sum_{\alpha=n}^{2p-n+1}k_{\alpha}\Gamma_{(2p+1)}^{n+\alpha}\nonumber \\
&&\Downarrow \nonumber \\
H^{({\uppercase\expandafter{\romannumeral1}},s)}_n(\mathbf{k}_{n-1,\parallel})=&&m_{n}\Gamma_{(2p+1)}^{2n-1}+\sum^{2p-n-s+1}_{\alpha=n}k_{\alpha}\Gamma_{(2p+1)}^{n+\alpha}\nonumber \\
&&+\sum^{s}_{\alpha=1}f_\alpha\Gamma_{(2p+1)}^{2p+2-\alpha}.
\end{eqnarray}
And for the symmetry class with chiral symmetry, the HOTPs with $Z_2$ properties can be obtained from  $H^{(\uppercase\expandafter{\romannumeral2})}$ in Eq.~\eqref{thH_2} through the conversion~\cite{Ryu_2010}:
\begin{eqnarray}\label{Z2withChiral}
H^{({\uppercase\expandafter{\romannumeral2}})}_n(\mathbf{k}_{n-1,\parallel})=&&m_n\Gamma_{(2p+1)}^{2n-1}+\sum_{\alpha=n}^{2p-n}k_{\alpha}\Gamma_{(2p+1)}^{n+\alpha}\nonumber \\
&&\Downarrow \nonumber \\
H^{({\uppercase\expandafter{\romannumeral2}},s)}_n(\mathbf{k}_{n-1,\parallel})=&&m_{n}\Gamma_{(2p+1)}^{2n-1}+\sum^{2p-n-s}_{\alpha=n}k_{\alpha}\Gamma_{(2p+1)}^{n+\alpha}\nonumber \\
&&+\sum^{s}_{\alpha=1}f_\alpha\Gamma_{(2p+1)}^{2p+1-\alpha}.
\end{eqnarray}
In Eqs.~(\ref{Z2withoutChiral}) and~(\ref{Z2withChiral}), $s=1,2$ indicates two descendants with $Z_2$ topology of $H^{(I)}$ or $H^{(II)}$, and $f_\alpha$ is chosen to be an odd function of the reduced momentum $\mathbf{k}_{n-1,\parallel}=(k_{n},k_{n+1},....,k_{d-s})$.
With these conversions,
the first (second) descendant falls in the same symmetry class as its parent Hamiltonian, but with one (two) spatial dimension lower
as one (two) momentum component is converted into $f_\alpha$.
%
%
The classification of these second-order topological phases with $Z_2$ property have also been listed in Table.~\ref{Classification}.
Corresponding lattice models can be obtained  by applying the transformation of Eq.~(\ref{trans}), while the dimension $d$ is changed to $(d-s)$.

We provide a $2$D D class second-order topological phase as an example, which is a first descendant of the lattice Hamiltonian of Eq.~\eqref{3dDsec},
\begin{eqnarray}\label{2dDsecZ2}
H^{({\rm D},1)}(\mathbf{k})=&&H^{({\rm D},1)}_1(\mathbf{k})+
H^{({\rm D},1)}_2(k_2), \nonumber \\
H^{({\rm D},1)}_1(\mathbf{k})=&&M_1(\mathbf{k})\sigma_x\tau_0+h_{1}(\mathbf{k})\sigma_y\tau_0, \nonumber \\
H^{({\rm D},1)}_2(k_2)=&&M_2(k_2)\sigma_z\tau_x+h_{2}(k_2)\sigma_z\tau_y,
+f_{1}(k_2)\sigma_z\tau_z,\nonumber\\
\end{eqnarray}
where $M_1=m_1+2-\cos k_1-\cos k_2$, $M_2=m_2+1-\cos k_2-\lambda_1\cos(2k_2)$, $h_1=\sin k_1$, $h_2=\sin k_2$ and $f_1=\sin(2k_2)$.
Here we introduce next-nearest neighbor hopping to generate $2k_2$ terms for demonstrating its $Z_2$ properties~\cite{Li2016}.

Analogous to its parent Hamiltonian, we can define $v_1$ and a 2-BIS $S_1^{m=2}$ for $H^{({\rm D},1)}_1(\mathbf{k})$, and a first-order BIS (1-BIS) $S_2^{m=1}$ for $H^{({\rm D},1)}_2(k_2)$.
When $S_2^{m=1}$ falls within the nontrivial region bounded by $S_1^{m=2}$, the boundary states of $H^{({\rm D},1)}_1(\mathbf{k})$ can capture the topology of $H^{({\rm D},1)}_2(k_2)$.
But we still need to check whether $H^{({\rm D},1)}_2(k_2)$ is topologically nontrivial or not, which can be characterized by a Berry phase~\cite{RevModPhys.82.1959} or the dynamical invariant at its highest-order BISs~\cite{ScienceBulletin671236}.
To define a topological invariant of $H^{({\rm D},1)}_2$ based on the BIS $S_2^{m=1}$, we introduce
an auxiliary Hamiltonian
\begin{eqnarray}\label{interpo}
\tilde{H}^{({\rm D},1)}_2(k_2,\theta)=&&\widetilde{M}_2(k_2,\theta)\sigma_z\tau_x+h_2(k_2)\sigma_z\tau_y \nonumber \\
&&+\widetilde{f}_{1}(k_2,\theta)\sigma_z\tau_z,
\end{eqnarray}
where 
$\widetilde{M}_2(k_2,\theta)=M_2(k_2)+\lambda_2(1-\cos\theta)$ and $\widetilde{f}_{1}(k_2,\theta)=\sin(2k_2+\theta)$.
This Hamiltonian preserves the same symmetries as ${H}^{({\rm D},1)}_2(k_2)$, and satisfies $\tilde{H}^{({\rm D},1)}_2(k_2,0)={H}^{({\rm D},1)}_2(k_2)$.
On the other hand, provided $\lambda_2$ is large enough, $\tilde{H}^{({\rm D},1)}_2(k_2,\theta=\pi)$ is topologically trivial, which is called the 'vacuum' projection.
Therefore, the topological difference between ${H}^{({\rm D},1)}_2(k_2)$ and vacuum can be captured by a topological invariant $\widetilde{v}_2$ defined for $\tilde{H}^{({\rm D},1)}_2(k_2,\theta)$ in the 2D parameter space of $(k_2,\theta)$~\cite{PhysRevB.78.195424,Ryu_2010}.
Finally, $\widetilde{v}_2$ can be obtained along a 1-BIS of $\tilde{H}^{({\rm D},1)}_2(k_2,\theta)$, $\tilde{S}_2^{m=1}: \widetilde{M}_2(k_2,\theta)=0$, which reproduces the 1-BIS $S_2^{m=1}$ of $H^{({\rm D},1)}_2(k_2)$ at $\theta=0$.
The $Z_2$ invariant for $H^{({\rm D},1)}_2(k_2)$ is thus defined as $$v_2=\mod(\tilde{v}_2,2),$$ which actually is independent from the details of interpolation between $\theta=0$ and $\pi$, but only depends on $H^{({\rm D},1)}_2(k_2)$.
%
%
%
\begin{figure}[!htp]
\includegraphics[width=1\columnwidth]{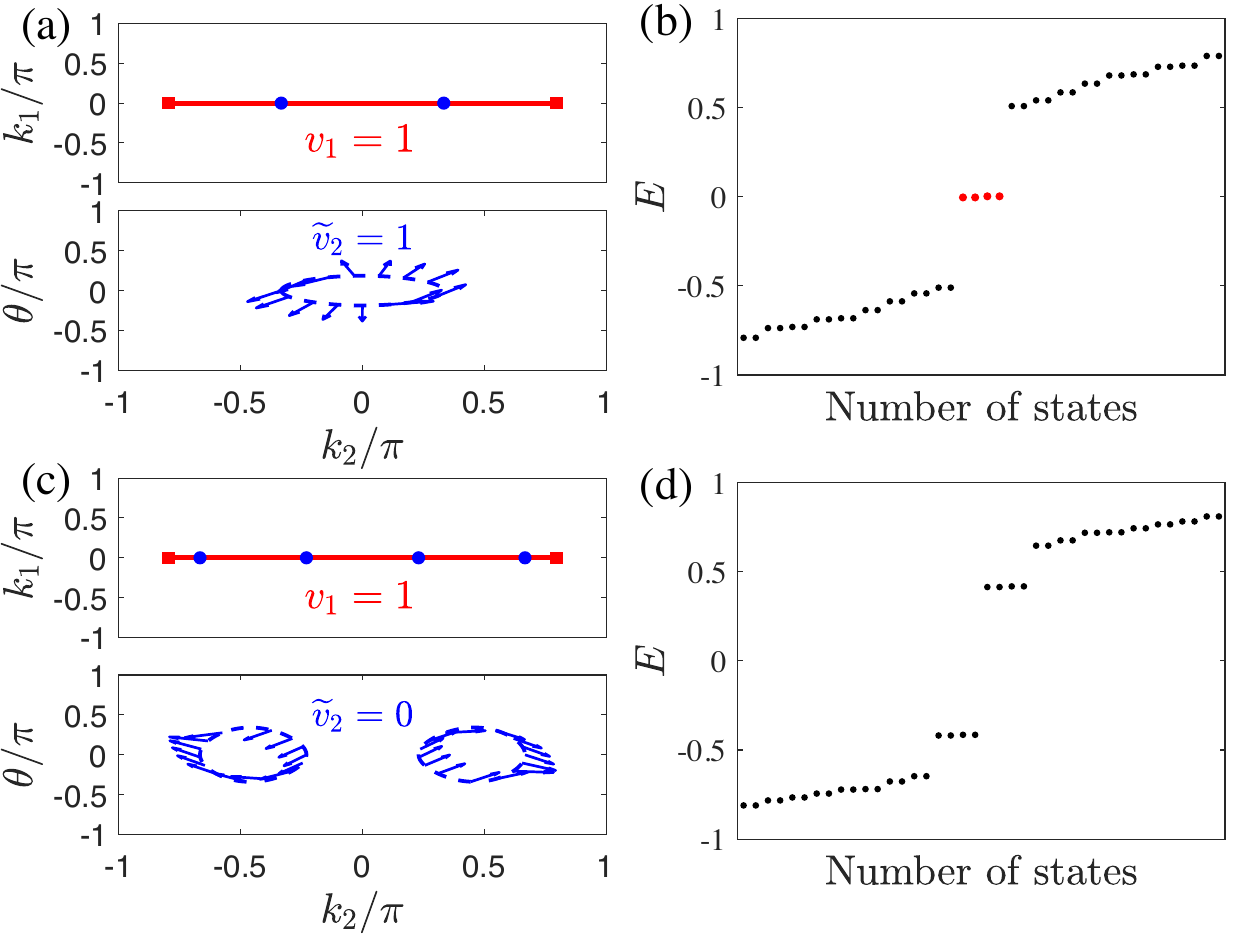}
\caption{\label{Z2EX}(color online) (a) Nested BISs for Hamiltonian $H^{({\rm D},1)}(\mathbf{k})$ of Eq.~\eqref{2dDsecZ2} and the vector $(h_2,\widetilde{f}_{1})$  of  Eq.~\eqref{interpo} (blue arrows, after normalization) along the BISs $\tilde{S}_2^{m=1}$ (dashed loops), with $\lambda_1=0$.
Here ${S}_1^{m=2}$  and ${S}_2^{m=1}$ are indicated by red squares and blue dots, respectively.
(b) Corresponding eigenenergies closed to zero under OBCs.
(c) and (d) demonstrate the same results as in (a) and (b), but with $\lambda_1=-2$.
Other parameters are $m_1=-1.8$, $m_2=-0.5$, and $\lambda_2=3$.
Red dots in (b) indicate zero-energy corner states.}
\end{figure}
In Fig.~\ref{Z2EX}(a), the two BISs of $H^{({\rm D},1)}(\mathbf{k})$ in Eq.~\eqref{2dDsecZ2} form a nested relation, and the topological invariant $\tilde{v}_2$ takes an odd value, so $v_2=1$.
As a result, four zero-energy corner states appear in the system, as displayed in Fig.~\ref{Z2EX}(b).

In contrast,
in Fig.~\ref{Z2EX}(c) we illustrate another situation where $\widetilde{v}_2=0$. The BIS $S_2^{m=1}$ also falls within the nontrivial region of $v_1$, yet the system is topologically trivial as $v_2=0$. Consistently, the energy spectrum in Fig.~\ref{Z2EX}(d) shows no zero-energy corner state in its band gap.
%
%
%
We note that the value of $\widetilde{v}_2$ may change when introducing a different $\tilde{H}^{({\rm D},1)}_2(k_2,\theta)$, but its parity reminds the same and predicts the $Z_2$ invariant $v_2$. An example with $\widetilde{v}_2=2$ is given in Appendix \ref{appB}.
%

\section{Asymmetric boundary states\label{Sec6}}
In the previous discussion, the higher-order topological boundary states are seen to distribute along
spatially symmetric hinges (e.g., with a $\mathcal{C}_4$ rotation symmetry in $k_2$-$k_3$ plane)
of 3D systems, and at the four corners of 2D systems. This is because the examples we consider are some minimal models with certain coincidental spatial symmetries, which are not necessary for constructing topological phases in the AZ classification.
For example, higher-order topological corner states can emerge in 2D and 3D lattices without any spatial symmetry, which host corner states with different configurations \cite{PhysRevB.98.205422}.
In this section, we will extend our discussion to several scenarios with more sophisticated nested BISs,
where asymmetric properties arise due to certain spatial-symmetry breaking.
%
In particular, crossed BISs give rise to asymmetric behaviors in different directions on the same surfaces of a 3D system, while certain non-Clifford terms can induce boundary states asymmetric between two opposite surfaces [e.g., $(100)$ and $(\overline{1}00)$].

\subsection{Crossed band inversion surfaces\label{sec61}}
\subsubsection{Asymmetric properties between $k_2$ and $k_3$}\label{sec:asymmetric_1}
\begin{figure}[!htp]
\includegraphics[width=1\columnwidth]{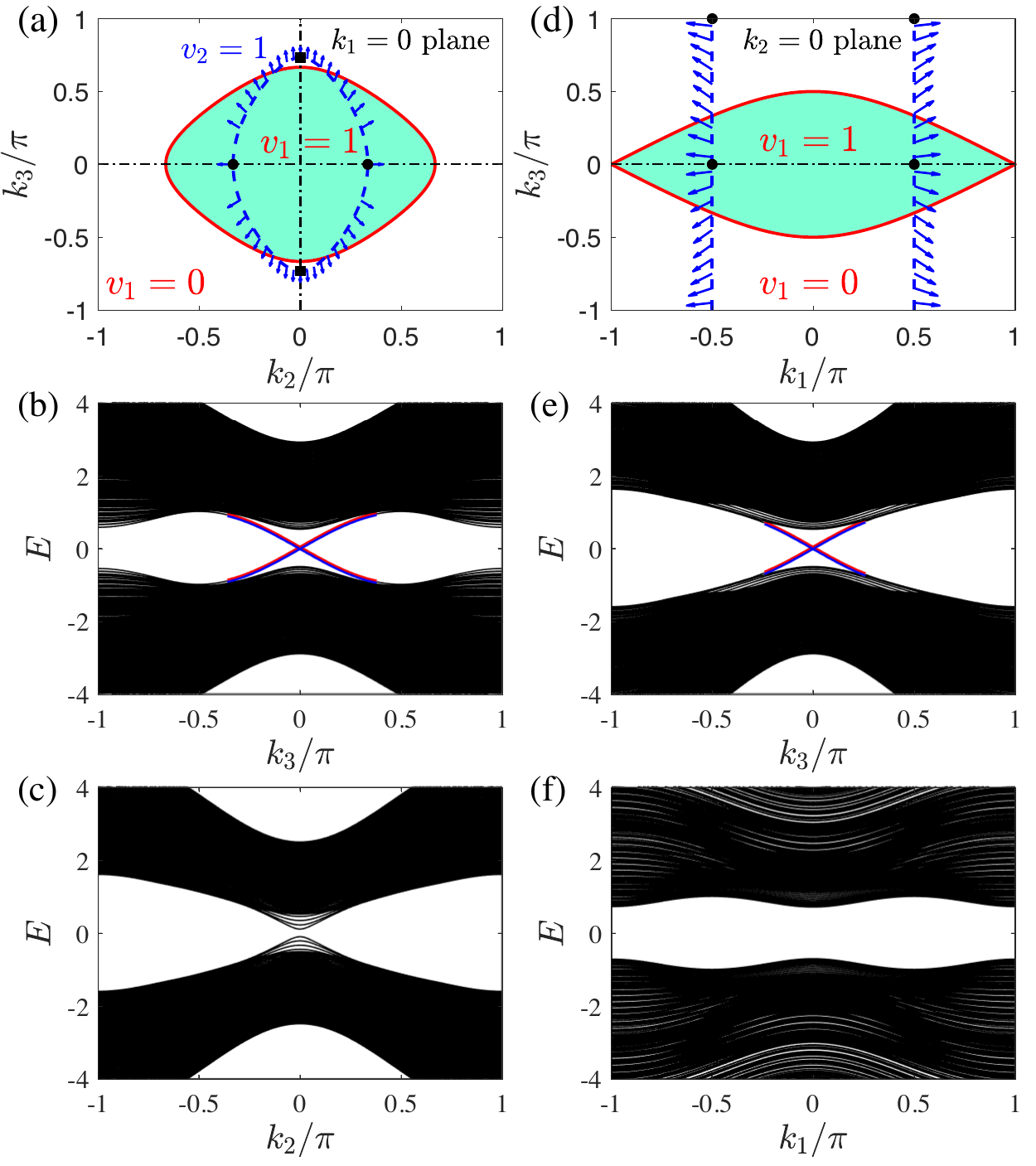}
\caption{\label{Dgamma}(color online) (a) Nested BISs for Hamiltonian $H^{(D)}(\mathbf{k})$ of Eq.~\eqref{3dDsec} with an asymmetric parameter $\gamma$, where $\gamma=0.3$ has been set. ${S}_1^{m=2}$, ${S}_2^{m=1}$, and the vector $(h_2,h_3)$ are indicated by red solid loop, blue dashed loop, and blue arrow, respectively.
2-BISs ${S}_2^{m=2,a}$ and ${S}_2^{m=2,b}$ of $H^{({\rm D})}_2(\mathbf{k}_{1,\parallel})$ are indicated by black dots and squares, respectively.
(b), (c) Energy spectra corresponding to (a) with PBC taken only in $r_3$- and $r_2$-directions, respectively. (d) The nested-BIS for Hamiltonian $H^{({\rm D})}(\mathbf{k})$ with the decomposition of Eq.~\eqref{3dDsectram}. (e), (f) Energy spectra corresponding to (b) with PBC taken only in $r_3$ and $r_1$ directions, respectively.
In all panels, $m_1=-1.5$ and $m_2=-0.5$ are set.}
\end{figure}
In the above sections, we have discussed several 3D systems where the BISs always enclose each other, which is ensured by a $\mathcal{C}_{4}$ rotation symmetry in the $k_2$-$k_3$ plane,
\begin{eqnarray}\label{C4_rot}
\mathcal{C}_{4}H^{({\rm D})}(k_1,k_2,k_3)\mathcal{C}^{-1}_{4}=H^{({\rm D})}(k_1,-k_3,k_2)
\end{eqnarray}
with $\mathcal{C}_{4}=e^{i\pi/4\sigma_0\tau_x}$.
%
%
To go beyond this scenario,
%
%
%
%
we consider a Hamiltonian similar to Eq.~\eqref{3dDsec}, but with an asymmetric parameter $\gamma$ in $M_2$:
\begin{eqnarray}\label{asymmetric_M2}
M_2(\mathbf{k}_{1,\parallel})=m_2+(1-\cos k_2)+\gamma(1-\cos k_3).
\end{eqnarray}
Thus, the $\mathcal{C}_{4}$ rotation symmetry is broken, and the second BIS $S_2^{m=1}$ is deformed and crosses the other BIS $S_1^{m=2}$, as shown in Fig.~\ref{Dgamma}(a).
%
%

To topologically characterize this asymmetric Hamiltonian with the nested-BIS method,
we further define two 2-BISs for  $H^{({\rm D})}_2(\mathbf{k}_{1,\parallel})$ as
$$S_2^{m=2,a}: M_2=h_3=0,~~S_2^{m=2,b}: M_2=h_2=0.$$
As shown in Fig. \ref{Dgamma}(a), $S_2^{m=2,a}$ falls within the region with a nonzero $v_1$, and it holds a nonzero topological invariant
$$v_2=\frac{1}{2}\sum_{S_2^{m=2,a}}{\rm Sgn}[\frac{\partial{M_2}}{\partial{k_2}}h_2]=1$$
for the parameters we choose \cite{PRXQuantum.2.020320,LI2021}.
Note that, by definition, this $v_2$ corresponds not to the winding of $(h_2,h_3)$ shown in Fig. \ref{Dgamma}(a), but to the winding of $(h_2,M_2)$ at $k_3=0$, with $k_2$ varying from $0$ to $2\pi$. Nevertheless, these two winding properties are equivalent, as they both reflect the 2D Chern topology of $H^{(D)}_2(\mathbf{k}_{1,\parallel})$.
Furthermore, a nonzero $v_2$ describes a nontrivial 1D topology along the $k_2$ direction at $k_3=0$, corresponding to a pair of chiral edge states of $H_2^{({\rm D})}(\mathbf{k}_{1,\parallel})$ when the OBC is taken along the $r_2$ direction.
Due to the nested relation between $S_2^{m=2,a}$ and $S_2^{m=1}$, such topological properties can be captured by the surface states of $H_1^{({\rm D})}(\mathbf{k}_{1,\parallel})$ and manifests as chiral-like hinge states in Fig. \ref{Dgamma}(b).
%

Similarly, a topological invariant defined for $S_2^{m=2,b}$ characterizes topological properties along the $k_3(r_3)$ direction. However, $S_2^{m=2,b}$ falls outside the nonzero region of $v_1$ enclosed by $S_1^{m=2}$, meaning that its topological properties (if any) cannot be captured by the surface states.
Therefore, the overall system shows a trivial second-order topology when the OBC is taken along the $r_2$ direction, consistent with the absence of chiral-like hinge states in Fig. \ref{Dgamma}(c).
Note that in both Figs. \ref{Dgamma}(b) and \ref{Dgamma}(c), we have also taken the OBC along the $r_1$ direction, and PBC along the third direction, so as to illustrate only the second-order topology associated to the two OBC directions in each case.
 %
%

\subsubsection{Asymmetric properties between $k_1$ and $k_3$}
As a matter of fact, the asymmetric behavior of hinge states and BISs can also be seen in the original Hamiltonian of Eq. \eqref{3dDsec},
where $\mathcal{C}_4$ rotation symmetry holds only in the $k_2$-$k_3$ plane, but not in the other two planes involving $k_1$.
%
To see this, we rewrite the Hamiltonian as
\begin{eqnarray}\label{3dDsectram}
H^{({\rm D})}(\mathbf{k})=&&H^{({\rm D})}_1(\mathbf{k})+
H^{({\rm D})}_2(k_1,k_3),\nonumber \\
H^{({\rm D})}_1(\mathbf{k})=&&\widetilde{M}_1(\mathbf{k})(\sigma_x\tau_0+\sigma_z\tau_x)/\sqrt{2}+h_{2}(\mathbf{k})\sigma_z\tau_y \nonumber \\
H^{({\rm D})}_2(k_1,k_3)=&&\widetilde{M}_2(k_1,k_3)(\sigma_x\tau_0-\sigma_z\tau_x)/\sqrt{2},\nonumber \\
&&+h_{1}(k_1)\sigma_y\tau_0
+h_{3}(k_3)\sigma_z\tau_z, \nonumber \\
\end{eqnarray}
where $\widetilde{M}_1=({M}_1+{M}_2)/\sqrt{2}=(m_1+m_2+5-\cos k_1-2\cos k_2-2\cos k_3)/\sqrt{2}$, $\widetilde{M}_2=({M}_1-{M}_2)/\sqrt{2}=(m_1-m_2+1-\cos k_1)/\sqrt{2}$,
and $h_i=\sin k_i$ with $i=1,2,3$ as for Eq.~\eqref{3dDsec}.
These terms also anti-communicate with each other, and now it is $k_2$ that appears only in $H^{(D)}_1(\mathbf{k})$.
Therefore, with this alternative expression of $H^{({\rm D})}(\mathbf{k})$, we can apply the nested-BIS method to analysis how the topological properties of $H^{({\rm D})}_2(k_1,k_3)$ are captured by the surface states under OBC along the $r_2$ direction.
%

%
To proceed further, we first define a 2-BIS
$S_1^{m=2}: H_1^{({\rm D})}=0$ of $H_1^{({\rm D})}$, and a 1-BIS $S_2^{m=1}$ of $H^{(D)}_2$. Similar to the previous asymmetric example,
here $S_2^{m=1}$ crosses $S_1^{m=2}$ as it is separated into two lines paralleling to $k_3$.
Therefore we need to further define a 2-BIS for $H^{(D)}_2$ as $S_2^{m=2,a}: \widetilde{M}_2=h_3=0$, which are two pairs of points at $k_3=0$ or $k_3=\pi$, as shown in Fig.~\ref{Dgamma}(d).
For each pair of the 2-BIS with the same $k_3$, a topological invariant can be obtained as
$$v_2(k_3=0/\pi)=\frac{1}{2}\sum_{S_2^{m=2,a},k_3=0/\pi}Sgn[\frac{\partial{\widetilde{M}_2}}{\partial{k_1}}h_1]=1,$$
meaning that a 2D system described by $H^{(D)}_2(k_1,k_3)$ holds counterpropagating edge states~\cite{PhysRevLett.112.026805,PhysRevB.90.155443,PhysRevB.101.235438} at momentum $k_3=0$ and $\pi$ when the $r_1$ direction takes an OBC.
However, only one pair of $S_2^{m=2,a}$ (with $k=0$)
falls within the region with nonzero $v_1$,
and its nontrivial topology is captured by the surface states of $H^{(D)}_1$,
manifested as a single pair of chiral-like hinge states (on each surface) under OBCs along $r_1$ and $r_2$ directions, as shown in Fig.~\ref{Dgamma}(e).
In contrast, there is no hinge state when $r_2$ and $r_3$ directions take OBCs, as shown in Fig.~\ref{Dgamma}(f).
Compared with the results in Figs.~\ref{secondEX}(a) and ~\ref{secondEX}(b) for the same system, we can see that this $3$D second-order topological phase holds asymmetric topological properties in 2D planes lacking a $\mathcal{C}_{4}$ rotation symmetry, i.e., the topological boundary states may exist only along certain hinges of these planes.

%
%
%
%
%
%
%

\subsection{Effects of non-Clifford operators\label{sec62}}
Finally, we extend the nested-BIS method beyond the Clifford algebra by introducing extra non-anticommuting terms.
%
%
In general, non-Clifford operators can be obtained as the product of several operators from the Clifford algebra.
For the HOTPs we consider, coefficients of these non-Clifford operators are further restricted by the symmetry class of the system.
%
%
%
Explicitly, we take the Hamiltonian of Eq. \eqref{3dDsec} as an example, and introduce non-Clifford terms
as products of the operators of the two mass terms $M_{1,2}$, and one of the rest of the three terms $h_{1,2,3}$, given by 
\begin{eqnarray}
g_1\bar{\Gamma}^{1}\equiv ig_1(\sigma_x\tau_0)(\sigma_z\tau_x)(\sigma_y\tau_0)=g_1\sigma_0\tau_x,\nonumber\\
g_2\bar{\Gamma}^{2}\equiv ig_2(\sigma_x\tau_0)(\sigma_z\tau_x)(\sigma_z\tau_y)=-g_2\sigma_x\tau_z,\nonumber\\
g_3\bar{\Gamma}^{3}\equiv ig_3(\sigma_x\tau_0)(\sigma_z\tau_x)(\sigma_z\tau_z)=g_3\sigma_x\tau_y,\nonumber
\end{eqnarray}
with $g_{1,2,3}$ some $\mathbf{k}$-independent parameters.
Obviously, these terms keep the particle-hole symmetry $\mathcal{C} \bar{\Gamma}^{1,2,3} \mathcal{C}^{-1}=-\bar{\Gamma}^{1,2,3}$ with $\mathcal{C}=\sigma_z\tau_z\mathcal{K}$, hence the system remains in the D class with $Z\times Z$ topology.
On the other hand, certain spatial symmetries will be broken with nonzero $g_{1,2,3}$, leading to asymmetric boundary states  between different hinges of the 3D system.
Specifically, the original Hamiltonian with $g_{1,2,3}=0$ satisfies the $\mathcal{C}_4$ rotation symmetry of Eq. \eqref{C4_rot}, and three chiral-mirror symmetries associated with the three directions:
\begin{eqnarray}
-H^{({\rm D})}(k_1,k_2,k_3)&=&\sigma_y\tau_0H^{({\rm D})}(-k_1,k_2,k_3)\sigma_y\tau_0\nonumber\\
&=&\sigma_z\tau_yH^{({\rm D})}(k_1,-k_2,k_3)\sigma_z\tau_y\nonumber\\
&=&\sigma_z\tau_zH^{({\rm D})}(k_1,k_2,-k_3)\sigma_z\tau_z.
\label{eq:chiral-mirror}
\end{eqnarray}
Consequently, topological boundary states must emerge along different hinges related by these symmetries.
A nonzero $g_1$ breaks the chiral-mirror symmetry along the $r_1$ direction and induces asymmetric behavior between $(100)$ and $(\bar{1}00)$ surfaces. Thus, it is referred to as a longitudinal non-Clifford term henceforth.
In contrast, the other two terms are referred to as mixed non-Clifford terms,
as a nonzero $g_2$ ($g_3$) breaks not only the chiral-mirror symmetry along the $r_2$ ($r_3$) direction but also the $\mathcal{C}_4$ rotation symmetry.
These two mixed non-Clifford terms can be mapped to each other through the $\mathcal{C}_4$ rotation operation, $\mathcal{C}_{4}\bar{\Gamma}^{2}\mathcal{C}_{4}^{-1}=-\bar{\Gamma}^{3}$ \footnote{The minus sign means that $g_2$ functions as $-g_3$ after the rotation. This is associated with the minus sign in defining $\bar{\Gamma}_2$.}, and we shall discuss only the first one in detail.
Further, we assume these terms are relatively weak compared with other parameters, otherwise the system may be driven to other first-order topological phases, e.g., a topological semimetal~\cite{RevModPhys.90.015001,RevModPhys.93.025002} or a weak topological insulator~\cite{PhysRevLett.98.106803,PhysRevB.75.121306,PhysRevB.79.195322}, as discussed in Appendix \ref{appC} in more detail. 

\begin{figure}[!htp]
\includegraphics[width=1\columnwidth]{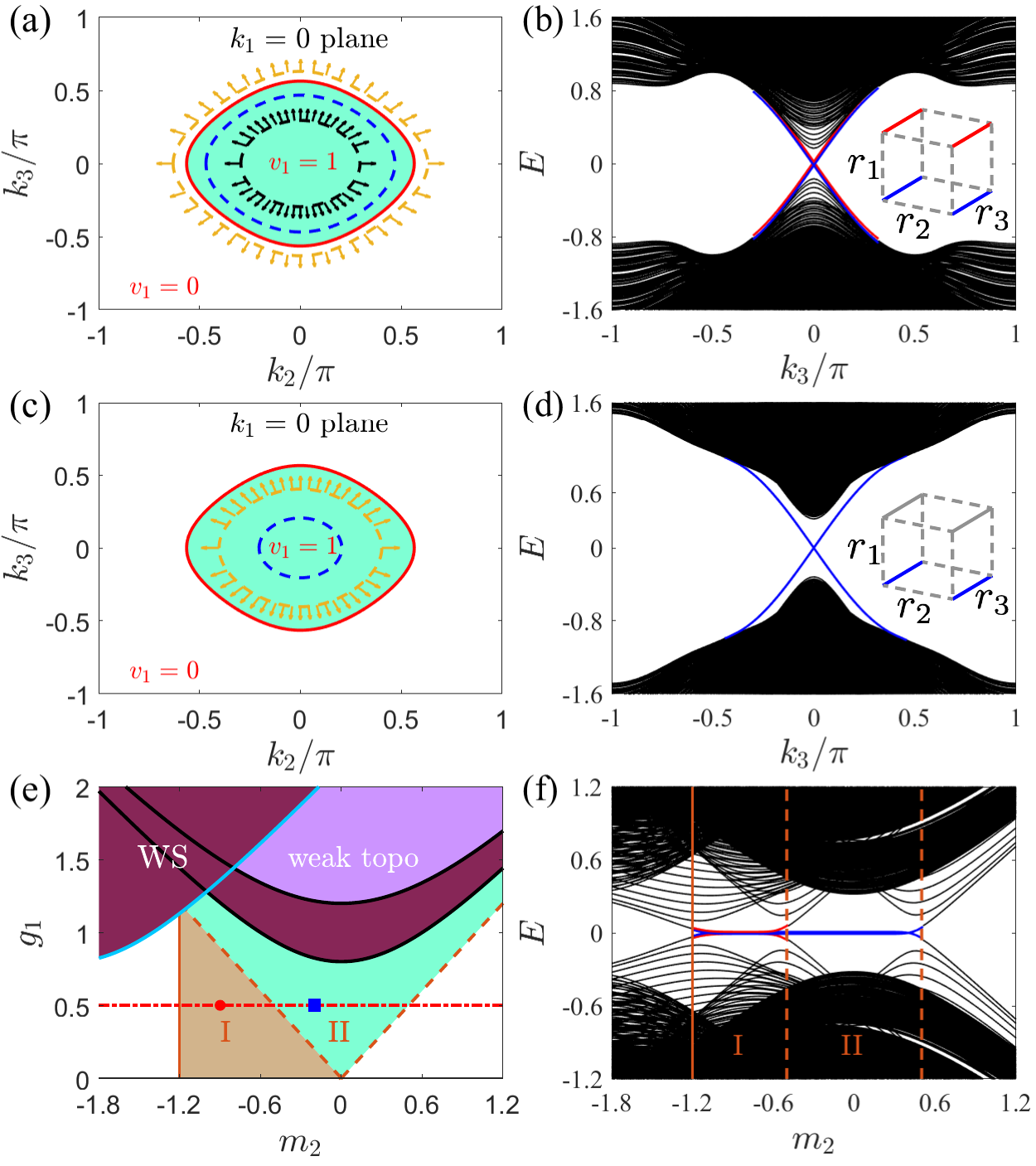}
\caption{\label{Dnoncli}(color online) (a) Nested BISs for the Hamiltonian $H^{(D)}(\mathbf{k})$ with the longitudinal non-Clifford term $g_1\sigma_0\tau_x$ of Eq.~\eqref{3dDsecnon}, with $m_1=-1.2$, $g_1=0.5$ and $m_2=-0.9$.
${S}_1^{m=2}$, ${S}_2^{m=1}$, and the vector $(h_2,h_3)$ are indicated by red solid loop, blue dashed loop, and blue arrows, respectively, and the extra BIS
$S_{+,2}^{m=1}$ ($S_{-,2}^{m=1}$) is marked by black (orange) dashed loops.
Arrows attached to BISs indicate the vector $(h_2,h_3)$. (b) Energy spectra corresponding to (a). Red and blue lines in the inset indicate the emergence of hinge states on $(100)$ and $(\overline{1}00)$ surfaces, respectively. (c), (d) The same results as in (a) and (b), but with $m_2=-0.2$. $S_{+,2}^{m=1}$ and hinge states on $(100)$ surface are seen to disappear.
(e) A phase diagram for Hamiltonian $H^{(D)}(\mathbf{k})$ of Eq.~\eqref{3dDsecnon} with $m_1=-1.2$. WS and ``weak topo" represent Weyl semimetallic and weak topological phases, which are discussed in the Appendix. Red dot (blue square) corresponds to the parameters used in (a) and (b) [(c) and (d)]. (f) Energy spectrum at $k_3=0$ along the red horizontal line ($g_1=0.5$) in (e), where $r_1$- and $r_2$ directions take OBCs.}
\end{figure}
%
%
%

\subsubsection{HOTPs with the longitudinal non-Clifford term}
First, we add the longitudinal non-Clifford term $g_1\sigma_0\tau_x$ to the Hamiltonian of Eq.~\eqref{3dDsec}, and rewrite it as:
\begin{eqnarray}\label{3dDsecnon}
H^{({\rm D})}(\mathbf{k})=&&H^{({\rm D})}_1(\mathbf{k})+
H^{({\rm D})}_2(\mathbf{k}_{1,\parallel}), \nonumber \\
H^{({\rm D})}_1(\mathbf{k})=&&M_1(\mathbf{k})\sigma_x\tau_0+h_{1}(\mathbf{k})\sigma_y\tau_0, \nonumber \\
H^{({\rm D})}_2(\mathbf{k}_{1,\parallel})=&&M_2(\mathbf{k}_{1,\parallel})\sigma_z\tau_x+g_1\sigma_0\tau_x \nonumber \\
&&+h_{2}(\mathbf{k}_{1,\parallel})\sigma_z\tau_y+h_{3}(\mathbf{k}_{1,\parallel})\sigma_z\tau_z.
\end{eqnarray}
In this decomposition, surface states of $H^{({\rm D})}_1(\mathbf{k})$ along the $r_1$ direction appear when the first topological invariant $v_1\neq0$ [defined in Eq. \eqref{windingDse}].
To see how the topological properties of $H^{({\rm D})}_2(\mathbf{k}_{1,\parallel})$ are captured by these surface states, we obtain effective 2D Hamiltonians for surface states through the projection of $P_{1,\pm}$,
\begin{eqnarray}\label{effH}
H^{({\rm D})}_{{\rm eff},\pm}=&&P_{1,\pm}H^{({\rm D})}_2P_{1,\pm} \nonumber \\
=&&P_{1,\pm}[(M_2\pm g_1)\sigma_z\tau_x+h_{2}\sigma_z\tau_y,
+h_{3}\sigma_z\tau_z]P_{1,\pm},\nonumber \\
P_{1,\pm}=&&[1\pm i(\sigma_y\tau_0)(\sigma_x\tau_0)]/2=(1\pm\sigma_z\tau_0)/2,
\end{eqnarray}
whose topology can be determined by their 1-BIS $S_{\pm,2}^{m=1}: {M}_2\pm g_1=0$.
Obviously the effective Hamiltonians for surface states on the $(100)$ and $(\bar{1}00)$ surfaces are different in the presence of the longitudinal non-Clifford term, which breaks the chiral-mirror symmetry $\sigma_y\tau_0H^{({\rm D})}(-k_1,k_2,k_3)\sigma_y\tau_0=-H^{({\rm D})}(k_1,k_2,k_3)$.
As a consequence, asymmetric behavior for these two surfaces are expected for the topological boundary states.

Interestingly, although the first-order topology of $H^{({\rm D})}_{{\rm eff},\pm}$ is associated with their BISs $S^{m=1}_{\pm,2}$,
it is the nested relation associated with $S^{m=1}_2$ for $H_2^{({\rm D})}$ that determines whether their topological properties are inherited by the surface states of $H_1^{({\rm D})}$, as shown by two typical examples in Figs. \ref{Dnoncli}(a)-\ref{Dnoncli}(d).
In Fig. \ref{Dnoncli}(a), both $H^{({\rm D})}_{{\rm eff},\pm}$ are topologically nontrivial,
yet their BISs $S^{m=1}_{\pm,2}$ (black and yellow dashed loops) fall outside and inside the nontrivial region of $v_1$, respectively.
Nevertheless,
$S^{m=1}_{2}$ falls within the nontrivial region of $v_1$,
and chiral-like hinge states emerge on both $(100)$ and $(\bar{1}00)$ surfaces, as shown by the spectrum with OBCs along $r_1$ and $r_2$ directions [Fig.~\ref{Dnoncli}(b)].
In Fig. \ref{Dnoncli}(c), $H^{({\rm D})}_{{\rm eff},+}$ becomes topologically trivial and its BIS $S^{m=1}_{+,2}$ disappears. Consistently, chiral-like hinge states exist only on $(\bar{1}00)$ surfaces, as shown in Fig.~\ref{Dnoncli}(d)  with the same boundary conditions.
We note that results with OBCs along $r_1$ and $r_3$ directions are identical, as the system possesses the $\mathcal{C}_{4}$ rotation symmetry in $k_2$-$k_3$ plane.

In Fig.~\ref{Dnoncli}(e), we display the phase diagram in $g_1$-$m_2$ parameter space with constant value of $m_1$.
The corresponding energy spectrum at $k_3=0$, namely the crossing point of the chiral-like hinge states, is displayed in Fig.~\ref{Dnoncli}(f) for $g_1=0.5$.
The two topologically nontrivial phases can thus be identified by the number of zero-energy states in the spectrum, denoted as phases I and II in these two panels.
The phase boundaries can also be classified into two types. The first one is marked by the orange solid line at $m_2=m_1=-1.2$, which is independent from $g_1$. It stands for the case where $S^{m=1}_2$, the BIS for $H_2^{({\rm D})}$, coincides with $S^{m=2}_1$, the BIS for $H_1^{({\rm D})}$ separating trivial and nontrivial regions of $v_1$. For $m_2<-1.2$, $S^{m=1}_2$ falls outside the nontrivial region of $v_1$ and the second-order topology becomes trivial.
On the other hand, the other types of phase boundaries are marked by the orange dashed lines, where one of $S^{m=1}_{\pm,2}$ vanishes, and the corresponding surface Hamiltonian $H^{({\rm D})}_{{\rm eff},\pm}$ becomes trivial.
Notably, in our system, $S^{m=1}_{2}$ will vanish before $S^{m=1}_{-,2}$ when increasing $m_2$.
However, such a transition does not involve any crossing of different BISs, and hence shall not change the inheriting relation of topology.
To conclude our results,
nontrivial topology of $H_{{\rm eff},\pm}^{({\rm D})}$ characterized by their own BISs $S^{m=1}_{\pm,2}$ are inherited by the surface states of $H_1^{({\rm D})}$, as long as the BIS $S^{m=1}_2$ for $H_2^{({\rm D})}$ falls, or even vanishes, within the nontrivial region of $v_1$.

Finally, we note that increasing $g_1$ can drive the system into a topological semimetal~\cite{RevModPhys.90.015001,RevModPhys.93.025002} or a weak topological insulator~\cite{PhysRevLett.98.106803,PhysRevB.75.121306,PhysRevB.79.195322}, as indicated in the phase diagram of Fig.~\ref{Dnoncli}(e).
In such cases, surface states protected by 2D first-order topology will appear, instead of the higher-order hinge states characterized by our current method of nested BISs (see Appendix \ref{appC} for more details).

\begin{figure}[!htp]
\includegraphics[width=1\columnwidth]{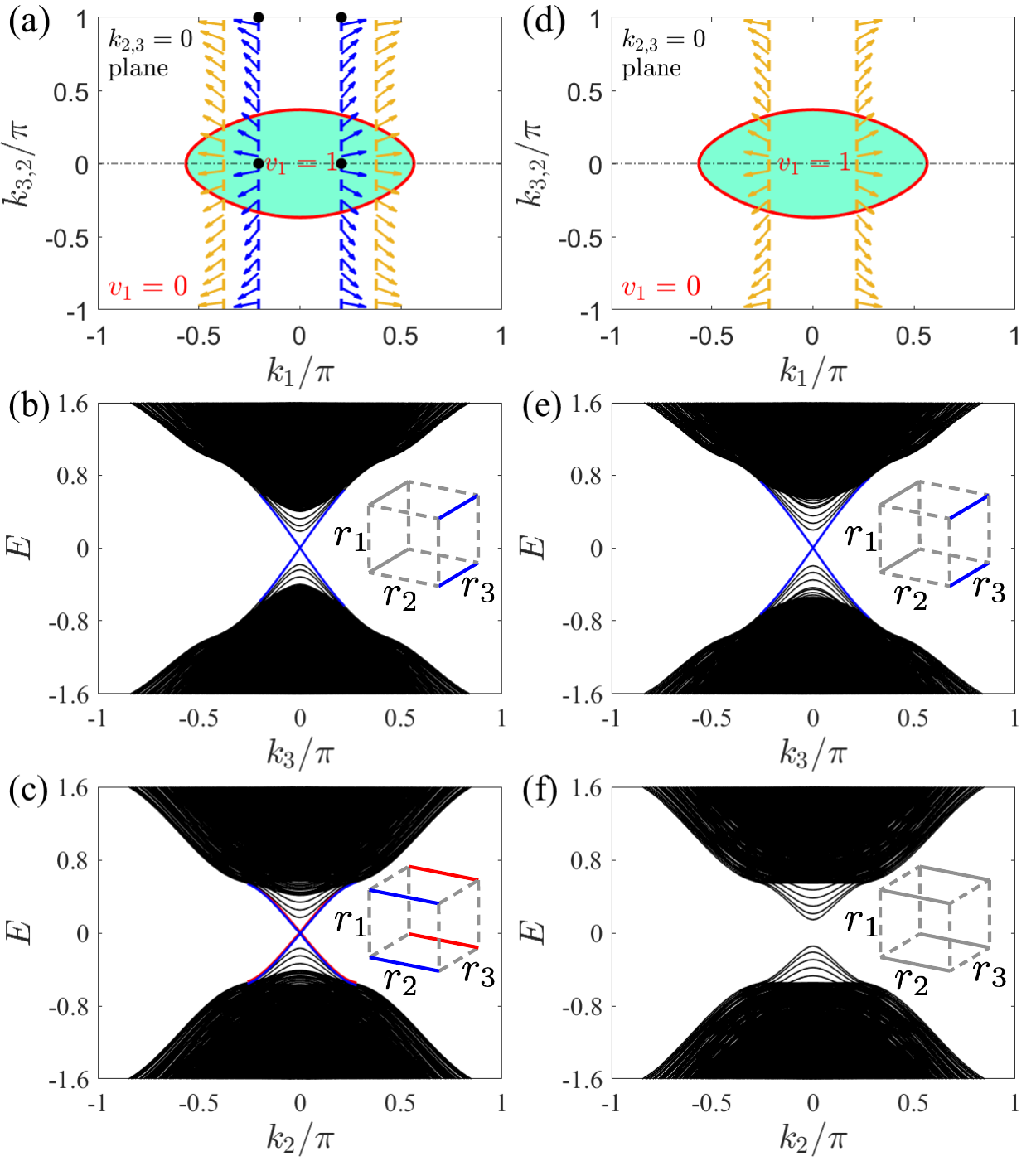}
\caption{\label{Dnonclig2}(color online) (a) Nested BIS for Hamiltonian $H^{(D)}(\mathbf{k})$ with the mixed non-Clifford term $-g_2\sigma_x\tau_z$, with $m_+\equiv m_1+m_2=1.2$, $m_-\equiv m_1-m_2=-0.2$ and $g_2=0.3$.
${S}_1^{m=2}$ and ${S}_2^{m=1}$ are indicated by red solid loop and blue dashed lines, respectively.
Black dots mark the 2-BIS ${S}_2^{m=2}$ of $H_2^{({\rm D})}$.
These BISs are identical for the two decomposition of Eqs. \eqref{3dDsectramg2} and \eqref{3dDsectramg3}
upon a transformation $k_2\leftrightarrow k_3$, and hence we put them in the same figure for convenience.
Yellow dash lines indicate $S^{m=1}_{-,2}$ in $k_1$-$k_3$ plane for the effective Hamiltonian in Eqs. \eqref{effHg2}, with $S^{m=1}_{+,2}$ vanishes for the chosen parameters.
For Eqs. \eqref{effHg3}, both $S^{m=1}_{\pm,2}$ in $k_1$-$k_2$ plane are identical to $S^{m=1}_2$ for $H_2^{({\rm D})}$ in Eqs. \eqref{3dDsectramg3}.
Yellow and blue arrows represent the vectors $(h_1,h_3)$ and  $(h_1,h_2)$ along their corresponding BISs, respectively.
(b), (c) Energy spectra corresponding to (a), with OBCs along $r_1/r_2$ and $r_1/r_3$ directions, respectively. Red and blue lines in the insets indicate the hinge states on $(010)$ and $(0\overline{1}0)$ [$(001)$ and $(00\overline{1})$] surface for (b) [(c)].
(d)-(f) The same results as in (a)-(c) but with $m_-=0.2$. The BIS $S^{m=1}_2$ vanishes in (d). Correspondingly, no hinge state exists in (f) with OBCs along $r_1/r_3$ directions.
}
\end{figure}

\begin{figure}[!htp]
\includegraphics[width=1\columnwidth]{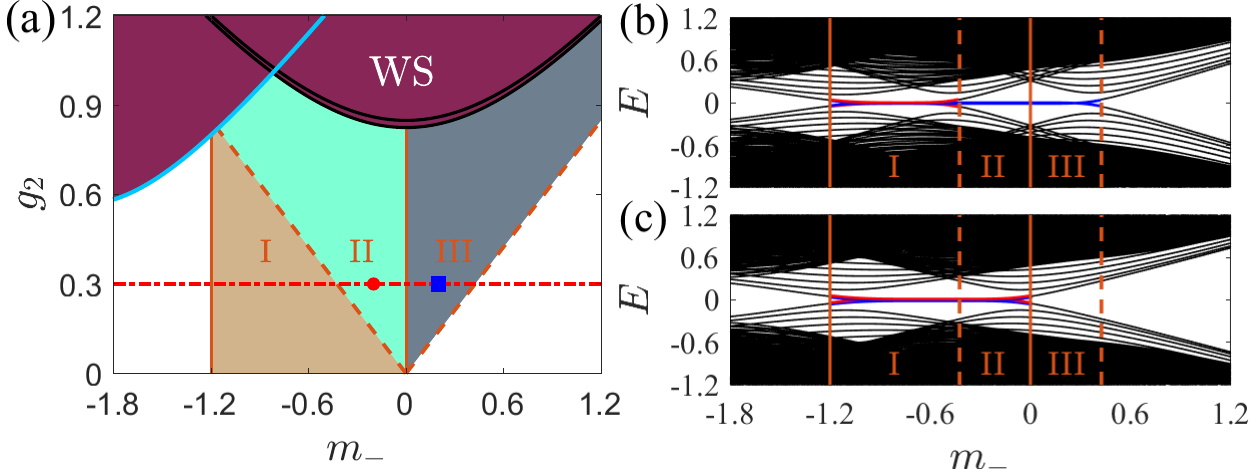}
\caption{\label{Dnonclig2phase}(color online) (a) A phase diagram for Hamiltonian $H^{({\rm D})}(\mathbf{k})$  of Eq.~\eqref{3dDsectramg2}, with the mixed non-Clifford term $-g_2\sigma_x\tau_z$ and $m_{+}=-1.2$. Parameters used in Fig.~\ref{Dnonclig2}(a)-(c) [(e), (f)] is marked with a red dot (blue square).
(b) Energy spectrum at $k_3=0$ along the red horizontal lines ($g_2=0.3$) in (a), with OBCs taken for $r_1$ and $r_2$ directions.
(c) Energy spectrum at $k_2=0$ along the red horizontal lines in (a), with OBCs taken for $r_1$ and $r_3$ directions.
For $m_-<-1/2$, $S_2^{m=2}$ falls outside $S_1^{m=2}$ for both cases, and the zero-energy states disappear in both (b) and (c).
At the two orange dashed lines, one of $S_{\pm,2}^m=1$ defined for $H_{{\rm eff},\pm}^{({\rm D})}$ in Eq. \eqref{effHg2} vanishes, and a pair of zero-energy states disappears in (b) when $m_-$ exceeds each dashed line.
When $m_->0$, $S_{\pm 2}^{m=2}$ (identical to $S_{2}^{m=2}$) for $H_{{\rm eff},\pm}^{({\rm D})}$ in Eq. \eqref{effHg3} vanishes, meaning these effective Hamiltonians for $(001)$ and $(00\bar{1})$ become topologically trivial, and zero-energy states disappear in (c).}
\end{figure}

\subsubsection{HOTPs with a mixed non-Clifford term}
Next we consider the effect of adding $-g_2\sigma_x\tau_z$, one of the two mixed non-Clifford terms, to the Hamiltonian of Eq.~\eqref{3dDsec}.
We find that in this case, it is more convenient to use the decomposition in Eq.~\eqref{3dDsectram} and rewrite the Hamiltonian as
\begin{eqnarray}\label{3dDsectramg2}
H^{({\rm D})}(\mathbf{k})=&&H^{({\rm D})}_1(\mathbf{k})+
H^{({\rm D})}_2(k_1,k_3),\nonumber\\
H^{({\rm D})}_1(\mathbf{k})=&&\widetilde{M}_1(\mathbf{k})(\sigma_x\tau_0+\sigma_z\tau_x)/\sqrt{2}+h_{2}(\mathbf{k})\sigma_z\tau_y, \nonumber \\
H^{({\rm D})}_2(k_1,k_3)=&&\widetilde{M}_2(k_1,k_3)(\sigma_x\tau_0-\sigma_z\tau_x)/\sqrt{2}-g_2\sigma_x\tau_z\nonumber \\
&&+h_{1}(k_1,k_3)\sigma_y\tau_0
+h_{3}(k_1,k_3)\sigma_z\tau_z,
\end{eqnarray}
where $H^{({\rm D})}_1(\mathbf{k})$ now determines the surface states along $r_2$-directions, i.e. on $(010)$ and $(0\overline{1}0)$ surfaces.
Regarding the (anti-)commuting relations between different components, this decomposition takes a similar form as the previous case with the longitudinal non-Clifford term, and hence the conclusion is also expected to apply here. However, due to the absence of a $\mathcal{C}_4$ rotation symmetry in the $k_1$-$k_3$ plane, the BISs may cross each other and induce asymmetric behaviors between hinge states along $r_1$ and $r_3$ directions, as discussed in Sec~\ref{sec61}.
To see this, we first write the effective Hamiltonians
 for the surface states on $(010)$ and $(0\overline{1}0)$ surfaces through the corresponding projection operators:
\begin{eqnarray}\label{effHg2}
H^{({\rm D})}_{{\rm eff},\pm}=&&P_{2,\pm}H^{({\rm D})}_2P_{2,\pm} \nonumber \\
=&&P_{2,\pm}[(\widetilde{M}_2\pm g_2)(\sigma_x\tau_0-\sigma_z\tau_x)/\sqrt{2}\nonumber \\
&&+h_{1}\sigma_y\tau_0,
+h_{3}\sigma_z\tau_z]P_{2,\pm},\nonumber \\
P_{2,\pm}=&&[1\pm i(\sigma_z\tau_y)(\sigma_x\tau_0+\sigma_z\tau_x)/\sqrt{2}]/2\nonumber \\
=&&[1\pm(\sigma_0\tau_z-\sigma_y\tau_y)/\sqrt{2}]/2.
\end{eqnarray}

Following the previous discussion, we need to consider a BIS for each of $H^{({\rm D})}_{{\rm eff},\pm}$, defined as
$$S_{\pm,1}^{m=2}: \widetilde{M}_2\pm g_2=0,$$
which characterizes topological properties of the effective Hamiltonians,
and another one for $H^{({\rm D})}_2$, which determines whether these topological properties are inherited by surface states along the $r_2$ direction and manifested as chiral-like hinge states.
Similarly to the cases in Sec. \ref{sec61}, the 1-BIS of $H^{({\rm D})}_2$  crosses the 2-BIS,
$$S^{m=2}_1: \widetilde{M}_1=h_2=0$$
defined for $H^{({\rm D})}_1$, as shown in Fig. \ref{Dnonclig2}(a) and (d).
Therefore, to apply the nested-BIS method, we shall follow our discussion in Sec. \ref{sec61} and consider a 2-BIS of $H^{({\rm D})}_2$:
$$S_{2}^{m=2}: \widetilde{M}_2=h_3=0.$$
We can see in Figs. \ref{Dnonclig2}(a) and \ref{Dnonclig2}(d) that $S_{2}^{m=2}$ falls and vanishes within the nontrivial regime of $\nu_1$, respectively, so the nontrivial topology of $H^{({\rm D})}_{{\rm eff},-}$ is manifested as chiral-like hinge states on the $(0\bar{1}0)$ surface as shown in Figs. \ref{Dnonclig2}(b) and \ref{Dnonclig2}(e).
For the parameters chosen in these figures, $H^{({\rm D})}_{{\rm eff},+}$ is topologically trivial as $S_{+,1}^{m=2}$ disappears, and no hinge state emerges on the $(010)$ surface.
The physics behind this asymmetric behavior is straightforward: a nonzero $g_2$ breaks the chiral-mirror symmetry along the $k_2$ direction [see Eq. \eqref{eq:chiral-mirror}], and may induce different topological properties on $(010)$ and $(0\overline{1}0)$ surfaces.

In contrast, the other two chiral-mirror symmetries along $k_1$ and $k_3$ directions are not violated by $g_2$, and symmetric behaviors are expected for these two directions. In Figs. \ref{Dnonclig2}(b) and \ref{Dnonclig2}(e), we already see that the hinge states are symmetric between $(100)$ and $(\bar{1}00)$ surfaces. To analyze hinge states on $(001)$ and $(00\bar{1})$ surfaces,
we consider a different decomposition and rewrite the Hamiltonian as
\begin{eqnarray}\label{3dDsectramg3}
H^{({\rm D})}(\mathbf{k})=&&H^{({\rm D})}_1(\mathbf{k})+
H^{({\rm D})}_2(k_1,k_2),\nonumber\\
H^{({\rm D})}_1(\mathbf{k})=&&\widetilde{M}_1(\mathbf{k})(\sigma_x\tau_0+\sigma_z\tau_x)/\sqrt{2}+h_{3}(\mathbf{k})\sigma_z\tau_z, \nonumber \\
H^{({\rm D})}_2(k_1,k_2)=&&\widetilde{M}_2(k_1,k_2)(\sigma_x\tau_0-\sigma_z\tau_x)/\sqrt{2}-g_2\sigma_x\tau_z\nonumber \\
&&+h_{1}(k_1,k_2)\sigma_y\tau_0
+h_{2}(k_1,k_2)\sigma_z\tau_y,
\end{eqnarray}
so $k_3$ is contained only in $H^{({\rm D})}_1(\mathbf{k})$. Next, the effective Hamiltonians for surface states on $(001)$ and $(00\bar{1})$ surfaces are given by projecting $H^{({\rm D})}_2$ on these two surfaces,
\begin{eqnarray}\label{effHg3}
H^{({\rm D})}_{{\rm eff},\pm}=&&P_{3,\pm}H^{({\rm D})}_2P_{3,\pm} \nonumber \\
=&&P_{3,\pm}[\widetilde{M}_2(\sigma_x\tau_0-\sigma_z\tau_x)/\sqrt{2}\nonumber \\
&&+h_{1}\sigma_y\tau_0
+h_{2}\sigma_z\tau_y]P_{3,\pm},\nonumber \\
P_{3,\pm}=&&[1\pm i(\sigma_z\tau_z)(\sigma_x\tau_0+\sigma_z\tau_x)/\sqrt{2}]/2\nonumber \\
=&&[1\mp(\sigma_y\tau_z+\sigma_0\tau_y)/\sqrt{2}]/2.
\end{eqnarray}

Comparing Eqs. \eqref{3dDsectramg3} and \eqref{effHg3} with Eqs. \eqref{3dDsectramg2} and \eqref{effHg2},
we can see that the two BISs defined for the latter case,
$$S^{m=2}_1: \widetilde{M}_1=h_3=0,~~
S_{2}^{m=2}: \widetilde{M}_2=h_2=0,$$
are identical to those of the previous case upon exchanging two momentum components, $k_2\leftrightarrow k_3$.
However, now that $g_2$ does not enter the two effective Hamiltonians in Eq. \eqref{effHg2}, their BISs become the same as $S_{2}^{m=2}$,
and appearance of chiral-like hinge states on $(001)$ and $(00\bar{1})$ surfaces are not affected by this mixed non-Clifford term.
This prediction is consistent with our numerical results for OBCs along $r_1$ and $r_3$ directions.
In Fig. \ref{Dnonclig2}(c), a pair of chiral-like hinge states is seen on each of the $(001)$ and $00\bar{1})$ surfaces, as $S_{2}^{m=2}$ falls within the nontrivial region of $v_1$ in Fig. \ref{Dnonclig2}(a).
On the other hand, Fig. \ref{Dnonclig2}(f) represents a topologically trivial case without any hinge state, as $S_{2}^{m=2}$ vanishes in Fig. \ref{Dnonclig2}(d), indicating trivial topology of $H^{({\rm D})}_{{\rm eff},\pm}$ in Eq. \eqref{effHg3}.
Compared with Figs. \ref{Dnonclig2}(b) and \ref{Dnonclig2}(e), it is also seen that hinge states behave differently under OBCs along $r_2$ and $r_3$ directions, as a consequence of breaking the $\mathcal{C}_4$ rotation symmetry by a nonzero $g_2$.

Combining these results, we obtain a phase diagram by analyzing the BISs of the system, as shown in Fig. \ref{Dnonclig2phase}(a). In phase I, chiral-like hinge states emerge on both surfaces along either $r_2$ or $r_3$ directions, analogous to phase I for the case with the longitudinal non-Clifford term.
Phases II and III are two topological phases where hinge states emerge asymmetrically between these two directions, as shown by the two examples in Fig. \ref{Dnonclig2}. To give a clear view of the topological phase transitions, we display the energy spectrum at $k_3=0$ under OBCs along $r_1$ and $r_2$ directions in Fig. \ref{Dnonclig2phase}(b), and that at $k_2=0$ under OBCs along $r_1$ and $r_3$ directions in Fig. \ref{Dnonclig2phase}(b).
The appearance and disappearance of topological hinge states are seen to match the topological transition predicted by BISs very well.
Also, when $g_2$ becomes larger, the system will enter semimetallic phases~\cite{RevModPhys.90.015001,RevModPhys.93.025002} as shown in Fig. \ref{Dnonclig2phase}(a), which can support surface states protected by 2D first-order topology (see Appendix \ref{appC} for more details).

\section{Discussion\label{Sec7}}
The concept of HOTPs has broadly extended our knowledge of topological phases of matter,
as many systems previously considered trivial according to the standard AZ symmetry classification
have been found to support higher-order topological states at boundaries of boundaries.
In this paper, we exhaustively investigate the emergence of HOTPs based on the AZ classification, and propose a universal scheme to construct HOTPs in each AZ class with the nested-BIS method.
An $n$th-order topological phase constructed in this way is topologically characterized by the geometric (nested) relation between $n$ BISs of the system, defined as where certain pseudospin components vanish in the BZ, and by $Z^n$, $2Z^n$, or $Z^{n-1}\times Z_2$ topological invariants, depending on which symmetry class the system belongs to.
These results are unveiled with both general minimal continuous Hamiltonians,
and several example lattice models with the continuous Hamiltonians taken as effective Hamiltonians at some high-symmetric points.
While the lattice examples considered here are either 2D or 3D, our scheme can apply to much more general scenarios without restriction of spatial dimension or order of topology.
To generalize our discussion,
we further consider cases with crossed BISs and/or non-Clifford operators, where higher-order boundary states become asymmetric due to the breaking of certain spatial symmetries by extra modulations to the Hamiltonian, allowing us to tune the configuration of higher-order boundary states in a flexible way.

%
%
%
%
%

In addition to offering a theoretical tool for investigating HOTPs from lower-order topology, our scheme is also useful for engineering HOTPs and probing their topological properties in various experiments.
The explicit construction using our scheme involves proper design of different pseudospin components, thus it is most applicable for quantum simulation with single or a few qubits,
with their parameters serving as a crystal momentum to form a synthetic BZ \cite{LI20211817,experimental2018ma,tan2019simulation,ji2020quantum,xin2020quantum,roushan2014observation}.
The BISs and corresponding topological invariants can be probed through measuring time-averaged pseudospin texture over long-time dynamics \cite{ZHANG20181385,PRXQuantum.2.020320,LI2021,PhysRevA.99.053606,PhysRevA.100.063624,PhysRevLett.125.183001,LU20202080,PhysRevResearch.3.013229,ScienceBulletin671236,2209.10394},
which has already been realized in several experimental platforms, such as superconducting qubits \cite{niu2021simulation} and ultracold atoms \cite{yi2019observing,Wang271}.
The asymmetric properties of higher-order boundary states also hint at potential applications in topological materials realizing higher-order hinge states \cite{PhysRevLett.124.156601,WANG2022788}.
Intriguingly, 3D topological insulator materials with noncollinear antiferromagnetic order~\cite{doi:10.1126/sciadv.aat0346} hold second-order topology and can be described by Hamiltonian with the a form like Eq.~\eqref{3dDsec}. A detailed discussion is given in Appendix \ref{appadd}. Similarly, the topological superconductors with $d$-wave pairing and $s_{\pm}$-wave paring~\cite{PhysRevLett.121.096803,PhysRevLett.121.186801} also can be characterized by our nested-BIS method and classed into DIII in Table.~\ref{Classification}.

Furthermore, we note that while the current study focuses only on gapped HOTPs,
we also expect our method, with proper modifications, to be applicable in the study of higher-order semimetallic phases~\cite{PhysRevB.98.241103,Wieder2020,PhysRevB.100.020509,PhysRevLett.125.266804,PhysRevB.101.205134,PhysRevB.101.220506,PhysRevLett.125.146401,PhysRevLett.127.066801}. 
Another encouraging future direction is to extend our method to HOTPs with interaction~\cite{PhysRevLett.127.176601,PhysRevB.98.235102}. 
A relevant study has shown that certain interacting topological phases can be characterized by topological indices defined at high-symmetric momenta~\cite{PhysRevB.85.165126}, which can be viewed as a type of symmetry-protected BISs, 
suggesting a promising route of extending our method to HOTPs with interaction.

%

\begin{acknowledgments}
This work is supported by the National Key R$\&$D Program of China (Grant No. 2018YFA0307500), the NSFC (Grants No. 12104519, No. 11874433, No. 12135018), and the Guangdong Basic and Applied Basic Research Foundation (2020A1515110773).
\end{acknowledgments}

\appendix

\section{Classification of general higher-order topological phases}\label{appA}
\begin{table*}
\centering
\caption{Classification of $n$th-order HOTPs based on the nested-BIS method.
The continuous Hamiltonians for topological phases characterized by $n$ $Z$-invariants are provided with $H^{({I})}$ in Eq.~\eqref{thH_1}, $H^{(II)}$ in Eq.~\eqref{thH_2}, $H^{(III)}$ in Eq.~\eqref{thH_3}, and $H^{(IV)}$ in Eq.~\eqref{thH_4} of the main text. Note that in order to support $n$th-order boundary states in a lattice model, the spatial dimension must be $d\geqslant n$.}
\label{Classificationhigher}
\begin{tabular}{|c||c|c|c||c|c|c|c|c|c|c|c|}
   \hline
    \multirow{2}{*}{Class}&\multicolumn{3}{c||}{Symmetry}& \multicolumn{8}{c|}{$d-n+1$ (mod $8$)}\\ \cline{2-12}
    & ~$\mathcal{T}$~ & ~$\mathcal{C}$~ & ~$\mathcal{S}$~ & 0 & 1 & 2 & 3 & 4 & 5& 6 & 7  \\
\hhline{|============|}
${\rm A}$   & 0 & 0 & 0 & $Z^n$ & 0 & $Z^n$ & 0 & $Z^n$&  0 & $Z^n$ & 0  \\
\hline
${\rm AIII}$   & 0 & 0 & $1$           & 0 & $Z^n$ & 0 & $Z^n$ & 0 & $Z^n$ &0 & $Z^n$  \\
\hline
${\rm AI}$ & $+$ & 0 & 0 & \tabincell{c}{$Z^n$\\$H^{(\uppercase\expandafter{\romannumeral1})}$} & 0 & 0 & 0& \tabincell{c}{$2Z^n$\\$H^{(\uppercase\expandafter{\romannumeral3})}$} & 0 & $Z^{n-1}\times Z_2$ & $Z^{n-1}\times Z_2$  \\
\hline
${\rm BDI}$   & $+$ & $+$ & $1$                               & $Z^{n-1}\times Z_2$ & \tabincell{c}{$Z^n$\\$H^{(\uppercase\expandafter{\romannumeral2})}$} & 0 & 0 & 0& \tabincell{c}{$2Z^n$\\$H^{(\uppercase\expandafter{\romannumeral4})}$} & 0 & $Z^{n-1}\times Z_2$ \\
\hline
{\rm D}                           & 0 & $+$ & 0          & $Z^{n-1}\times Z_2$ & $Z^{n-1}\times Z_2$ & \tabincell{c}{$Z^n$\\$H^{(\uppercase\expandafter{\romannumeral1})}$} & 0 & 0& 0 & \tabincell{c}{$2Z^n$\\$H^{(\uppercase\expandafter{\romannumeral3})}$}  & 0\\
\hline
${\rm DIII}$      & $-$ & $+$ & $1$     & 0 & $Z^{n-1}\times Z_2$ & $Z^{n-1}\times Z_2$ & \tabincell{c}{$Z^n$\\$H^{(\uppercase\expandafter{\romannumeral2})}$} & 0& 0 & 0  & \tabincell{c}{$2Z^n$\\$H^{(\uppercase\expandafter{\romannumeral4})}$}\\
\hline
${\rm AII}$      & $-$ & 0 & 0        &\tabincell{c}{$2Z^n$\\$H^{(\uppercase\expandafter{\romannumeral3})}$} & 0 & $Z^{n-1}\times Z_2$ & $Z^{n-1}\times Z_2$ & \tabincell{c}{$Z^n$\\$H^{(\uppercase\expandafter{\romannumeral1})}$}& 0 & 0 & 0 \\
\hline
${\rm CII}$      & $-$ & $-$ & $1$      & 0 & \tabincell{c}{$2Z^n$\\$H^{(\uppercase\expandafter{\romannumeral4})}$} & 0 & $Z^{n-1}\times Z_2$ & $Z^{n-1}\times Z_2$& \tabincell{c}{$Z^n$\\$H^{(\uppercase\expandafter{\romannumeral2})}$} & 0  & 0 \\
\hline
${\rm C}$                               & 0 & $-$ & 0         & 0 & 0 & \tabincell{c}{$2Z^n$\\$H^{(\uppercase\expandafter{\romannumeral3})}$} & 0 & $Z^{n-1}\times Z_2$& $Z^{n-1}\times Z_2$ & \tabincell{c}{$Z^n$\\$H^{(\uppercase\expandafter{\romannumeral1})}$} & 0 \\
\hline
${\rm CI}$      & $+$ & $-$ & $1$            & 0 & 0 & 0& \tabincell{c}{$2Z^n$\\$H^{(\uppercase\expandafter{\romannumeral4})}$} & 0& $Z^{n-1}\times Z_2$ & $Z^{n-1}\times Z_2$  & \tabincell{c}{$Z^n$\\$H^{(\uppercase\expandafter{\romannumeral2})}$}\\
\hline
\end{tabular}
\end{table*}

In the main text, we have provided a table of symmetry classification for second-order topological phases.
In this appendix, we discuss the behavior when the order of topology increases and give a general classification of the HOTPs phases based on the nested-BIS method.

As described in Sec.~\ref{Sec32} of the main text, a $d$D Hamiltonian $H^{(d,n)}(\mathbf{k})$ supporting $n$-order HOTPs with $Z^n$ topological invariant contains $J=d+n$ anti-commuting terms, including several purely real matrices coupled to crystal momentum.
A corresponding anti-unitary symmetry operator $\hat{A}$ is given by the product of these real matrices and the complex conjugate operator $\mathcal{K}$, such as in Eqs.\eqref{squareA1} and \eqref{antiuA3} of the main text.
In addition,
if this Hamiltonian supports chiral symmetry with the symmetry operator $\hat{S}$,
another anti-unitary symmetry operator can be obtained as $\hat{A}\hat{S}$,
which is also a product of real matrices since $\hat{S}$ is purely real in our construction.
Therefore all anti-unitary symmetry operators in our models are given by the product of the complex conjugate operator $\mathcal{K}$  and several purely real matrices, which may be coupled only to crystal momentum in the Hamiltonian.
%
%

Without loss of generality, we assume $H^{(d,n)}(\mathbf{k})$ holds an anti-unitary symmetry
\begin{eqnarray}
\hat{A}H^{(d,n)}(\mathbf{k})\hat{A}^{-1}=(-1)^{a}H^{(d,n)}(-\mathbf{k})\label{eq:app_sym1},
\end{eqnarray}
with $\hat{A}^2=(-1)^{a(a-1)/2}$,
where $\hat{A}$ is given by the product of $a$ purely real matrices and $\mathcal{K}$.
Next, we choose a momentum component of the Hamiltonian and transform it into a mass term, i.e.
$k_{\beta}\Gamma_{(2p+1)}^{2\alpha+1}\rightarrow m\Gamma_{(2p+1)}^{2\alpha+1}$.
In this way we obtain a $(d-1)$D Hamiltonian $H^{(d-1,n+1)}(\mathbf{k})$ supporting $(n+1)$-order HOTPs.
Obviously, this mass term breaks the above anti-unitary symmetry,
$\hat{A}m\Gamma_{(2p+1)}^{2\alpha+1}\hat{A}^{-1}\neq(-1)^{a}m\Gamma_{(2p+1)}^{2\alpha+1}$.
Instead, we can define another anti-unitary symmetry $\hat{A}'=\Gamma_{(2p+1)}^{2\alpha+1}\hat{A}$, equivalent to removing $\Gamma_{(2p+1)}^{2\alpha+1}$ from $\hat{A}$. That is, the operator $\hat{A}'$ is given by of $(a-1)$ purely real matrices, satisfying
$$\hat{A'}m\Gamma_{(2p+1)}^{2\alpha+1}\hat{A'}^{-1}=(-1)^{a-1}m\Gamma_{(2p+1)}^{2\alpha+1}.$$
Furthermore,
the rest of the Hamiltonian does not contain $\Gamma_{(2p+1)}^{2\alpha+1}$ and hence also satisfies a condition similar to Eq. \eqref{eq:app_sym1}, leading to an anti-unitary symmetry for the whole system, described by
\begin{eqnarray}
\hat{A'}H^{(d-1,n+1)}(\mathbf{k})\hat{A'}^{-1}=(-1)^{a-1}H^{(d-1,n+1)}(\mathbf{-k}).\nonumber\\\label{eq:app_sym2}
\end{eqnarray}
Therefore, if $\hat{A}$ describes a particle-hole symmetry $\mathcal{C}$ with $\hat{A}^2=\pm1$ (time-reversal symmetry $\mathcal{T}$ with $\hat{A}^2=\pm1$) for $H^{(d,n)}(\mathbf{k})$,
$\hat{A}'$ describes a time-reversal symmetry $\mathcal{T}$ with $[\hat{A}']^2=\pm1$ (particle-hole symmetry $\mathcal{C}$ with $[\hat{A}']^2=\mp1$ ) for $H^{(d-1,n+1)}(\mathbf{k})$.

To conclude, compared with $H^{(d,n)}(\mathbf{k})$ with a $Z^n$ topological invariant, $H^{(d-1,n+1)}(\mathbf{k})$ describes a system with one order higher of topology, and is shifted one column to the left (as the spatial dimension is reduced by $1$) and up two rows (due to the changing of symmetry condition) in the symmetry classification table. Finally,
HOTPs with $Z_2$ topological properties can be derived from a parent Hamiltonian indexed by $Z^n$, as discussed in Sec.~\ref{Sec5} of the main text. Therefore, they also obey the same shifting rule as for the $Z^n$ systems. Therefore, a general symmetry classification table is obtained for arbitrary orders of topology in arbitrary spatial dimensions, as shown in Table~\ref{Classificationhigher}.

\section{An example of 3D second-order topological insulators}\label{appadd}
In this appendix, we show how our method can apply to a model describing 3D materials with noncollinear antiferromagnetic order, 
which holds second-order topology with chiral hinge states~\cite{doi:10.1126/sciadv.aat0346}. Its Hamiltonian reads
\begin{eqnarray}\label{3DTI}
		H_c(\mathbf{k})=&&(M+t\sum_{i=x,y,z}\cos k_i)\sigma_0\tau_z\nonumber \\&&+\Delta_1\sum_{i=x,y,z}\sin k_i\sigma_i\tau_x+\Delta_2(\cos k_x-\cos k_y)\sigma_0\tau_y, \nonumber\\
\end{eqnarray}
where the particle-hole symmetry operator is given by 
$\mathcal{C}=\sigma_y\tau_y\mathcal{K}$.
	To apply the nested-BIS method, we rewrite this Hamiltonian as
\begin{eqnarray}\label{3DTIRE}
		H_{c}(\mathbf{k})=&&H_{c1}(\mathbf{k})+H_{c2}(\mathbf{k}),\nonumber\\
		H_{c1}(\mathbf{k})=&&M'_{1}(\mathbf{k})\sigma_0\tau'_z+h_{1}(\mathbf{k})\sigma_x\tau_x, \nonumber \\
		H_{c2}(\mathbf{k})=&&M'_2(\mathbf{k}_{1,\parallel})\sigma_0\tau'_y+h_{2}(\mathbf{k}_{1,\parallel})\sigma_x\tau_y \nonumber \\
		&&+h_{3}(\mathbf{k}_{1,\parallel})\sigma_x\tau_z,
\end{eqnarray}
where 
\begin{eqnarray}
&\tau'_z=(t\tau_z+\Delta_2\tau_y)/\sqrt{t^2+\Delta_2^2},\nonumber\\
&\tau'_y=(\Delta_2\tau_z-t\tau_y)/\sqrt{t^2+\Delta_2^2},\nonumber
\end{eqnarray}
and
\begin{eqnarray}
&M'_1=m_1'+t_{1x}\cos k_x+t_{1y}\cos k_y+t_{1z}\cos k_z,\nonumber\\
&M'_2=m_2'+t_{2y}\cos k_y+t_{2z}\cos k_z,\nonumber\\
&h_1=\Delta_1\sin k_x,~~h_2=\Delta_1\sin k_y,~~h_3=\Delta_1\sin k_z\nonumber
\end{eqnarray}
with
\begin{eqnarray}
&m_1'=\frac{Mt}{\sqrt{t^2+\Delta_2^2}},~~m_2'=\frac{M\Delta_2}{\sqrt{t^2+\Delta_2^2}},\nonumber\\
&t_{1x}=\sqrt{t^2+\Delta_2^2},~~
t_{1y}=\frac{(t^2-\Delta_2^2)}{\sqrt{t^2+\Delta_2^2}},~~
t_{1z}=\frac{t^2}{\sqrt{t^2+\Delta_2^2}},\nonumber\\
&t_{2y}=\frac{2t\Delta_2}{\sqrt{t^2+\Delta_2^2}},~~
t_{2z}=\frac{t\Delta_2}{\sqrt{t^2+\Delta_2^2}}.\nonumber
\end{eqnarray}
This Hamiltonian takes the same form as Eq.~(\ref{3dDsec}) of the main text, but only with richer parameters. Thus, its topological properties can be directly analyzed with our nested BIS method. 

\section{Properties of the $Z_2$ invariants}\label{appB}

In Sec.~\ref{Sec5} of the main text, we have discussed the HOTP with a $Z_2$ topological invariant, where the interpolation of Eq.~(\ref{interpo}) has been introduced to define the topological properties of $H_2$ in Eq.~\eqref{2dDsecZ2}.
To reveal the $Z_2$ properties of topological invariants ${v}_2$, we provide another interpolation:
\begin{eqnarray}\label{interpo2}
\tilde{H}^{({\rm D},1)}_2(k_2,\theta)=&&\widetilde{M}_2(k_2,\theta)\sigma_z\tau_x+\widetilde{h}_{2}(k_2,\theta)\sigma_z\tau_y \nonumber \\
&&+\widetilde{f}_{1}(k_2,\theta)\sigma_z\tau_z.
\end{eqnarray}
\begin{figure}[!htp]
\includegraphics[width=1\columnwidth]{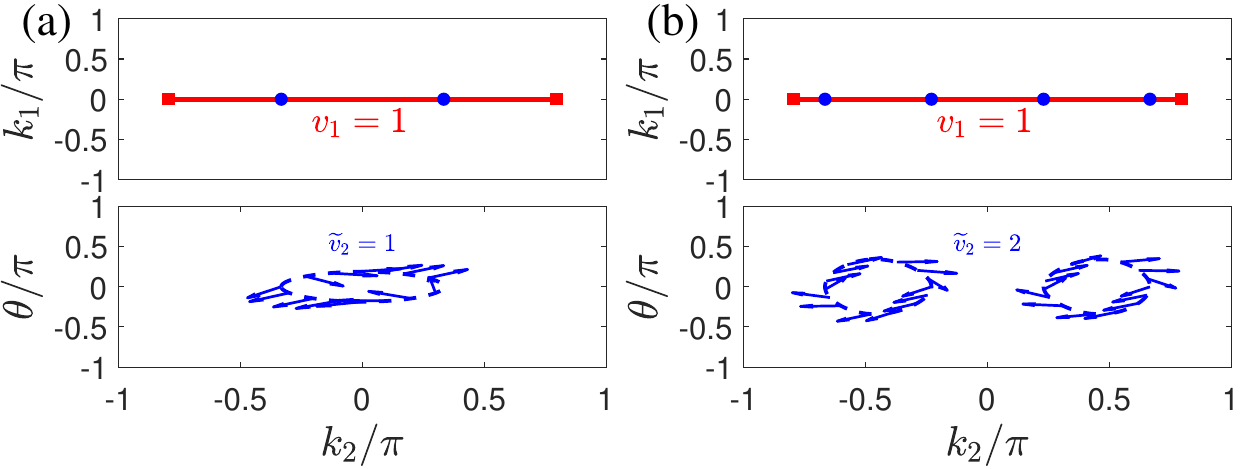}
\caption{\label{sub_wind}(color online) Nested BISs for Hamiltonian $ H^{({\rm D},1)}(\mathbf{k})$ of Eq.~\eqref{2dDsecZ2} of main text and the vector $(\widetilde{h}_{2},\widetilde{f}_{1})$ along the BIS $\tilde{S}_2^{m=1}$ for interpolation of Eq.~(\ref{interpo2}), where $b=3$ is set.  Here ${S}_1^{m=2}$  and ${S}_2^{m=1}$ are indicated by red squares and blue dots, respectively. In all panels, $m_1=-1.8$, $m_2=-0.5$, and $\lambda_2=3$ have been set, and $\lambda_1=0$ ($\lambda_1=-2$) for (a) [(b)].}
\end{figure}
Compared with  Eq.~(\ref{interpo}), here we keep the same form of $\widetilde{M}_2$, but introduce
a different $\widetilde{f}_{1}(k_2,\theta)$ and
a $\theta$-dependant $\widetilde{h}_{2}(k_2,\theta)$,
\begin{eqnarray}\label{interpoBB}
\widetilde{M}_2(k_2,\theta)&=&M_2(k_2)+\lambda_2(1-\cos\theta),\nonumber \\
\widetilde{h}_{2}(k_2,\theta)&=&\sin(k_2)\cos(2\theta)+2\sin(2\theta), \nonumber \\
\widetilde{f}_{1}(k_2,\theta)&=&\sin(2k_2)+\sin(\theta).
\end{eqnarray}
This Hamiltonian preserves the same symmetries as ${H}^{({\rm D},1)}_2(k_2)$, the original Hamiltonian in Eq.~\eqref{2dDsecZ2},
and satisfies $\tilde{H}^{({\rm D},1)}_2(k_2,0)={H}^{({\rm D},1)}_2(k_2)$.
Under this interpolation, $\widetilde{v}_2$ dose not change for the parameters used in Figs.~\ref{Z2EX}(a) and \ref{Z2EX}(b), as shown in Fig.~\ref{sub_wind}(a).
But for the trivial cases described in Figs.~\ref{Z2EX}(c) and \ref{Z2EX}(d), the topological invariant becomes $\tilde{v}_2=2$, as shown in Fig.~\ref{sub_wind}(b).
This example demonstrates that the $\widetilde{v}_2$ may change for different interpolations, but the obtained $Z_2$ topological invariant $v_2=\mod(\tilde{v}_2,2)$ for $H^{({\rm D},1)}_2(k_2)$ is unchanged.
%

%
%
%
\section{First-order topological phases induced by the non-Clifford terms}\label{appC}

\begin{figure}
\includegraphics[width=1\columnwidth]{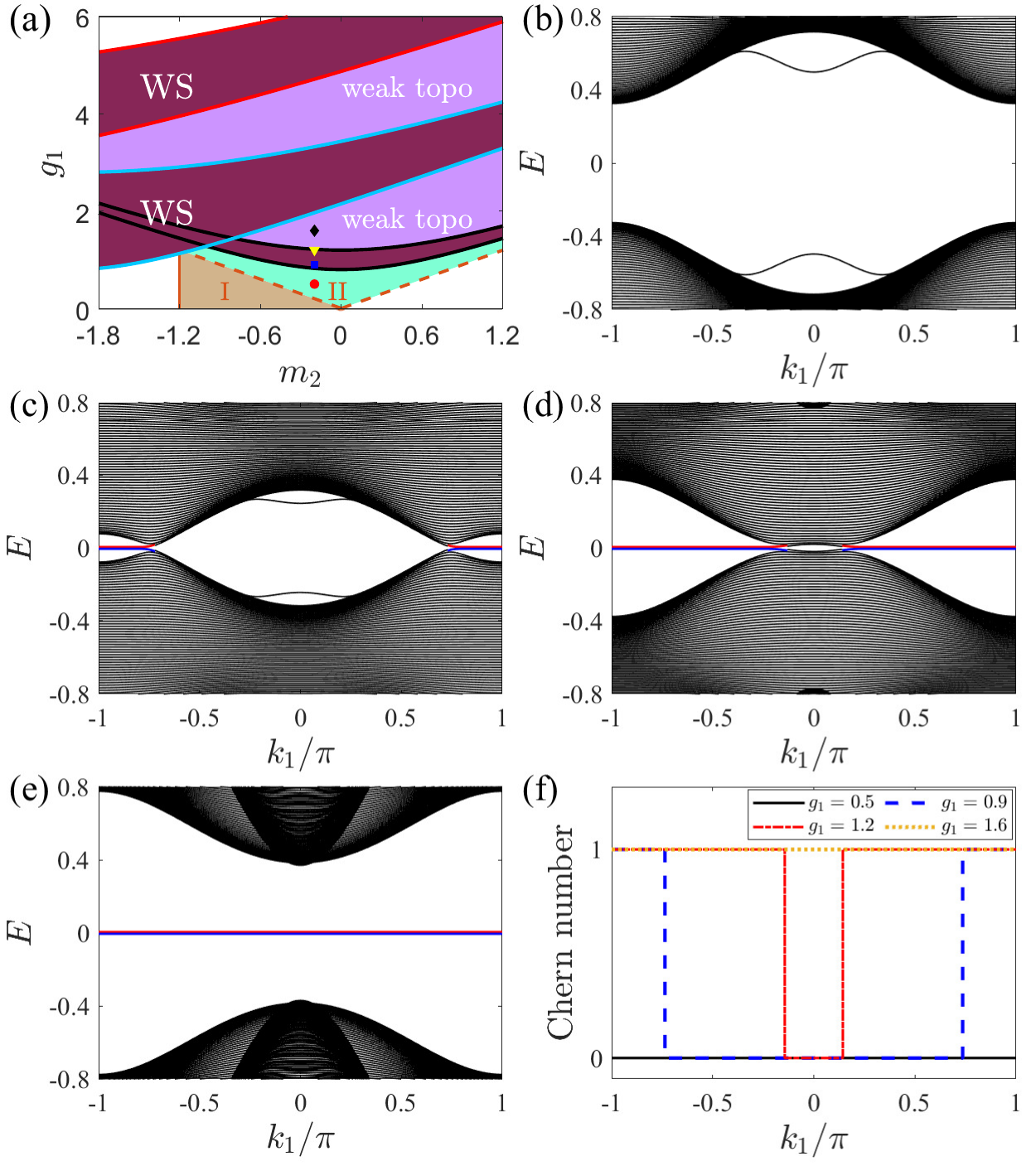}
\caption{\label{sub_energy}(color online) (a) The phase diagram for Hamiltonian $H^{(D)}(\mathbf{k})$ with the longitudinal non-clifford term $g_1\sigma_0\tau_x$ of Eq.~\eqref{3dDsecnon} and $m_1=-1.2$. "WS" and "weak topo" represent the Weyl semimetallic and weak topological phases. (b) to (e) Energy spectra under the OBC along the $r_3$ direction and PBC along other two directions, where the blue and red lines indicates the surface states.
Parameters are (b) $g_1=0.5$, (c) $0.9$, (d) $1.2$, (e) $1.6$, corresponding to red dot, blue square, yellow triangle and black diamond in (a), respectively.
Other parameters $m_2=-0.2$ and $k_2=0$ for all these panels.
Blue and red lines indicate the surface states in $(001)$ and $(00\overline{1})$ surfaces, respectively. (f) The Chern number calculated in $k_2$-$k_3$ space as a function of $k_1$ for different cases described in (b) to (e).}
\end{figure}

In this appendix, we will discuss the phases with first-order boundary states induced by the non-Clifford terms.
More specifically, the longitude non-Clifford term can induce a Weyl semimetallic~\cite{RevModPhys.90.015001,RevModPhys.93.025002} and weak topological phases~\cite{PhysRevLett.98.106803,PhysRevB.75.121306,PhysRevB.79.195322}, while the mixed non-Clifford terms only induce the first one.

We first consider cases with the longitude non-Clifford term, $g_1\sigma_0\tau_x$.
Figure~\ref{sub_energy}(a) displays its phase diagram in a larger parameter space of $g_1$.
By increasing $g_1$, different pairs of Weyl points will emerge and annihilate along the solid lines in the figure at different crystal momenta.
%
Specifically, a pair of Weyl points appears at $(\pi,0,0)$ when $g_1$ reaches the lower black solid line, and annihilates at $(0,0,0)$ when $g_1$ reaches the higher black solid line.
Another two pairs of Weyl points appear at $(0,\pi,0)$ and $(0,0,\pi)$ for the lower light-blue line, and annihilate at $(\pi,\pi,0)$ and $(\pi,0,\pi)$ for the higher light-blue solid line.
Finally, the lower (higher) red solid line indicates one pair of Weyl points appear (annihilate) at $(0,\pi,\pi)$ [$(\pi,\pi,\pi)$].
Combining these information, we obtained several Weyl semimetallic phases, as indicated by the claret color in the phase diagram.
%
In addition, two weak topological phases holding gapped bulk and zero-energy first-order surface states are seen to emerge between these gapless phases.

To investigate their topological properties, we display the energy spectrum for different parameter in Figs.~\ref{sub_energy}(b) and \ref{sub_energy}(e), where we set OBC along the $r_3$ direction and a PBC along the other two directions.
As expected and discussed in the main text, HOTPs with small $g_1$ do not hold topologically protected $1$st-order boundary states, as shown in Fig.~\ref{sub_energy}(b) for parameters at the red dot in Fig.~\ref{sub_energy}(a).
Figures \ref{sub_energy}(c) and \ref{sub_energy}(d) illustrate spectra of the semimetallic phases corresponding to the blue square and yellow triangle in Fig.~\ref{sub_energy}(a) respectively, hosting a pair of Weyl points at $(k_{1}^{\pm},0,0)$ with $k_{1}^{\pm}=\pm\arccos(\frac{(m_1+1)^2+m_2^2+1-g_1^2}{2(m_1+1)})$.
%
Zero-energy first-order boundary states are also seen connecting the two Weyl points.
%
Finally, further increasing $g_1$ has the pair of Weyl points annihilating at $(0,0,0)$, driving the system into the weak topological phase with zero-energy first-order boundary states for every $k_1$~\cite{PhysRevLett.98.106803,PhysRevB.75.121306,PhysRevB.79.195322}, as displayed in Fig.~\ref{sub_energy}(e).
%
%

For both of these two first-order topological phases, the topological properties can be described by a Chern number, with the original 3D system taken as pieces of 2D systems of $k_{2,3}$, and $k_1\in(-\pi,\pi]$ taken as a parameter.
In Fig. \ref{sub_energy}(f), we display the Chern number of the two occupied bands (with negative eigenenergies) as a function of $k_1$.
We can see that the Chern number is zero for every $k_1$ for the HOTP, representing a trivial first-order topology of this phase.
In contrast, the Chern number is nonzero for $k_1\in(-\pi,k_{1}^{-})\bigcup(k_{1}^{+},\pi)$ and zero for $k_1\in(k_{1}^{-},k_{1}^{+})$ for the Weyl semimetallic phase, associated with the values of $k_1$ with zero-energy first-order boundary states in Figs. \ref{sub_energy}(c) and \ref{sub_energy}(d).
%
%
After the Weyl points annihilate at $(0,0,0)$, the Chern number becomes nonzero for every $k_1$, corresponding to the zero-energy first-order boundary states shown in Fig.~\ref{sub_energy}(e).
We also note that due to the $\mathcal{C}_{4}$ rotation symmetry in the $k_2$-$k_3$ plane,
the results are identical when the $r_2$ direction takes an OBC and the other two directions take a PBC. 
%

\begin{figure}[!htp]
\includegraphics[width=1\columnwidth]{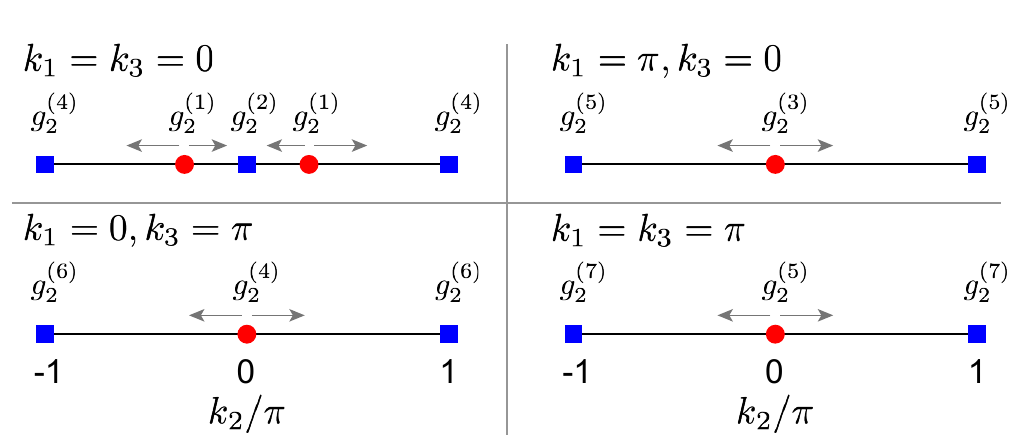}
\caption{\label{sub_energyg2}(color online) The creation, annihilation and evolution of Weyl points along high-symmetric lines in the BZ, for the Hamiltonian $H^{(D)}(\mathbf{k})$ with a non-Clifford term $-g_2\sigma_x\tau_z$  [i.e., Eq.~\eqref{3dDsectramg2} in the main text]. Red dots (blue squares) indicate the creation (annihilation) of Weyl points at $g_2=g_2^{(i)}$, as given in Eq.~\eqref{parag2}. Gray arrows show the moving directions of Weyl points when increasing $g_2$. Here $m_+=-1.2$ and $m_-=0.2$ are set. }
\end{figure} 

For the case with a mixed non-Clifford term $-g_2\sigma_x\tau_z$, increasing $g_2$ is found to induce only semimetallic phases, as shown in the phase diagram in Fig. \ref{Dnonclig2phase}(a) in the main text.
Specifically, multiple pairs of Weyl points emerge and annihilate at different crystal momenta with different values of $g_2=g_2^{(i)}$, with $i=1,2,...7$ and
\begin{eqnarray}\label{parag2}
g_2^{(1)}=&&\sqrt{[2-(m_++2)^2+m_-^2]/2},\nonumber \\
g_2^{(2)}=&&\sqrt{(m_+^2+m_-^2)/2},\nonumber \\
g_2^{(3)}=&&\sqrt{[(m_++2)^2+(m_-+2)^2]/2},\nonumber \\
g_2^{(4)}=&&\sqrt{[(m_++4)^2+m_-^2]/2},\nonumber \\
g_2^{(5)}=&&\sqrt{[(m_++6)^2+(m_-+2)^2]/2},\nonumber \\
g_2^{(6)}=&&\sqrt{[(m_++8)^2+m_-^2]/2},\nonumber \\
g_2^{(7)}=&&\sqrt{[(m_++10)^2+(m_-+2)^2]/2}.\nonumber
\end{eqnarray}
The lower black, higher black, and light blue solid lines in Fig. \ref{Dnonclig2phase}(a) are given by $g_2=g_2^{(1)}$, $g_2^{(2)}$, and $g_2^{(3)}$, respectively.
These Weyl points are found to exist only along four high-symmetric lines with $k_{1,3}=0$ or $\pi$, and their emergence, annihilation, and moving directions with increasing $g_2$ are shown in Fig. \ref{sub_energyg2}.
To conclude, at least one pair of Weyl points exist for $g_2\in(g_2^{(1)},g_2^{(7)})$, and all Weyl points annihilate and disappear when $g_2>g_2^{(7)}$, resulting in a gaped phase without nontrivial topology.

%


\bigskip

\begin{thebibliography}{94}%
\makeatletter
\providecommand \@ifxundefined [1]{%
 \@ifx{#1\undefined}
}%
\providecommand \@ifnum [1]{%
 \ifnum #1\expandafter \@firstoftwo
 \else \expandafter \@secondoftwo
 \fi
}%
\providecommand \@ifx [1]{%
 \ifx #1\expandafter \@firstoftwo
 \else \expandafter \@secondoftwo
 \fi
}%
\providecommand \natexlab [1]{#1}%
\providecommand \enquote  [1]{``#1''}%
\providecommand \bibnamefont  [1]{#1}%
\providecommand \bibfnamefont [1]{#1}%
\providecommand \citenamefont [1]{#1}%
\providecommand \href@noop [0]{\@secondoftwo}%
\providecommand \href [0]{\begingroup \@sanitize@url \@href}%
\providecommand \@href[1]{\@@startlink{#1}\@@href}%
\providecommand \@@href[1]{\endgroup#1\@@endlink}%
\providecommand \@sanitize@url [0]{\catcode `\\12\catcode `\$12\catcode
  `\&12\catcode `\#12\catcode `\^12\catcode `\_12\catcode `\%12\relax}%
\providecommand \@@startlink[1]{}%
\providecommand \@@endlink[0]{}%
\providecommand \url  [0]{\begingroup\@sanitize@url \@url }%
\providecommand \@url [1]{\endgroup\@href {#1}{\urlprefix }}%
\providecommand \urlprefix  [0]{URL }%
\providecommand \Eprint [0]{\href }%
\providecommand \doibase [0]{http://dx.doi.org/}%
\providecommand \selectlanguage [0]{\@gobble}%
\providecommand \bibinfo  [0]{\@secondoftwo}%
\providecommand \bibfield  [0]{\@secondoftwo}%
\providecommand \translation [1]{[#1]}%
\providecommand \BibitemOpen [0]{}%
\providecommand \bibitemStop [0]{}%
\providecommand \bibitemNoStop [0]{.\EOS\space}%
\providecommand \EOS [0]{\spacefactor3000\relax}%
\providecommand \BibitemShut  [1]{\csname bibitem#1\endcsname}%
\let\auto@bib@innerbib\@empty
\bibitem [{\citenamefont {Hasan}\ and\ \citenamefont
  {Kane}(2010)}]{RevModPhys.82.3045}%
  \BibitemOpen
  \bibfield  {author} {\bibinfo {author} {\bibfnamefont {M.~Z.}\ \bibnamefont
  {Hasan}}\ and\ \bibinfo {author} {\bibfnamefont {C.~L.}\ \bibnamefont
  {Kane}},\ }\bibfield  {title} {\enquote {\bibinfo {title} {Colloquium:
  Topological insulators},}\ }\href {\doibase 10.1103/RevModPhys.82.3045}
  {\bibfield  {journal} {\bibinfo  {journal} {Rev. Mod. Phys.}\ }\textbf
  {\bibinfo {volume} {82}},\ \bibinfo {pages} {3045--3067} (\bibinfo {year}
  {2010})}\BibitemShut {NoStop}%
\bibitem [{\citenamefont {Qi}\ and\ \citenamefont
  {Zhang}(2011)}]{RevModPhys.83.1057}%
  \BibitemOpen
  \bibfield  {author} {\bibinfo {author} {\bibfnamefont {X.-L.}\ \bibnamefont
  {Qi}}\ and\ \bibinfo {author} {\bibfnamefont {S.-C.}\ \bibnamefont {Zhang}},\
  }\bibfield  {title} {\enquote {\bibinfo {title} {Topological insulators and
  superconductors},}\ }\href {\doibase 10.1103/RevModPhys.83.1057} {\bibfield
  {journal} {\bibinfo  {journal} {Rev. Mod. Phys.}\ }\textbf {\bibinfo {volume}
  {83}},\ \bibinfo {pages} {1057--1110} (\bibinfo {year} {2011})}\BibitemShut
  {NoStop}%
\bibitem [{\citenamefont {Benalcazar}\ \emph
  {et~al.}(2017{\natexlab{a}})\citenamefont {Benalcazar}, \citenamefont
  {Bernevig},\ and\ \citenamefont {Hughes}}]{Benalcazar61}%
  \BibitemOpen
  \bibfield  {author} {\bibinfo {author} {\bibfnamefont {W.~A.}\ \bibnamefont
  {Benalcazar}}, \bibinfo {author} {\bibfnamefont {B.~Andrei}\ \bibnamefont
  {Bernevig}}, \ and\ \bibinfo {author} {\bibfnamefont {Taylor~L.}\
  \bibnamefont {Hughes}},\ }\bibfield  {title} {\enquote {\bibinfo {title}
  {Quantized electric multipole insulators},}\ }\href {\doibase
  10.1126/science.aah6442} {\bibfield  {journal} {\bibinfo  {journal}
  {Science}\ }\textbf {\bibinfo {volume} {357}},\ \bibinfo {pages} {61--66}
  (\bibinfo {year} {2017}{\natexlab{a}})}\BibitemShut {NoStop}%
\bibitem [{\citenamefont {Benalcazar}\ \emph
  {et~al.}(2017{\natexlab{b}})\citenamefont {Benalcazar}, \citenamefont
  {Bernevig},\ and\ \citenamefont {Hughes}}]{PhysRevB.96.245115}%
  \BibitemOpen
  \bibfield  {author} {\bibinfo {author} {\bibfnamefont {W.~A.}\ \bibnamefont
  {Benalcazar}}, \bibinfo {author} {\bibfnamefont {B.~A.}\ \bibnamefont
  {Bernevig}}, \ and\ \bibinfo {author} {\bibfnamefont {T.~L.}\ \bibnamefont
  {Hughes}},\ }\bibfield  {title} {\enquote {\bibinfo {title} {Electric
  multipole moments, topological multipole moment pumping, and chiral hinge
  states in crystalline insulators},}\ }\href {\doibase
  10.1103/PhysRevB.96.245115} {\bibfield  {journal} {\bibinfo  {journal} {Phys.
  Rev. B}\ }\textbf {\bibinfo {volume} {96}},\ \bibinfo {pages} {245115}
  (\bibinfo {year} {2017}{\natexlab{b}})}\BibitemShut {NoStop}%
\bibitem [{\citenamefont {Langbehn}\ \emph {et~al.}(2017)\citenamefont
  {Langbehn}, \citenamefont {Trifunovic}, \citenamefont {von Oppen},\ and\
  \citenamefont {Brouwer}}]{PhysRevLett.119.246401}%
  \BibitemOpen
  \bibfield  {author} {\bibinfo {author} {\bibfnamefont {Y.}~\bibnamefont
  {Langbehn}, \bibfnamefont {J.and~Peng}}, \bibinfo {author} {\bibfnamefont
  {L.}~\bibnamefont {Trifunovic}}, \bibinfo {author} {\bibfnamefont
  {F.}~\bibnamefont {von Oppen}}, \ and\ \bibinfo {author} {\bibfnamefont
  {P.~W.}\ \bibnamefont {Brouwer}},\ }\bibfield  {title} {\enquote {\bibinfo
  {title} {Reflection-symmetric second-order topological insulators and
  superconductors},}\ }\href {\doibase 10.1103/PhysRevLett.119.246401}
  {\bibfield  {journal} {\bibinfo  {journal} {Phys. Rev. Lett.}\ }\textbf
  {\bibinfo {volume} {119}},\ \bibinfo {pages} {246401} (\bibinfo {year}
  {2017})}\BibitemShut {NoStop}%
\bibitem [{\citenamefont {Song}\ \emph {et~al.}(2017)\citenamefont {Song},
  \citenamefont {Fang},\ and\ \citenamefont {Fang}}]{PhysRevLett.119.246402}%
  \BibitemOpen
  \bibfield  {author} {\bibinfo {author} {\bibfnamefont {Z.}~\bibnamefont
  {Song}}, \bibinfo {author} {\bibfnamefont {Z.}~\bibnamefont {Fang}}, \ and\
  \bibinfo {author} {\bibfnamefont {C.}~\bibnamefont {Fang}},\ }\bibfield
  {title} {\enquote {\bibinfo {title} {$(d\ensuremath{-}2)$-dimensional edge
  states of rotation symmetry protected topological states},}\ }\href {\doibase
  10.1103/PhysRevLett.119.246402} {\bibfield  {journal} {\bibinfo  {journal}
  {Phys. Rev. Lett.}\ }\textbf {\bibinfo {volume} {119}},\ \bibinfo {pages}
  {246402} (\bibinfo {year} {2017})}\BibitemShut {NoStop}%
\bibitem [{\citenamefont {Geier}\ \emph {et~al.}(2018)\citenamefont {Geier},
  \citenamefont {Trifunovic}, \citenamefont {Hoskam},\ and\ \citenamefont
  {Brouwer}}]{PhysRevB.97.205135}%
  \BibitemOpen
  \bibfield  {author} {\bibinfo {author} {\bibfnamefont {M.}~\bibnamefont
  {Geier}}, \bibinfo {author} {\bibfnamefont {L.}~\bibnamefont {Trifunovic}},
  \bibinfo {author} {\bibfnamefont {M.}~\bibnamefont {Hoskam}}, \ and\ \bibinfo
  {author} {\bibfnamefont {P.~W.}\ \bibnamefont {Brouwer}},\ }\bibfield
  {title} {\enquote {\bibinfo {title} {Second-order topological insulators and
  superconductors with an order-two crystalline symmetry},}\ }\href {\doibase
  10.1103/PhysRevB.97.205135} {\bibfield  {journal} {\bibinfo  {journal} {Phys.
  Rev. B}\ }\textbf {\bibinfo {volume} {97}},\ \bibinfo {pages} {205135}
  (\bibinfo {year} {2018})}\BibitemShut {NoStop}%
\bibitem [{\citenamefont {Trifunovic}\ and\ \citenamefont
  {Brouwer}(2019)}]{PhysRevX.9.011012}%
  \BibitemOpen
  \bibfield  {author} {\bibinfo {author} {\bibfnamefont {L.}~\bibnamefont
  {Trifunovic}}\ and\ \bibinfo {author} {\bibfnamefont {P.~W.}\ \bibnamefont
  {Brouwer}},\ }\bibfield  {title} {\enquote {\bibinfo {title} {Higher-order
  bulk-boundary correspondence for topological crystalline phases},}\ }\href
  {\doibase 10.1103/PhysRevX.9.011012} {\bibfield  {journal} {\bibinfo
  {journal} {Phys. Rev. X}\ }\textbf {\bibinfo {volume} {9}},\ \bibinfo {pages}
  {011012} (\bibinfo {year} {2019})}\BibitemShut {NoStop}%
\bibitem [{\citenamefont {Ezawa}(2020)}]{ezawa2020edge}%
  \BibitemOpen
  \bibfield  {author} {\bibinfo {author} {\bibfnamefont {Motohiko}\
  \bibnamefont {Ezawa}},\ }\bibfield  {title} {\enquote {\bibinfo {title}
  {Edge-corner correspondence: Boundary-obstructed topological phases with
  chiral symmetry},}\ }\href {\doibase 10.1103/PhysRevB.102.121405} {\bibfield
  {journal} {\bibinfo  {journal} {Phys. Rev. B}\ }\textbf {\bibinfo {volume}
  {102}},\ \bibinfo {pages} {121405} (\bibinfo {year} {2020})}\BibitemShut
  {NoStop}%
\bibitem [{\citenamefont {Asaga}\ and\ \citenamefont
  {Fukui}(2020)}]{asaga2020boundary}%
  \BibitemOpen
  \bibfield  {author} {\bibinfo {author} {\bibfnamefont {Koichi}\ \bibnamefont
  {Asaga}}\ and\ \bibinfo {author} {\bibfnamefont {Takahiro}\ \bibnamefont
  {Fukui}},\ }\bibfield  {title} {\enquote {\bibinfo {title}
  {Boundary-obstructed topological phases of a massive dirac fermion in a
  magnetic field},}\ }\href {\doibase 10.1103/PhysRevB.102.155102} {\bibfield
  {journal} {\bibinfo  {journal} {Phys. Rev. B}\ }\textbf {\bibinfo {volume}
  {102}},\ \bibinfo {pages} {155102} (\bibinfo {year} {2020})}\BibitemShut
  {NoStop}%
\bibitem [{\citenamefont {Wu}\ \emph {et~al.}(2020{\natexlab{a}})\citenamefont
  {Wu}, \citenamefont {Benalcazar}, \citenamefont {Li}, \citenamefont
  {Thomale}, \citenamefont {Liu},\ and\ \citenamefont {Hu}}]{wu2020boundary}%
  \BibitemOpen
  \bibfield  {author} {\bibinfo {author} {\bibfnamefont {Xianxin}\ \bibnamefont
  {Wu}}, \bibinfo {author} {\bibfnamefont {Wladimir~A.}\ \bibnamefont
  {Benalcazar}}, \bibinfo {author} {\bibfnamefont {Yinxiang}\ \bibnamefont
  {Li}}, \bibinfo {author} {\bibfnamefont {Ronny}\ \bibnamefont {Thomale}},
  \bibinfo {author} {\bibfnamefont {Chao-Xing}\ \bibnamefont {Liu}}, \ and\
  \bibinfo {author} {\bibfnamefont {Jiangping}\ \bibnamefont {Hu}},\ }\bibfield
   {title} {\enquote {\bibinfo {title} {Boundary-obstructed topological
  high-${\mathit{t}}_{c}$ superconductivity in iron pnictides},}\ }\href
  {\doibase 10.1103/PhysRevX.10.041014} {\bibfield  {journal} {\bibinfo
  {journal} {Phys. Rev. X}\ }\textbf {\bibinfo {volume} {10}},\ \bibinfo
  {pages} {041014} (\bibinfo {year} {2020}{\natexlab{a}})}\BibitemShut
  {NoStop}%
\bibitem [{\citenamefont {Tiwari}\ \emph {et~al.}(2020)\citenamefont {Tiwari},
  \citenamefont {Jahin},\ and\ \citenamefont {Wang}}]{tiwari2020chiral}%
  \BibitemOpen
  \bibfield  {author} {\bibinfo {author} {\bibfnamefont {Apoorv}\ \bibnamefont
  {Tiwari}}, \bibinfo {author} {\bibfnamefont {Ammar}\ \bibnamefont {Jahin}}, \
  and\ \bibinfo {author} {\bibfnamefont {Yuxuan}\ \bibnamefont {Wang}},\
  }\bibfield  {title} {\enquote {\bibinfo {title} {Chiral dirac
  superconductors: Second-order and boundary-obstructed topology},}\ }\href
  {\doibase 10.1103/PhysRevResearch.2.043300} {\bibfield  {journal} {\bibinfo
  {journal} {Phys. Rev. Research}\ }\textbf {\bibinfo {volume} {2}},\ \bibinfo
  {pages} {043300} (\bibinfo {year} {2020})}\BibitemShut {NoStop}%
\bibitem [{\citenamefont {Khalaf}\ \emph {et~al.}(2021)\citenamefont {Khalaf},
  \citenamefont {Benalcazar}, \citenamefont {Hughes},\ and\ \citenamefont
  {Queiroz}}]{khalaf2021boundary}%
  \BibitemOpen
  \bibfield  {author} {\bibinfo {author} {\bibfnamefont {Eslam}\ \bibnamefont
  {Khalaf}}, \bibinfo {author} {\bibfnamefont {Wladimir~A.}\ \bibnamefont
  {Benalcazar}}, \bibinfo {author} {\bibfnamefont {Taylor~L.}\ \bibnamefont
  {Hughes}}, \ and\ \bibinfo {author} {\bibfnamefont {Raquel}\ \bibnamefont
  {Queiroz}},\ }\bibfield  {title} {\enquote {\bibinfo {title}
  {Boundary-obstructed topological phases},}\ }\href {\doibase
  10.1103/PhysRevResearch.3.013239} {\bibfield  {journal} {\bibinfo  {journal}
  {Phys. Rev. Research}\ }\textbf {\bibinfo {volume} {3}},\ \bibinfo {pages}
  {013239} (\bibinfo {year} {2021})}\BibitemShut {NoStop}%
\bibitem [{\citenamefont {Ezawa}(2018)}]{PhysRevLett.120.026801}%
  \BibitemOpen
  \bibfield  {author} {\bibinfo {author} {\bibfnamefont {M.}~\bibnamefont
  {Ezawa}},\ }\bibfield  {title} {\enquote {\bibinfo {title} {Higher-order
  topological insulators and semimetals on the breathing kagome and pyrochlore
  lattices},}\ }\href {\doibase 10.1103/PhysRevLett.120.026801} {\bibfield
  {journal} {\bibinfo  {journal} {Phys. Rev. Lett.}\ }\textbf {\bibinfo
  {volume} {120}},\ \bibinfo {pages} {026801} (\bibinfo {year}
  {2018})}\BibitemShut {NoStop}%
\bibitem [{\citenamefont {Khalaf}(2018)}]{PhysRevB.97.205136}%
  \BibitemOpen
  \bibfield  {author} {\bibinfo {author} {\bibfnamefont {E.}~\bibnamefont
  {Khalaf}},\ }\bibfield  {title} {\enquote {\bibinfo {title} {Higher-order
  topological insulators and superconductors protected by inversion
  symmetry},}\ }\href {\doibase 10.1103/PhysRevB.97.205136} {\bibfield
  {journal} {\bibinfo  {journal} {Phys. Rev. B}\ }\textbf {\bibinfo {volume}
  {97}},\ \bibinfo {pages} {205136} (\bibinfo {year} {2018})}\BibitemShut
  {NoStop}%
\bibitem [{\citenamefont {Kunst}\ \emph {et~al.}(2018)\citenamefont {Kunst},
  \citenamefont {van Miert},\ and\ \citenamefont
  {Bergholtz}}]{PhysRevB.97.241405}%
  \BibitemOpen
  \bibfield  {author} {\bibinfo {author} {\bibfnamefont {F.~K.}\ \bibnamefont
  {Kunst}}, \bibinfo {author} {\bibfnamefont {G.}~\bibnamefont {van Miert}}, \
  and\ \bibinfo {author} {\bibfnamefont {E.~J.}\ \bibnamefont {Bergholtz}},\
  }\bibfield  {title} {\enquote {\bibinfo {title} {Lattice models with exactly
  solvable topological hinge and corner states},}\ }\href {\doibase
  10.1103/PhysRevB.97.241405} {\bibfield  {journal} {\bibinfo  {journal} {Phys.
  Rev. B}\ }\textbf {\bibinfo {volume} {97}},\ \bibinfo {pages} {241405}
  (\bibinfo {year} {2018})}\BibitemShut {NoStop}%
\bibitem [{\citenamefont {Xie}\ \emph {et~al.}(2018)\citenamefont {Xie},
  \citenamefont {Wang}, \citenamefont {Wang}, \citenamefont {Zhu},
  \citenamefont {Jiang}, \citenamefont {Lu},\ and\ \citenamefont
  {Chen}}]{PhysRevB.98.205147}%
  \BibitemOpen
  \bibfield  {author} {\bibinfo {author} {\bibfnamefont {B.-Y.}\ \bibnamefont
  {Xie}}, \bibinfo {author} {\bibfnamefont {H.-F.}\ \bibnamefont {Wang}},
  \bibinfo {author} {\bibfnamefont {H.-X.}\ \bibnamefont {Wang}}, \bibinfo
  {author} {\bibfnamefont {X.-Y.}\ \bibnamefont {Zhu}}, \bibinfo {author}
  {\bibfnamefont {J.-H.}\ \bibnamefont {Jiang}}, \bibinfo {author}
  {\bibfnamefont {M.-H.}\ \bibnamefont {Lu}}, \ and\ \bibinfo {author}
  {\bibfnamefont {Y.-F.}\ \bibnamefont {Chen}},\ }\bibfield  {title} {\enquote
  {\bibinfo {title} {Second-order photonic topological insulator with corner
  states},}\ }\href {\doibase 10.1103/PhysRevB.98.205147} {\bibfield  {journal}
  {\bibinfo  {journal} {Phys. Rev. B}\ }\textbf {\bibinfo {volume} {98}},\
  \bibinfo {pages} {205147} (\bibinfo {year} {2018})}\BibitemShut {NoStop}%
\bibitem [{\citenamefont {Li}\ \emph {et~al.}(2018)\citenamefont {Li},
  \citenamefont {Umer},\ and\ \citenamefont {Gong}}]{PhysRevB.98.205422}%
  \BibitemOpen
  \bibfield  {author} {\bibinfo {author} {\bibfnamefont {L.}~\bibnamefont
  {Li}}, \bibinfo {author} {\bibfnamefont {M.}~\bibnamefont {Umer}}, \ and\
  \bibinfo {author} {\bibfnamefont {J.}~\bibnamefont {Gong}},\ }\bibfield
  {title} {\enquote {\bibinfo {title} {Direct prediction of corner state
  configurations from edge winding numbers in two- and three-dimensional
  chiral-symmetric lattice systems},}\ }\href {\doibase
  10.1103/PhysRevB.98.205422} {\bibfield  {journal} {\bibinfo  {journal} {Phys.
  Rev. B}\ }\textbf {\bibinfo {volume} {98}},\ \bibinfo {pages} {205422}
  (\bibinfo {year} {2018})}\BibitemShut {NoStop}%
\bibitem [{\citenamefont {Serra-Garcia}\ \emph {et~al.}(2019)\citenamefont
  {Serra-Garcia}, \citenamefont {S\"usstrunk},\ and\ \citenamefont
  {Huber}}]{PhysRevB.99.020304}%
  \BibitemOpen
  \bibfield  {author} {\bibinfo {author} {\bibfnamefont {Marc}\ \bibnamefont
  {Serra-Garcia}}, \bibinfo {author} {\bibfnamefont {Roman}\ \bibnamefont
  {S\"usstrunk}}, \ and\ \bibinfo {author} {\bibfnamefont {Sebastian~D.}\
  \bibnamefont {Huber}},\ }\bibfield  {title} {\enquote {\bibinfo {title}
  {Observation of quadrupole transitions and edge mode topology in an lc
  circuit network},}\ }\href {\doibase 10.1103/PhysRevB.99.020304} {\bibfield
  {journal} {\bibinfo  {journal} {Phys. Rev. B}\ }\textbf {\bibinfo {volume}
  {99}},\ \bibinfo {pages} {020304} (\bibinfo {year} {2019})}\BibitemShut
  {NoStop}%
\bibitem [{\citenamefont {Serra-Garcia}\ \emph {et~al.}(2018)\citenamefont
  {Serra-Garcia}, \citenamefont {Peri}, \citenamefont {Susstrunk},
  \citenamefont {Bilal}, \citenamefont {Larsen}, \citenamefont {Villanueva},\
  and\ \citenamefont {Huber}}]{Serra-Garcia2018}%
  \BibitemOpen
  \bibfield  {author} {\bibinfo {author} {\bibfnamefont {M.}~\bibnamefont
  {Serra-Garcia}}, \bibinfo {author} {\bibfnamefont {V.}~\bibnamefont {Peri}},
  \bibinfo {author} {\bibfnamefont {R.}~\bibnamefont {Susstrunk}}, \bibinfo
  {author} {\bibfnamefont {O.~R.}\ \bibnamefont {Bilal}}, \bibinfo {author}
  {\bibfnamefont {T.}~\bibnamefont {Larsen}}, \bibinfo {author} {\bibfnamefont
  {L.~G.}\ \bibnamefont {Villanueva}}, \ and\ \bibinfo {author} {\bibfnamefont
  {S.~D.}\ \bibnamefont {Huber}},\ }\bibfield  {title} {\enquote {\bibinfo
  {title} {Observation of a phononic quadrupole topological insulator},}\
  }\href {\doibase 10.1038/nature25156} {\bibfield  {journal} {\bibinfo
  {journal} {Nature(London)}\ }\textbf {\bibinfo {volume} {555}},\ \bibinfo
  {pages} {342} (\bibinfo {year} {2018})}\BibitemShut {NoStop}%
\bibitem [{\citenamefont {Schindler}\ \emph
  {et~al.}(2018{\natexlab{a}})\citenamefont {Schindler}, \citenamefont {Wang},
  \citenamefont {Vergniory}, \citenamefont {Cook}, \citenamefont {Murani},
  \citenamefont {Sengupta}, \citenamefont {Kasumov}, \citenamefont {Deblock},
  \citenamefont {Jeon}, \citenamefont {Drozdov}, \citenamefont {Bouchiat},
  \citenamefont {Gu\'{e}ron}, \citenamefont {Yazdani}, \citenamefont
  {Bernevig},\ and\ \citenamefont {Neupert}}]{Schindler2018}%
  \BibitemOpen
  \bibfield  {author} {\bibinfo {author} {\bibfnamefont {F.}~\bibnamefont
  {Schindler}}, \bibinfo {author} {\bibfnamefont {Z.}~\bibnamefont {Wang}},
  \bibinfo {author} {\bibfnamefont {M.~G.}\ \bibnamefont {Vergniory}}, \bibinfo
  {author} {\bibfnamefont {A.~M.}\ \bibnamefont {Cook}}, \bibinfo {author}
  {\bibfnamefont {A.}~\bibnamefont {Murani}}, \bibinfo {author} {\bibfnamefont
  {S.}~\bibnamefont {Sengupta}}, \bibinfo {author} {\bibfnamefont {A.~Y.}\
  \bibnamefont {Kasumov}}, \bibinfo {author} {\bibfnamefont {R.}~\bibnamefont
  {Deblock}}, \bibinfo {author} {\bibfnamefont {S.}~\bibnamefont {Jeon}},
  \bibinfo {author} {\bibfnamefont {I.}~\bibnamefont {Drozdov}}, \bibinfo
  {author} {\bibfnamefont {H.}~\bibnamefont {Bouchiat}}, \bibinfo {author}
  {\bibfnamefont {S.}~\bibnamefont {Gu\'{e}ron}}, \bibinfo {author}
  {\bibfnamefont {A.}~\bibnamefont {Yazdani}}, \bibinfo {author} {\bibfnamefont
  {B.~A.}\ \bibnamefont {Bernevig}}, \ and\ \bibinfo {author} {\bibfnamefont
  {T.}~\bibnamefont {Neupert}},\ }\bibfield  {title} {\enquote {\bibinfo
  {title} {Higher-order topology in bismuth},}\ }\href {\doibase
  10.1038/s41567-018-0224-7} {\bibfield  {journal} {\bibinfo  {journal} {Nat.
  Phys.}\ }\textbf {\bibinfo {volume} {15}},\ \bibinfo {pages} {918} (\bibinfo
  {year} {2018}{\natexlab{a}})}\BibitemShut {NoStop}%
\bibitem [{\citenamefont {Peterson}\ \emph {et~al.}(2018)\citenamefont
  {Peterson}, \citenamefont {Benalcazar}, \citenamefont {Hughes},\ and\
  \citenamefont {Bahl}}]{Peterson2018}%
  \BibitemOpen
  \bibfield  {author} {\bibinfo {author} {\bibfnamefont {C.~W.}\ \bibnamefont
  {Peterson}}, \bibinfo {author} {\bibfnamefont {W.~A.}\ \bibnamefont
  {Benalcazar}}, \bibinfo {author} {\bibfnamefont {T.~L.}\ \bibnamefont
  {Hughes}}, \ and\ \bibinfo {author} {\bibfnamefont {G.}~\bibnamefont
  {Bahl}},\ }\bibfield  {title} {\enquote {\bibinfo {title} {A quantized
  microwave quadrupole insulator with topologically protected corner states},}\
  }\href {\doibase 10.1038/nature25777} {\bibfield  {journal} {\bibinfo
  {journal} {Nature(London)}\ }\textbf {\bibinfo {volume} {555}},\ \bibinfo
  {pages} {346} (\bibinfo {year} {2018})}\BibitemShut {NoStop}%
\bibitem [{\citenamefont {Liu}\ \emph {et~al.}(2019)\citenamefont {Liu},
  \citenamefont {Zhang}, \citenamefont {Ai}, \citenamefont {Gong},
  \citenamefont {Kawabata}, \citenamefont {Ueda},\ and\ \citenamefont
  {Nori}}]{PhysRevLett.122.076801}%
  \BibitemOpen
  \bibfield  {author} {\bibinfo {author} {\bibfnamefont {T.}~\bibnamefont
  {Liu}}, \bibinfo {author} {\bibfnamefont {Y.-R.}\ \bibnamefont {Zhang}},
  \bibinfo {author} {\bibfnamefont {Q.}~\bibnamefont {Ai}}, \bibinfo {author}
  {\bibfnamefont {Z.}~\bibnamefont {Gong}}, \bibinfo {author} {\bibfnamefont
  {K.}~\bibnamefont {Kawabata}}, \bibinfo {author} {\bibfnamefont
  {M.}~\bibnamefont {Ueda}}, \ and\ \bibinfo {author} {\bibfnamefont
  {F.}~\bibnamefont {Nori}},\ }\bibfield  {title} {\enquote {\bibinfo {title}
  {Second-order topological phases in non-hermitian systems},}\ }\href
  {\doibase 10.1103/PhysRevLett.122.076801} {\bibfield  {journal} {\bibinfo
  {journal} {Phys. Rev. Lett.}\ }\textbf {\bibinfo {volume} {122}},\ \bibinfo
  {pages} {076801} (\bibinfo {year} {2019})}\BibitemShut {NoStop}%
\bibitem [{\citenamefont {Lee}\ \emph {et~al.}(2019)\citenamefont {Lee},
  \citenamefont {Li},\ and\ \citenamefont {Gong}}]{PhysRevLett.123.016805}%
  \BibitemOpen
  \bibfield  {author} {\bibinfo {author} {\bibfnamefont {C.~H.}\ \bibnamefont
  {Lee}}, \bibinfo {author} {\bibfnamefont {L.}~\bibnamefont {Li}}, \ and\
  \bibinfo {author} {\bibfnamefont {J.}~\bibnamefont {Gong}},\ }\bibfield
  {title} {\enquote {\bibinfo {title} {Hybrid higher-order skin-topological
  modes in nonreciprocal systems},}\ }\href {\doibase
  10.1103/PhysRevLett.123.016805} {\bibfield  {journal} {\bibinfo  {journal}
  {Phys. Rev. Lett.}\ }\textbf {\bibinfo {volume} {123}},\ \bibinfo {pages}
  {016805} (\bibinfo {year} {2019})}\BibitemShut {NoStop}%
\bibitem [{\citenamefont {Peng}\ and\ \citenamefont
  {Refael}(2019)}]{PhysRevLett.123.016806}%
  \BibitemOpen
  \bibfield  {author} {\bibinfo {author} {\bibfnamefont {Y.}~\bibnamefont
  {Peng}}\ and\ \bibinfo {author} {\bibfnamefont {G.}~\bibnamefont {Refael}},\
  }\bibfield  {title} {\enquote {\bibinfo {title} {Floquet second-order
  topological insulators from nonsymmorphic space-time symmetries},}\ }\href
  {\doibase 10.1103/PhysRevLett.123.016806} {\bibfield  {journal} {\bibinfo
  {journal} {Phys. Rev. Lett.}\ }\textbf {\bibinfo {volume} {123}},\ \bibinfo
  {pages} {016806} (\bibinfo {year} {2019})}\BibitemShut {NoStop}%
\bibitem [{\citenamefont {Zeng}\ \emph {et~al.}(2019)\citenamefont {Zeng},
  \citenamefont {Stanescu}, \citenamefont {Zhang}, \citenamefont {Scarola},\
  and\ \citenamefont {Tewari}}]{PhysRevLett.123.060402}%
  \BibitemOpen
  \bibfield  {author} {\bibinfo {author} {\bibfnamefont {C.}~\bibnamefont
  {Zeng}}, \bibinfo {author} {\bibfnamefont {T.~D.}\ \bibnamefont {Stanescu}},
  \bibinfo {author} {\bibfnamefont {C.}~\bibnamefont {Zhang}}, \bibinfo
  {author} {\bibfnamefont {V.~W.}\ \bibnamefont {Scarola}}, \ and\ \bibinfo
  {author} {\bibfnamefont {S.}~\bibnamefont {Tewari}},\ }\bibfield  {title}
  {\enquote {\bibinfo {title} {Majorana corner modes with solitons in an
  attractive hubbard-hofstadter model of cold atom optical lattices},}\ }\href
  {\doibase 10.1103/PhysRevLett.123.060402} {\bibfield  {journal} {\bibinfo
  {journal} {Phys. Rev. Lett.}\ }\textbf {\bibinfo {volume} {123}},\ \bibinfo
  {pages} {060402} (\bibinfo {year} {2019})}\BibitemShut {NoStop}%
\bibitem [{\citenamefont {Luo}\ and\ \citenamefont
  {Zhang}(2019)}]{PhysRevLett.123.073601}%
  \BibitemOpen
  \bibfield  {author} {\bibinfo {author} {\bibfnamefont {X.-W.}\ \bibnamefont
  {Luo}}\ and\ \bibinfo {author} {\bibfnamefont {C.}~\bibnamefont {Zhang}},\
  }\bibfield  {title} {\enquote {\bibinfo {title} {Higher-order topological
  corner states induced by gain and loss},}\ }\href {\doibase
  10.1103/PhysRevLett.123.073601} {\bibfield  {journal} {\bibinfo  {journal}
  {Phys. Rev. Lett.}\ }\textbf {\bibinfo {volume} {123}},\ \bibinfo {pages}
  {073601} (\bibinfo {year} {2019})}\BibitemShut {NoStop}%
\bibitem [{\citenamefont {Zhang}\ \emph
  {et~al.}(2019{\natexlab{a}})\citenamefont {Zhang}, \citenamefont {Cole},
  \citenamefont {Wu},\ and\ \citenamefont
  {Das~Sarma}}]{PhysRevLett.123.167001}%
  \BibitemOpen
  \bibfield  {author} {\bibinfo {author} {\bibfnamefont {R.-X.}\ \bibnamefont
  {Zhang}}, \bibinfo {author} {\bibfnamefont {W.~S.}\ \bibnamefont {Cole}},
  \bibinfo {author} {\bibfnamefont {X.}~\bibnamefont {Wu}}, \ and\ \bibinfo
  {author} {\bibfnamefont {S.}~\bibnamefont {Das~Sarma}},\ }\bibfield  {title}
  {\enquote {\bibinfo {title} {Higher-order topology and nodal topological
  superconductivity in fe(se,te) heterostructures},}\ }\href {\doibase
  10.1103/PhysRevLett.123.167001} {\bibfield  {journal} {\bibinfo  {journal}
  {Phys. Rev. Lett.}\ }\textbf {\bibinfo {volume} {123}},\ \bibinfo {pages}
  {167001} (\bibinfo {year} {2019}{\natexlab{a}})}\BibitemShut {NoStop}%
\bibitem [{\citenamefont {Wang}\ \emph {et~al.}(2019)\citenamefont {Wang},
  \citenamefont {Wieder}, \citenamefont {Li}, \citenamefont {Yan},\ and\
  \citenamefont {Bernevig}}]{PhysRevLett.123.186401}%
  \BibitemOpen
  \bibfield  {author} {\bibinfo {author} {\bibfnamefont {Z.}~\bibnamefont
  {Wang}}, \bibinfo {author} {\bibfnamefont {B.~J.}\ \bibnamefont {Wieder}},
  \bibinfo {author} {\bibfnamefont {J.}~\bibnamefont {Li}}, \bibinfo {author}
  {\bibfnamefont {B.}~\bibnamefont {Yan}}, \ and\ \bibinfo {author}
  {\bibfnamefont {B.~A.}\ \bibnamefont {Bernevig}},\ }\bibfield  {title}
  {\enquote {\bibinfo {title} {Higher-order topology, monopole nodal lines, and
  the origin of large fermi arcs in transition metal dichalcogenides
  $x{\mathrm{te}}_{2}$ ($x=\mathrm{Mo},\mathrm{W}$)},}\ }\href {\doibase
  10.1103/PhysRevLett.123.186401} {\bibfield  {journal} {\bibinfo  {journal}
  {Phys. Rev. Lett.}\ }\textbf {\bibinfo {volume} {123}},\ \bibinfo {pages}
  {186401} (\bibinfo {year} {2019})}\BibitemShut {NoStop}%
\bibitem [{\citenamefont {Sheng}\ \emph {et~al.}(2019)\citenamefont {Sheng},
  \citenamefont {Chen}, \citenamefont {Liu}, \citenamefont {Chen},
  \citenamefont {Yu}, \citenamefont {Zhao},\ and\ \citenamefont
  {Yang}}]{PhysRevLett.123.256402}%
  \BibitemOpen
  \bibfield  {author} {\bibinfo {author} {\bibfnamefont {X.-L.}\ \bibnamefont
  {Sheng}}, \bibinfo {author} {\bibfnamefont {C.}~\bibnamefont {Chen}},
  \bibinfo {author} {\bibfnamefont {H.}~\bibnamefont {Liu}}, \bibinfo {author}
  {\bibfnamefont {Z.}~\bibnamefont {Chen}}, \bibinfo {author} {\bibfnamefont
  {Z.-M.}\ \bibnamefont {Yu}}, \bibinfo {author} {\bibfnamefont {Y.~X.}\
  \bibnamefont {Zhao}}, \ and\ \bibinfo {author} {\bibfnamefont {S.~A.}\
  \bibnamefont {Yang}},\ }\bibfield  {title} {\enquote {\bibinfo {title}
  {Two-dimensional second-order topological insulator in graphdiyne},}\ }\href
  {\doibase 10.1103/PhysRevLett.123.256402} {\bibfield  {journal} {\bibinfo
  {journal} {Phys. Rev. Lett.}\ }\textbf {\bibinfo {volume} {123}},\ \bibinfo
  {pages} {256402} (\bibinfo {year} {2019})}\BibitemShut {NoStop}%
\bibitem [{\citenamefont {Li}\ \emph {et~al.}(2020{\natexlab{a}})\citenamefont
  {Li}, \citenamefont {Fu}, \citenamefont {Hu}, \citenamefont {Li},\ and\
  \citenamefont {Shen}}]{PhysRevLett.125.166801}%
  \BibitemOpen
  \bibfield  {author} {\bibinfo {author} {\bibfnamefont {C.-A.}\ \bibnamefont
  {Li}}, \bibinfo {author} {\bibfnamefont {B.}~\bibnamefont {Fu}}, \bibinfo
  {author} {\bibfnamefont {Z.-A.}\ \bibnamefont {Hu}}, \bibinfo {author}
  {\bibfnamefont {J.}~\bibnamefont {Li}}, \ and\ \bibinfo {author}
  {\bibfnamefont {S.-Q.}\ \bibnamefont {Shen}},\ }\bibfield  {title} {\enquote
  {\bibinfo {title} {Topological phase transitions in disordered electric
  quadrupole insulators},}\ }\href {\doibase 10.1103/PhysRevLett.125.166801}
  {\bibfield  {journal} {\bibinfo  {journal} {Phys. Rev. Lett.}\ }\textbf
  {\bibinfo {volume} {125}},\ \bibinfo {pages} {166801} (\bibinfo {year}
  {2020}{\natexlab{a}})}\BibitemShut {NoStop}%
\bibitem [{\citenamefont {C\ifmmode \u{a}\else \u{a}\fi{}lug\ifmmode~\u{a}\else
  \u{a}\fi{}ru}\ \emph {et~al.}(2019)\citenamefont {C\ifmmode \u{a}\else
  \u{a}\fi{}lug\ifmmode~\u{a}\else \u{a}\fi{}ru}, \citenamefont {Juri\ifmmode
  \check{c}\else \v{c}\fi{}i\ifmmode~\acute{c}\else \'{c}\fi{}},\ and\
  \citenamefont {Roy}}]{PhysRevB.99.041301}%
  \BibitemOpen
  \bibfield  {author} {\bibinfo {author} {\bibfnamefont {Dumitru}\ \bibnamefont
  {C\ifmmode \u{a}\else \u{a}\fi{}lug\ifmmode~\u{a}\else \u{a}\fi{}ru}},
  \bibinfo {author} {\bibfnamefont {Vladimir}\ \bibnamefont {Juri\ifmmode
  \check{c}\else \v{c}\fi{}i\ifmmode~\acute{c}\else \'{c}\fi{}}}, \ and\
  \bibinfo {author} {\bibfnamefont {Bitan}\ \bibnamefont {Roy}},\ }\bibfield
  {title} {\enquote {\bibinfo {title} {Higher-order topological phases: A
  general principle of construction},}\ }\href {\doibase
  10.1103/PhysRevB.99.041301} {\bibfield  {journal} {\bibinfo  {journal} {Phys.
  Rev. B}\ }\textbf {\bibinfo {volume} {99}},\ \bibinfo {pages} {041301}
  (\bibinfo {year} {2019})}\BibitemShut {NoStop}%
\bibitem [{\citenamefont {Li}\ \emph {et~al.}(2020{\natexlab{b}})\citenamefont
  {Li}, \citenamefont {Wang}, \citenamefont {Li}, \citenamefont {Zheng},
  \citenamefont {Brinkman}, \citenamefont {Yu},\ and\ \citenamefont
  {Liao}}]{PhysRevLett.124.156601}%
  \BibitemOpen
  \bibfield  {author} {\bibinfo {author} {\bibfnamefont {C.-Z.}\ \bibnamefont
  {Li}}, \bibinfo {author} {\bibfnamefont {A.-Q.}\ \bibnamefont {Wang}},
  \bibinfo {author} {\bibfnamefont {C.}~\bibnamefont {Li}}, \bibinfo {author}
  {\bibfnamefont {W.-Z.}\ \bibnamefont {Zheng}}, \bibinfo {author}
  {\bibfnamefont {A.}~\bibnamefont {Brinkman}}, \bibinfo {author}
  {\bibfnamefont {D.-P.}\ \bibnamefont {Yu}}, \ and\ \bibinfo {author}
  {\bibfnamefont {Z.-M.}\ \bibnamefont {Liao}},\ }\bibfield  {title} {\enquote
  {\bibinfo {title} {Reducing electronic transport dimension to topological
  hinge states by increasing geometry size of dirac semimetal josephson
  junctions},}\ }\href {\doibase 10.1103/PhysRevLett.124.156601} {\bibfield
  {journal} {\bibinfo  {journal} {Phys. Rev. Lett.}\ }\textbf {\bibinfo
  {volume} {124}},\ \bibinfo {pages} {156601} (\bibinfo {year}
  {2020}{\natexlab{b}})}\BibitemShut {NoStop}%
\bibitem [{\citenamefont {Wang}\ \emph {et~al.}(2022)\citenamefont {Wang},
  \citenamefont {Xiang}, \citenamefont {Zhao},\ and\ \citenamefont
  {Liao}}]{WANG2022788}%
  \BibitemOpen
  \bibfield  {author} {\bibinfo {author} {\bibfnamefont {A.-Q.}\ \bibnamefont
  {Wang}}, \bibinfo {author} {\bibfnamefont {P.-Z.}\ \bibnamefont {Xiang}},
  \bibinfo {author} {\bibfnamefont {T.-Y.}\ \bibnamefont {Zhao}}, \ and\
  \bibinfo {author} {\bibfnamefont {Z.-M.}\ \bibnamefont {Liao}},\ }\bibfield
  {title} {\enquote {\bibinfo {title} {Topological nature of higher-order hinge
  states revealed by spin transport},}\ }\href {\doibase
  https://doi.org/10.1016/j.scib.2022.02.003} {\bibfield  {journal} {\bibinfo
  {journal} {Sci. Bull.}\ }\textbf {\bibinfo {volume} {67}},\ \bibinfo {pages}
  {788--793} (\bibinfo {year} {2022})}\BibitemShut {NoStop}%
\bibitem [{\citenamefont {Wang}\ \emph
  {et~al.}(2021{\natexlab{a}})\citenamefont {Wang}, \citenamefont {Ke},
  \citenamefont {Chang}, \citenamefont {Lu}, \citenamefont {Gao}, \citenamefont
  {Lee},\ and\ \citenamefont {Jin}}]{PhysRevB.104.224303}%
  \BibitemOpen
  \bibfield  {author} {\bibinfo {author} {\bibfnamefont {Yao}\ \bibnamefont
  {Wang}}, \bibinfo {author} {\bibfnamefont {Yongguan}\ \bibnamefont {Ke}},
  \bibinfo {author} {\bibfnamefont {Yi-Jun}\ \bibnamefont {Chang}}, \bibinfo
  {author} {\bibfnamefont {Yong-Heng}\ \bibnamefont {Lu}}, \bibinfo {author}
  {\bibfnamefont {Jun}\ \bibnamefont {Gao}}, \bibinfo {author} {\bibfnamefont
  {Chaohong}\ \bibnamefont {Lee}}, \ and\ \bibinfo {author} {\bibfnamefont
  {Xian-Min}\ \bibnamefont {Jin}},\ }\bibfield  {title} {\enquote {\bibinfo
  {title} {Constructing higher-order topological states in higher
  dimensions},}\ }\href {\doibase 10.1103/PhysRevB.104.224303} {\bibfield
  {journal} {\bibinfo  {journal} {Phys. Rev. B}\ }\textbf {\bibinfo {volume}
  {104}},\ \bibinfo {pages} {224303} (\bibinfo {year}
  {2021}{\natexlab{a}})}\BibitemShut {NoStop}%
\bibitem [{\citenamefont {Benalcazar}\ and\ \citenamefont
  {Cerjan}(2022)}]{PhysRevLett.128.127601}%
  \BibitemOpen
  \bibfield  {author} {\bibinfo {author} {\bibfnamefont {W.~A.}\ \bibnamefont
  {Benalcazar}}\ and\ \bibinfo {author} {\bibfnamefont {A.}~\bibnamefont
  {Cerjan}},\ }\bibfield  {title} {\enquote {\bibinfo {title} {Chiral-symmetric
  higher-order topological phases of matter},}\ }\href {\doibase
  10.1103/PhysRevLett.128.127601} {\bibfield  {journal} {\bibinfo  {journal}
  {Phys. Rev. Lett.}\ }\textbf {\bibinfo {volume} {128}},\ \bibinfo {pages}
  {127601} (\bibinfo {year} {2022})}\BibitemShut {NoStop}%
\bibitem [{\citenamefont {Li}\ \emph {et~al.}(2021)\citenamefont {Li},
  \citenamefont {Zhu},\ and\ \citenamefont {Gong}}]{LI2021}%
  \BibitemOpen
  \bibfield  {author} {\bibinfo {author} {\bibfnamefont {L.}~\bibnamefont
  {Li}}, \bibinfo {author} {\bibfnamefont {W.}~\bibnamefont {Zhu}}, \ and\
  \bibinfo {author} {\bibfnamefont {J.}~\bibnamefont {Gong}},\ }\bibfield
  {title} {\enquote {\bibinfo {title} {Direct dynamical characterization of
  higher-order topological phases with nested band inversion surfaces},}\
  }\href {\doibase https://doi.org/10.1016/j.scib.2021.04.006} {\bibfield
  {journal} {\bibinfo  {journal} {Sci. Bull.}\ }\textbf {\bibinfo {volume}
  {66}},\ \bibinfo {pages} {1502--1510} (\bibinfo {year} {2021})}\BibitemShut
  {NoStop}%
\bibitem [{\citenamefont {Lei}\ \emph {et~al.}(2022)\citenamefont {Lei},
  \citenamefont {Li},\ and\ \citenamefont {Deng}}]{2209.14811}%
  \BibitemOpen
  \bibfield  {author} {\bibinfo {author} {\bibfnamefont {Z.}~\bibnamefont
  {Lei}}, \bibinfo {author} {\bibfnamefont {L.}~\bibnamefont {Li}}, \ and\
  \bibinfo {author} {\bibfnamefont {Y.}~\bibnamefont {Deng}},\ }\href@noop {}
  {\enquote {\bibinfo {title} {Tunable symmetry-protected higher-order
  topological states with fermionic atoms in bilayer optical lattices},}\ }
  (\bibinfo {year} {2022}),\ \Eprint {http://arxiv.org/abs/arXiv:2209.14811}
  {arXiv:2209.14811} \BibitemShut {NoStop}%
\bibitem [{\citenamefont {Zirnbauer}(1996)}]{doi:10.1063/1.531675}%
  \BibitemOpen
  \bibfield  {author} {\bibinfo {author} {\bibfnamefont {M.~R.}\ \bibnamefont
  {Zirnbauer}},\ }\bibfield  {title} {\enquote {\bibinfo {title} {Riemannian
  symmetric superspaces and their origin in random-matrix theory},}\ }\href
  {\doibase 10.1063/1.531675} {\bibfield  {journal} {\bibinfo  {journal} {J.
  Math. Phys. (N.Y.)}\ }\textbf {\bibinfo {volume} {37}},\ \bibinfo {pages}
  {4986--5018} (\bibinfo {year} {1996})}\BibitemShut {NoStop}%
\bibitem [{\citenamefont {Altland}\ and\ \citenamefont
  {Zirnbauer}(1997)}]{PhysRevB.55.1142}%
  \BibitemOpen
  \bibfield  {author} {\bibinfo {author} {\bibfnamefont {A.}~\bibnamefont
  {Altland}}\ and\ \bibinfo {author} {\bibfnamefont {M.~R.}\ \bibnamefont
  {Zirnbauer}},\ }\bibfield  {title} {\enquote {\bibinfo {title} {Nonstandard
  symmetry classes in mesoscopic normal-superconducting hybrid structures},}\
  }\href {\doibase 10.1103/PhysRevB.55.1142} {\bibfield  {journal} {\bibinfo
  {journal} {Phys. Rev. B}\ }\textbf {\bibinfo {volume} {55}},\ \bibinfo
  {pages} {1142--1161} (\bibinfo {year} {1997})}\BibitemShut {NoStop}%
\bibitem [{\citenamefont {Ryu}\ \emph {et~al.}(2010)\citenamefont {Ryu},
  \citenamefont {Schnyder}, \citenamefont {Furusaki},\ and\ \citenamefont
  {Ludwig}}]{Ryu_2010}%
  \BibitemOpen
  \bibfield  {author} {\bibinfo {author} {\bibfnamefont {S.}~\bibnamefont
  {Ryu}}, \bibinfo {author} {\bibfnamefont {A.~P.}\ \bibnamefont {Schnyder}},
  \bibinfo {author} {\bibfnamefont {A.}~\bibnamefont {Furusaki}}, \ and\
  \bibinfo {author} {\bibfnamefont {A.~W.~W.}\ \bibnamefont {Ludwig}},\
  }\bibfield  {title} {\enquote {\bibinfo {title} {Topological insulators and
  superconductors: tenfold way and dimensional hierarchy},}\ }\href {\doibase
  10.1088/1367-2630/12/6/065010} {\bibfield  {journal} {\bibinfo  {journal}
  {New J. Phys.}\ }\textbf {\bibinfo {volume} {12}},\ \bibinfo {pages} {065010}
  (\bibinfo {year} {2010})}\BibitemShut {NoStop}%
\bibitem [{\citenamefont {Chiu}\ \emph {et~al.}(2016)\citenamefont {Chiu},
  \citenamefont {Teo}, \citenamefont {Schnyder},\ and\ \citenamefont
  {Ryu}}]{RevModPhys.88.035005}%
  \BibitemOpen
  \bibfield  {author} {\bibinfo {author} {\bibfnamefont {C.-K.}\ \bibnamefont
  {Chiu}}, \bibinfo {author} {\bibfnamefont {J.~C.~Y.}\ \bibnamefont {Teo}},
  \bibinfo {author} {\bibfnamefont {A.~P.}\ \bibnamefont {Schnyder}}, \ and\
  \bibinfo {author} {\bibfnamefont {S.}~\bibnamefont {Ryu}},\ }\bibfield
  {title} {\enquote {\bibinfo {title} {Classification of topological quantum
  matter with symmetries},}\ }\href {\doibase 10.1103/RevModPhys.88.035005}
  {\bibfield  {journal} {\bibinfo  {journal} {Rev. Mod. Phys.}\ }\textbf
  {\bibinfo {volume} {88}},\ \bibinfo {pages} {035005} (\bibinfo {year}
  {2016})}\BibitemShut {NoStop}%
\bibitem [{\citenamefont {Kitaev}(2001)}]{Kitaev_2001}%
  \BibitemOpen
  \bibfield  {author} {\bibinfo {author} {\bibfnamefont {A~Yu}\ \bibnamefont
  {Kitaev}},\ }\bibfield  {title} {\enquote {\bibinfo {title} {Unpaired
  majorana fermions in quantum wires},}\ }\href {\doibase
  10.1070/1063-7869/44/10s/s29} {\bibfield  {journal} {\bibinfo  {journal}
  {Phys. Usp.}\ }\textbf {\bibinfo {volume} {44}},\ \bibinfo {pages} {131--136}
  (\bibinfo {year} {2001})}\BibitemShut {NoStop}%
\bibitem [{\citenamefont {Elliott}\ and\ \citenamefont
  {Franz}(2015)}]{RevModPhys.87.137}%
  \BibitemOpen
  \bibfield  {author} {\bibinfo {author} {\bibfnamefont {S.~R.}\ \bibnamefont
  {Elliott}}\ and\ \bibinfo {author} {\bibfnamefont {M.}~\bibnamefont
  {Franz}},\ }\bibfield  {title} {\enquote {\bibinfo {title} {Colloquium:
  Majorana fermions in nuclear, particle, and solid-state physics},}\ }\href
  {\doibase 10.1103/RevModPhys.87.137} {\bibfield  {journal} {\bibinfo
  {journal} {Rev. Mod. Phys.}\ }\textbf {\bibinfo {volume} {87}},\ \bibinfo
  {pages} {137--163} (\bibinfo {year} {2015})}\BibitemShut {NoStop}%
\bibitem [{\citenamefont {Okuma}\ \emph {et~al.}(2019)\citenamefont {Okuma},
  \citenamefont {Sato},\ and\ \citenamefont {Shiozaki}}]{PhysRevB.99.085127}%
  \BibitemOpen
  \bibfield  {author} {\bibinfo {author} {\bibfnamefont {N.}~\bibnamefont
  {Okuma}}, \bibinfo {author} {\bibfnamefont {M.}~\bibnamefont {Sato}}, \ and\
  \bibinfo {author} {\bibfnamefont {K.}~\bibnamefont {Shiozaki}},\ }\bibfield
  {title} {\enquote {\bibinfo {title} {Topological classification under
  nonmagnetic and magnetic point group symmetry: Application of real-space
  atiyah-hirzebruch spectral sequence to higher-order topology},}\ }\href
  {\doibase 10.1103/PhysRevB.99.085127} {\bibfield  {journal} {\bibinfo
  {journal} {Phys. Rev. B}\ }\textbf {\bibinfo {volume} {99}},\ \bibinfo
  {pages} {085127} (\bibinfo {year} {2019})}\BibitemShut {NoStop}%
\bibitem [{\citenamefont {Rasmussen}\ and\ \citenamefont
  {Lu}(2020)}]{PhysRevB.101.085137}%
  \BibitemOpen
  \bibfield  {author} {\bibinfo {author} {\bibfnamefont {A.}~\bibnamefont
  {Rasmussen}}\ and\ \bibinfo {author} {\bibfnamefont {Y.-M.}\ \bibnamefont
  {Lu}},\ }\bibfield  {title} {\enquote {\bibinfo {title} {Classification and
  construction of higher-order symmetry-protected topological phases of
  interacting bosons},}\ }\href {\doibase 10.1103/PhysRevB.101.085137}
  {\bibfield  {journal} {\bibinfo  {journal} {Phys. Rev. B}\ }\textbf {\bibinfo
  {volume} {101}},\ \bibinfo {pages} {085137} (\bibinfo {year}
  {2020})}\BibitemShut {NoStop}%
\bibitem [{\citenamefont {Zhang}\ \emph {et~al.}(2018)\citenamefont {Zhang},
  \citenamefont {Zhang}, \citenamefont {Niu},\ and\ \citenamefont
  {Liu}}]{ZHANG20181385}%
  \BibitemOpen
  \bibfield  {author} {\bibinfo {author} {\bibfnamefont {L.}~\bibnamefont
  {Zhang}}, \bibinfo {author} {\bibfnamefont {L.}~\bibnamefont {Zhang}},
  \bibinfo {author} {\bibfnamefont {S.}~\bibnamefont {Niu}}, \ and\ \bibinfo
  {author} {\bibfnamefont {X.-J.}\ \bibnamefont {Liu}},\ }\bibfield  {title}
  {\enquote {\bibinfo {title} {Dynamical classification of topological quantum
  phases},}\ }\href {\doibase https://doi.org/10.1016/j.scib.2018.09.018}
  {\bibfield  {journal} {\bibinfo  {journal} {Sci. Bull.}\ }\textbf {\bibinfo
  {volume} {63}},\ \bibinfo {pages} {1385--1391} (\bibinfo {year}
  {2018})}\BibitemShut {NoStop}%
\bibitem [{\citenamefont {Yu}\ \emph {et~al.}(2021)\citenamefont {Yu},
  \citenamefont {Ji}, \citenamefont {Zhang}, \citenamefont {Wang},
  \citenamefont {Wu},\ and\ \citenamefont {Liu}}]{PRXQuantum.2.020320}%
  \BibitemOpen
  \bibfield  {author} {\bibinfo {author} {\bibfnamefont {X.-L.}\ \bibnamefont
  {Yu}}, \bibinfo {author} {\bibfnamefont {W.}~\bibnamefont {Ji}}, \bibinfo
  {author} {\bibfnamefont {L.}~\bibnamefont {Zhang}}, \bibinfo {author}
  {\bibfnamefont {Y.}~\bibnamefont {Wang}}, \bibinfo {author} {\bibfnamefont
  {J.}~\bibnamefont {Wu}}, \ and\ \bibinfo {author} {\bibfnamefont {X.-J.}\
  \bibnamefont {Liu}},\ }\bibfield  {title} {\enquote {\bibinfo {title}
  {Quantum dynamical characterization and simulation of topological phases with
  high-order band inversion surfaces},}\ }\href {\doibase
  10.1103/PRXQuantum.2.020320} {\bibfield  {journal} {\bibinfo  {journal} {PRX
  Quantum}\ }\textbf {\bibinfo {volume} {2}},\ \bibinfo {pages} {020320}
  (\bibinfo {year} {2021})}\BibitemShut {NoStop}%
\bibitem [{\citenamefont {Li}\ and\ \citenamefont {Gong}(2021)}]{LI20211817}%
  \BibitemOpen
  \bibfield  {author} {\bibinfo {author} {\bibfnamefont {Linhu}\ \bibnamefont
  {Li}}\ and\ \bibinfo {author} {\bibfnamefont {Jiangbin}\ \bibnamefont
  {Gong}},\ }\bibfield  {title} {\enquote {\bibinfo {title} {Probing
  higher-order band topology via spin texture measurements: quantum
  simulation},}\ }\href {\doibase https://doi.org/10.1016/j.scib.2021.05.025}
  {\bibfield  {journal} {\bibinfo  {journal} {Science Bulletin}\ }\textbf
  {\bibinfo {volume} {66}},\ \bibinfo {pages} {1817--1818} (\bibinfo {year}
  {2021})}\BibitemShut {NoStop}%
\bibitem [{\citenamefont {Mong}\ and\ \citenamefont
  {Shivamoggi}(2011)}]{PhysRevB.83.125109}%
  \BibitemOpen
  \bibfield  {author} {\bibinfo {author} {\bibfnamefont {R.~S.~K.}\
  \bibnamefont {Mong}}\ and\ \bibinfo {author} {\bibfnamefont {V.}~\bibnamefont
  {Shivamoggi}},\ }\bibfield  {title} {\enquote {\bibinfo {title} {Edge states
  and the bulk-boundary correspondence in dirac hamiltonians},}\ }\href
  {\doibase 10.1103/PhysRevB.83.125109} {\bibfield  {journal} {\bibinfo
  {journal} {Phys. Rev. B}\ }\textbf {\bibinfo {volume} {83}},\ \bibinfo
  {pages} {125109} (\bibinfo {year} {2011})}\BibitemShut {NoStop}%
\bibitem [{\citenamefont {Li}\ \emph {et~al.}(2017)\citenamefont {Li},
  \citenamefont {Yap}, \citenamefont {Ara\'ujo},\ and\ \citenamefont
  {Gong}}]{PhysRevB.96.235424}%
  \BibitemOpen
  \bibfield  {author} {\bibinfo {author} {\bibfnamefont {L.}~\bibnamefont
  {Li}}, \bibinfo {author} {\bibfnamefont {H.~H.}\ \bibnamefont {Yap}},
  \bibinfo {author} {\bibfnamefont {M.~A.~N.}\ \bibnamefont {Ara\'ujo}}, \ and\
  \bibinfo {author} {\bibfnamefont {J.}~\bibnamefont {Gong}},\ }\bibfield
  {title} {\enquote {\bibinfo {title} {Engineering topological phases with a
  three-dimensional nodal-loop semimetal},}\ }\href {\doibase
  10.1103/PhysRevB.96.235424} {\bibfield  {journal} {\bibinfo  {journal} {Phys.
  Rev. B}\ }\textbf {\bibinfo {volume} {96}},\ \bibinfo {pages} {235424}
  (\bibinfo {year} {2017})}\BibitemShut {NoStop}%
\bibitem [{Note1()}]{Note1}%
  \BibitemOpen
  \bibinfo {note} {A chiral symmetry seem to emerge for $H_2^{(I)}(\protect
  \mathbf {k}_{1,\parallel })$ due to the absence of $\Gamma ^{1,2}_{(2p+1)}$,
  yet the Hilbert space of this effective Hamiltonian can be reduced to
  $2^{p-1}$ dimension, where the chiral symmetry is ruled out.}\BibitemShut
  {Stop}%
\bibitem [{\citenamefont {Li}\ and\ \citenamefont
  {Ara\'ujo}(2016)}]{PhysRevB.94.165117}%
  \BibitemOpen
  \bibfield  {author} {\bibinfo {author} {\bibfnamefont {Linhu}\ \bibnamefont
  {Li}}\ and\ \bibinfo {author} {\bibfnamefont {Miguel A.~N.}\ \bibnamefont
  {Ara\'ujo}},\ }\bibfield  {title} {\enquote {\bibinfo {title} {Topological
  insulating phases from two-dimensional nodal loop semimetals},}\ }\href
  {\doibase 10.1103/PhysRevB.94.165117} {\bibfield  {journal} {\bibinfo
  {journal} {Phys. Rev. B}\ }\textbf {\bibinfo {volume} {94}},\ \bibinfo
  {pages} {165117} (\bibinfo {year} {2016})}\BibitemShut {NoStop}%
\bibitem [{\citenamefont {Schindler}\ \emph
  {et~al.}(2018{\natexlab{b}})\citenamefont {Schindler}, \citenamefont {Cook},
  \citenamefont {Vergniory}, \citenamefont {Wang}, \citenamefont {Parkin},
  \citenamefont {Bernevig},\ and\ \citenamefont
  {Neupert}}]{doi:10.1126/sciadv.aat0346}%
  \BibitemOpen
  \bibfield  {author} {\bibinfo {author} {\bibfnamefont {F.}~\bibnamefont
  {Schindler}}, \bibinfo {author} {\bibfnamefont {A.~M.}\ \bibnamefont {Cook}},
  \bibinfo {author} {\bibfnamefont {M.~G.}\ \bibnamefont {Vergniory}}, \bibinfo
  {author} {\bibfnamefont {Z.}~\bibnamefont {Wang}}, \bibinfo {author}
  {\bibfnamefont {S.~S.~P.}\ \bibnamefont {Parkin}}, \bibinfo {author}
  {\bibfnamefont {B.~A.}\ \bibnamefont {Bernevig}}, \ and\ \bibinfo {author}
  {\bibfnamefont {T.}~\bibnamefont {Neupert}},\ }\bibfield  {title} {\enquote
  {\bibinfo {title} {Higher-order topological insulators},}\ }\href {\doibase
  10.1126/sciadv.aat0346} {\bibfield  {journal} {\bibinfo  {journal} {Sci.
  Adv.}\ }\textbf {\bibinfo {volume} {4}},\ \bibinfo {pages} {eaat0346}
  (\bibinfo {year} {2018}{\natexlab{b}})}\BibitemShut {NoStop}%
\bibitem [{\citenamefont {Qi}\ \emph {et~al.}(2008)\citenamefont {Qi},
  \citenamefont {Hughes},\ and\ \citenamefont {Zhang}}]{PhysRevB.78.195424}%
  \BibitemOpen
  \bibfield  {author} {\bibinfo {author} {\bibfnamefont {X.-L.}\ \bibnamefont
  {Qi}}, \bibinfo {author} {\bibfnamefont {T.~L.}\ \bibnamefont {Hughes}}, \
  and\ \bibinfo {author} {\bibfnamefont {S.-C.}\ \bibnamefont {Zhang}},\
  }\bibfield  {title} {\enquote {\bibinfo {title} {Topological field theory of
  time-reversal invariant insulators},}\ }\href {\doibase
  10.1103/PhysRevB.78.195424} {\bibfield  {journal} {\bibinfo  {journal} {Phys.
  Rev. B}\ }\textbf {\bibinfo {volume} {78}},\ \bibinfo {pages} {195424}
  (\bibinfo {year} {2008})}\BibitemShut {NoStop}%
\bibitem [{\citenamefont {Li}\ \emph {et~al.}(2016)\citenamefont {Li},
  \citenamefont {Yang},\ and\ \citenamefont {Chen}}]{Li2016}%
  \BibitemOpen
  \bibfield  {author} {\bibinfo {author} {\bibfnamefont {L.}~\bibnamefont
  {Li}}, \bibinfo {author} {\bibfnamefont {C.}~\bibnamefont {Yang}}, \ and\
  \bibinfo {author} {\bibfnamefont {S.}~\bibnamefont {Chen}},\ }\bibfield
  {title} {\enquote {\bibinfo {title} {Topological invariants for phase
  transition points of one-dimensional z2 topological systems},}\ }\href
  {\doibase 10.1140/epjb/e2016-70325-x} {\bibfield  {journal} {\bibinfo
  {journal} {Eur. Phys. J. B}\ }\textbf {\bibinfo {volume} {89}},\ \bibinfo
  {pages} {195} (\bibinfo {year} {2016})}\BibitemShut {NoStop}%
\bibitem [{\citenamefont {Xiao}\ \emph {et~al.}(2010)\citenamefont {Xiao},
  \citenamefont {Chang},\ and\ \citenamefont {Niu}}]{RevModPhys.82.1959}%
  \BibitemOpen
  \bibfield  {author} {\bibinfo {author} {\bibfnamefont {D.}~\bibnamefont
  {Xiao}}, \bibinfo {author} {\bibfnamefont {M.-C.}\ \bibnamefont {Chang}}, \
  and\ \bibinfo {author} {\bibfnamefont {Q.}~\bibnamefont {Niu}},\ }\bibfield
  {title} {\enquote {\bibinfo {title} {Berry phase effects on electronic
  properties},}\ }\href {\doibase 10.1103/RevModPhys.82.1959} {\bibfield
  {journal} {\bibinfo  {journal} {Rev. Mod. Phys.}\ }\textbf {\bibinfo {volume}
  {82}},\ \bibinfo {pages} {1959--2007} (\bibinfo {year} {2010})}\BibitemShut
  {NoStop}%
\bibitem [{\citenamefont {Zhang}\ \emph {et~al.}(2022)\citenamefont {Zhang},
  \citenamefont {Jia},\ and\ \citenamefont {Liu}}]{ScienceBulletin671236}%
  \BibitemOpen
  \bibfield  {author} {\bibinfo {author} {\bibfnamefont {L.}~\bibnamefont
  {Zhang}}, \bibinfo {author} {\bibfnamefont {W.}~\bibnamefont {Jia}}, \ and\
  \bibinfo {author} {\bibfnamefont {X.-J.}\ \bibnamefont {Liu}},\ }\bibfield
  {title} {\enquote {\bibinfo {title} {Universal topological quench dynamics
  for z2 topological phases},}\ }\href
  {http://www.sciengine.com/publisher/Science China Press/journal/Science
  Bulletin/67/12/10.1016/j.scib.2022.04.019, doi =} {\bibfield  {journal}
  {\bibinfo  {journal} {Sci. Bull.}\ }\textbf {\bibinfo {volume} {67}},\
  \bibinfo {pages} {1236--1242} (\bibinfo {year} {2022})}\BibitemShut {NoStop}%
\bibitem [{\citenamefont {Lababidi}\ \emph {et~al.}(2014)\citenamefont
  {Lababidi}, \citenamefont {Satija},\ and\ \citenamefont
  {Zhao}}]{PhysRevLett.112.026805}%
  \BibitemOpen
  \bibfield  {author} {\bibinfo {author} {\bibfnamefont {Mahmoud}\ \bibnamefont
  {Lababidi}}, \bibinfo {author} {\bibfnamefont {Indubala~I.}\ \bibnamefont
  {Satija}}, \ and\ \bibinfo {author} {\bibfnamefont {Erhai}\ \bibnamefont
  {Zhao}},\ }\bibfield  {title} {\enquote {\bibinfo {title}
  {Counter-propagating edge modes and topological phases of a kicked quantum
  hall system},}\ }\href {\doibase 10.1103/PhysRevLett.112.026805} {\bibfield
  {journal} {\bibinfo  {journal} {Phys. Rev. Lett.}\ }\textbf {\bibinfo
  {volume} {112}},\ \bibinfo {pages} {026805} (\bibinfo {year}
  {2014})}\BibitemShut {NoStop}%
\bibitem [{\citenamefont {Yoshimura}\ \emph {et~al.}(2014)\citenamefont
  {Yoshimura}, \citenamefont {Imura}, \citenamefont {Fukui},\ and\
  \citenamefont {Hatsugai}}]{PhysRevB.90.155443}%
  \BibitemOpen
  \bibfield  {author} {\bibinfo {author} {\bibfnamefont {Y.}~\bibnamefont
  {Yoshimura}}, \bibinfo {author} {\bibfnamefont {K.-I.}\ \bibnamefont
  {Imura}}, \bibinfo {author} {\bibfnamefont {T.}~\bibnamefont {Fukui}}, \ and\
  \bibinfo {author} {\bibfnamefont {Y.}~\bibnamefont {Hatsugai}},\ }\bibfield
  {title} {\enquote {\bibinfo {title} {Characterizing weak topological
  properties: Berry phase point of view},}\ }\href {\doibase
  10.1103/PhysRevB.90.155443} {\bibfield  {journal} {\bibinfo  {journal} {Phys.
  Rev. B}\ }\textbf {\bibinfo {volume} {90}},\ \bibinfo {pages} {155443}
  (\bibinfo {year} {2014})}\BibitemShut {NoStop}%
\bibitem [{\citenamefont {Umer}\ \emph {et~al.}(2020)\citenamefont {Umer},
  \citenamefont {Bomantara},\ and\ \citenamefont {Gong}}]{PhysRevB.101.235438}%
  \BibitemOpen
  \bibfield  {author} {\bibinfo {author} {\bibfnamefont {Muhammad}\
  \bibnamefont {Umer}}, \bibinfo {author} {\bibfnamefont {Raditya~Weda}\
  \bibnamefont {Bomantara}}, \ and\ \bibinfo {author} {\bibfnamefont
  {Jiangbin}\ \bibnamefont {Gong}},\ }\bibfield  {title} {\enquote {\bibinfo
  {title} {Counterpropagating edge states in floquet topological insulating
  phases},}\ }\href {\doibase 10.1103/PhysRevB.101.235438} {\bibfield
  {journal} {\bibinfo  {journal} {Phys. Rev. B}\ }\textbf {\bibinfo {volume}
  {101}},\ \bibinfo {pages} {235438} (\bibinfo {year} {2020})}\BibitemShut
  {NoStop}%
\bibitem [{Note2()}]{Note2}%
  \BibitemOpen
  \bibinfo {note} {The minus sign means that $g_2$ functions as $-g_3$ after
  the rotation. This is associated with the minus sign in defining $\protect
  \mathaccentV {bar}016{\Gamma }_2$.}\BibitemShut {Stop}%
\bibitem [{\citenamefont {Armitage}\ \emph {et~al.}(2018)\citenamefont
  {Armitage}, \citenamefont {Mele},\ and\ \citenamefont
  {Vishwanath}}]{RevModPhys.90.015001}%
  \BibitemOpen
  \bibfield  {author} {\bibinfo {author} {\bibfnamefont {N.~P.}\ \bibnamefont
  {Armitage}}, \bibinfo {author} {\bibfnamefont {E.~J.}\ \bibnamefont {Mele}},
  \ and\ \bibinfo {author} {\bibfnamefont {A.}~\bibnamefont {Vishwanath}},\
  }\bibfield  {title} {\enquote {\bibinfo {title} {Weyl and dirac semimetals in
  three-dimensional solids},}\ }\href {\doibase 10.1103/RevModPhys.90.015001}
  {\bibfield  {journal} {\bibinfo  {journal} {Rev. Mod. Phys.}\ }\textbf
  {\bibinfo {volume} {90}},\ \bibinfo {pages} {015001} (\bibinfo {year}
  {2018})}\BibitemShut {NoStop}%
\bibitem [{\citenamefont {Lv}\ \emph {et~al.}(2021)\citenamefont {Lv},
  \citenamefont {Qian},\ and\ \citenamefont {Ding}}]{RevModPhys.93.025002}%
  \BibitemOpen
  \bibfield  {author} {\bibinfo {author} {\bibfnamefont {B.~Q.}\ \bibnamefont
  {Lv}}, \bibinfo {author} {\bibfnamefont {T.}~\bibnamefont {Qian}}, \ and\
  \bibinfo {author} {\bibfnamefont {H.}~\bibnamefont {Ding}},\ }\bibfield
  {title} {\enquote {\bibinfo {title} {Experimental perspective on
  three-dimensional topological semimetals},}\ }\href {\doibase
  10.1103/RevModPhys.93.025002} {\bibfield  {journal} {\bibinfo  {journal}
  {Rev. Mod. Phys.}\ }\textbf {\bibinfo {volume} {93}},\ \bibinfo {pages}
  {025002} (\bibinfo {year} {2021})}\BibitemShut {NoStop}%
\bibitem [{\citenamefont {Fu}\ \emph {et~al.}(2007)\citenamefont {Fu},
  \citenamefont {Kane},\ and\ \citenamefont {Mele}}]{PhysRevLett.98.106803}%
  \BibitemOpen
  \bibfield  {author} {\bibinfo {author} {\bibfnamefont {L.}~\bibnamefont
  {Fu}}, \bibinfo {author} {\bibfnamefont {C.~L.}\ \bibnamefont {Kane}}, \ and\
  \bibinfo {author} {\bibfnamefont {E.~J.}\ \bibnamefont {Mele}},\ }\bibfield
  {title} {\enquote {\bibinfo {title} {Topological insulators in three
  dimensions},}\ }\href {\doibase 10.1103/PhysRevLett.98.106803} {\bibfield
  {journal} {\bibinfo  {journal} {Phys. Rev. Lett.}\ }\textbf {\bibinfo
  {volume} {98}},\ \bibinfo {pages} {106803} (\bibinfo {year}
  {2007})}\BibitemShut {NoStop}%
\bibitem [{\citenamefont {Moore}\ and\ \citenamefont
  {Balents}(2007)}]{PhysRevB.75.121306}%
  \BibitemOpen
  \bibfield  {author} {\bibinfo {author} {\bibfnamefont {J.~E.}\ \bibnamefont
  {Moore}}\ and\ \bibinfo {author} {\bibfnamefont {L.}~\bibnamefont
  {Balents}},\ }\bibfield  {title} {\enquote {\bibinfo {title} {Topological
  invariants of time-reversal-invariant band structures},}\ }\href {\doibase
  10.1103/PhysRevB.75.121306} {\bibfield  {journal} {\bibinfo  {journal} {Phys.
  Rev. B}\ }\textbf {\bibinfo {volume} {75}},\ \bibinfo {pages} {121306}
  (\bibinfo {year} {2007})}\BibitemShut {NoStop}%
\bibitem [{\citenamefont {Roy}(2009)}]{PhysRevB.79.195322}%
  \BibitemOpen
  \bibfield  {author} {\bibinfo {author} {\bibfnamefont {R.}~\bibnamefont
  {Roy}},\ }\bibfield  {title} {\enquote {\bibinfo {title} {Topological phases
  and the quantum spin hall effect in three dimensions},}\ }\href {\doibase
  10.1103/PhysRevB.79.195322} {\bibfield  {journal} {\bibinfo  {journal} {Phys.
  Rev. B}\ }\textbf {\bibinfo {volume} {79}},\ \bibinfo {pages} {195322}
  (\bibinfo {year} {2009})}\BibitemShut {NoStop}%
\bibitem [{\citenamefont {Ma}\ \emph {et~al.}(2018)\citenamefont {Ma},
  \citenamefont {Zhou}, \citenamefont {Zhang}, \citenamefont {Li},
  \citenamefont {Cheng}, \citenamefont {Geng}, \citenamefont {Rong},
  \citenamefont {Shi}, \citenamefont {Gong},\ and\ \citenamefont
  {Du}}]{experimental2018ma}%
  \BibitemOpen
  \bibfield  {author} {\bibinfo {author} {\bibfnamefont {Wenchao}\ \bibnamefont
  {Ma}}, \bibinfo {author} {\bibfnamefont {Longwen}\ \bibnamefont {Zhou}},
  \bibinfo {author} {\bibfnamefont {Qi}~\bibnamefont {Zhang}}, \bibinfo
  {author} {\bibfnamefont {Min}\ \bibnamefont {Li}}, \bibinfo {author}
  {\bibfnamefont {Chunyang}\ \bibnamefont {Cheng}}, \bibinfo {author}
  {\bibfnamefont {Jianpei}\ \bibnamefont {Geng}}, \bibinfo {author}
  {\bibfnamefont {Xing}\ \bibnamefont {Rong}}, \bibinfo {author} {\bibfnamefont
  {Fazhan}\ \bibnamefont {Shi}}, \bibinfo {author} {\bibfnamefont {Jiangbin}\
  \bibnamefont {Gong}}, \ and\ \bibinfo {author} {\bibfnamefont {Jiangfeng}\
  \bibnamefont {Du}},\ }\bibfield  {title} {\enquote {\bibinfo {title}
  {Experimental observation of a generalized thouless pump with a single
  spin},}\ }\href {\doibase 10.1103/PhysRevLett.120.120501} {\bibfield
  {journal} {\bibinfo  {journal} {Phys. Rev. Lett.}\ }\textbf {\bibinfo
  {volume} {120}},\ \bibinfo {pages} {120501} (\bibinfo {year}
  {2018})}\BibitemShut {NoStop}%
\bibitem [{\citenamefont {Tan}\ \emph {et~al.}(2019)\citenamefont {Tan},
  \citenamefont {Zhao}, \citenamefont {Liu}, \citenamefont {Xue}, \citenamefont
  {Yu}, \citenamefont {Wang},\ and\ \citenamefont {Yu}}]{tan2019simulation}%
  \BibitemOpen
  \bibfield  {author} {\bibinfo {author} {\bibfnamefont {Xinsheng}\
  \bibnamefont {Tan}}, \bibinfo {author} {\bibfnamefont {YX}~\bibnamefont
  {Zhao}}, \bibinfo {author} {\bibfnamefont {Qiang}\ \bibnamefont {Liu}},
  \bibinfo {author} {\bibfnamefont {Guangming}\ \bibnamefont {Xue}}, \bibinfo
  {author} {\bibfnamefont {Hai-Feng}\ \bibnamefont {Yu}}, \bibinfo {author}
  {\bibfnamefont {ZD}~\bibnamefont {Wang}}, \ and\ \bibinfo {author}
  {\bibfnamefont {Yang}\ \bibnamefont {Yu}},\ }\bibfield  {title} {\enquote
  {\bibinfo {title} {Simulation and manipulation of tunable weyl-semimetal
  bands using superconducting quantum circuits},}\ }\href@noop {} {\bibfield
  {journal} {\bibinfo  {journal} {Physical review letters}\ }\textbf {\bibinfo
  {volume} {122}},\ \bibinfo {pages} {010501} (\bibinfo {year}
  {2019})}\BibitemShut {NoStop}%
\bibitem [{\citenamefont {Ji}\ \emph {et~al.}(2020)\citenamefont {Ji},
  \citenamefont {Zhang}, \citenamefont {Wang}, \citenamefont {Zhang},
  \citenamefont {Guo}, \citenamefont {Chai}, \citenamefont {Rong},
  \citenamefont {Shi}, \citenamefont {Liu}, \citenamefont {Wang} \emph
  {et~al.}}]{ji2020quantum}%
  \BibitemOpen
  \bibfield  {author} {\bibinfo {author} {\bibfnamefont {Wentao}\ \bibnamefont
  {Ji}}, \bibinfo {author} {\bibfnamefont {Lin}\ \bibnamefont {Zhang}},
  \bibinfo {author} {\bibfnamefont {Mengqi}\ \bibnamefont {Wang}}, \bibinfo
  {author} {\bibfnamefont {Long}\ \bibnamefont {Zhang}}, \bibinfo {author}
  {\bibfnamefont {Yuhang}\ \bibnamefont {Guo}}, \bibinfo {author}
  {\bibfnamefont {Zihua}\ \bibnamefont {Chai}}, \bibinfo {author}
  {\bibfnamefont {Xing}\ \bibnamefont {Rong}}, \bibinfo {author} {\bibfnamefont
  {Fazhan}\ \bibnamefont {Shi}}, \bibinfo {author} {\bibfnamefont {Xiong-Jun}\
  \bibnamefont {Liu}}, \bibinfo {author} {\bibfnamefont {Ya}~\bibnamefont
  {Wang}},  \emph {et~al.},\ }\bibfield  {title} {\enquote {\bibinfo {title}
  {Quantum simulation for three-dimensional chiral topological insulator},}\
  }\href@noop {} {\bibfield  {journal} {\bibinfo  {journal} {Physical Review
  Letters}\ }\textbf {\bibinfo {volume} {125}},\ \bibinfo {pages} {020504}
  (\bibinfo {year} {2020})}\BibitemShut {NoStop}%
\bibitem [{\citenamefont {Xin}\ \emph {et~al.}(2020)\citenamefont {Xin},
  \citenamefont {Li}, \citenamefont {Fan}, \citenamefont {Zhu}, \citenamefont
  {Zhang}, \citenamefont {Nie}, \citenamefont {Li}, \citenamefont {Liu},\ and\
  \citenamefont {Lu}}]{xin2020quantum}%
  \BibitemOpen
  \bibfield  {author} {\bibinfo {author} {\bibfnamefont {Tao}\ \bibnamefont
  {Xin}}, \bibinfo {author} {\bibfnamefont {Yishan}\ \bibnamefont {Li}},
  \bibinfo {author} {\bibfnamefont {Yu-ang}\ \bibnamefont {Fan}}, \bibinfo
  {author} {\bibfnamefont {Xuanran}\ \bibnamefont {Zhu}}, \bibinfo {author}
  {\bibfnamefont {Yingjie}\ \bibnamefont {Zhang}}, \bibinfo {author}
  {\bibfnamefont {Xinfang}\ \bibnamefont {Nie}}, \bibinfo {author}
  {\bibfnamefont {Jun}\ \bibnamefont {Li}}, \bibinfo {author} {\bibfnamefont
  {Qihang}\ \bibnamefont {Liu}}, \ and\ \bibinfo {author} {\bibfnamefont
  {Dawei}\ \bibnamefont {Lu}},\ }\bibfield  {title} {\enquote {\bibinfo {title}
  {Quantum phases of three-dimensional chiral topological insulators on a spin
  quantum simulator},}\ }\href@noop {} {\bibfield  {journal} {\bibinfo
  {journal} {Physical Review Letters}\ }\textbf {\bibinfo {volume} {125}},\
  \bibinfo {pages} {090502} (\bibinfo {year} {2020})}\BibitemShut {NoStop}%
\bibitem [{\citenamefont {Roushan}\ \emph {et~al.}(2014)\citenamefont
  {Roushan}, \citenamefont {Neill}, \citenamefont {Chen}, \citenamefont
  {Kolodrubetz}, \citenamefont {Quintana}, \citenamefont {Leung}, \citenamefont
  {Fang}, \citenamefont {Barends}, \citenamefont {Campbell}, \citenamefont
  {Chen} \emph {et~al.}}]{roushan2014observation}%
  \BibitemOpen
  \bibfield  {author} {\bibinfo {author} {\bibfnamefont {Pedram}\ \bibnamefont
  {Roushan}}, \bibinfo {author} {\bibfnamefont {C}~\bibnamefont {Neill}},
  \bibinfo {author} {\bibfnamefont {Yu}~\bibnamefont {Chen}}, \bibinfo {author}
  {\bibfnamefont {M}~\bibnamefont {Kolodrubetz}}, \bibinfo {author}
  {\bibfnamefont {C}~\bibnamefont {Quintana}}, \bibinfo {author} {\bibfnamefont
  {N}~\bibnamefont {Leung}}, \bibinfo {author} {\bibfnamefont {M}~\bibnamefont
  {Fang}}, \bibinfo {author} {\bibfnamefont {R}~\bibnamefont {Barends}},
  \bibinfo {author} {\bibfnamefont {B}~\bibnamefont {Campbell}}, \bibinfo
  {author} {\bibfnamefont {Z}~\bibnamefont {Chen}},  \emph {et~al.},\
  }\bibfield  {title} {\enquote {\bibinfo {title} {Observation of topological
  transitions in interacting quantum circuits},}\ }\href@noop {} {\bibfield
  {journal} {\bibinfo  {journal} {Nature}\ }\textbf {\bibinfo {volume} {515}},\
  \bibinfo {pages} {241--244} (\bibinfo {year} {2014})}\BibitemShut {NoStop}%
\bibitem [{\citenamefont {Zhang}\ \emph
  {et~al.}(2019{\natexlab{b}})\citenamefont {Zhang}, \citenamefont {Zhang},\
  and\ \citenamefont {Liu}}]{PhysRevA.99.053606}%
  \BibitemOpen
  \bibfield  {author} {\bibinfo {author} {\bibfnamefont {L.}~\bibnamefont
  {Zhang}}, \bibinfo {author} {\bibfnamefont {L.}~\bibnamefont {Zhang}}, \ and\
  \bibinfo {author} {\bibfnamefont {X.-J.}\ \bibnamefont {Liu}},\ }\bibfield
  {title} {\enquote {\bibinfo {title} {Dynamical detection of topological
  charges},}\ }\href {\doibase 10.1103/PhysRevA.99.053606} {\bibfield
  {journal} {\bibinfo  {journal} {Phys. Rev. A}\ }\textbf {\bibinfo {volume}
  {99}},\ \bibinfo {pages} {053606} (\bibinfo {year}
  {2019}{\natexlab{b}})}\BibitemShut {NoStop}%
\bibitem [{\citenamefont {Zhang}\ \emph
  {et~al.}(2019{\natexlab{c}})\citenamefont {Zhang}, \citenamefont {Zhang},\
  and\ \citenamefont {Liu}}]{PhysRevA.100.063624}%
  \BibitemOpen
  \bibfield  {author} {\bibinfo {author} {\bibfnamefont {L.}~\bibnamefont
  {Zhang}}, \bibinfo {author} {\bibfnamefont {L.}~\bibnamefont {Zhang}}, \ and\
  \bibinfo {author} {\bibfnamefont {X.-J.}\ \bibnamefont {Liu}},\ }\bibfield
  {title} {\enquote {\bibinfo {title} {Characterizing topological phases by
  quantum quenches: A general theory},}\ }\href {\doibase
  10.1103/PhysRevA.100.063624} {\bibfield  {journal} {\bibinfo  {journal}
  {Phys. Rev. A}\ }\textbf {\bibinfo {volume} {100}},\ \bibinfo {pages}
  {063624} (\bibinfo {year} {2019}{\natexlab{c}})}\BibitemShut {NoStop}%
\bibitem [{\citenamefont {Zhang}\ \emph {et~al.}(2020)\citenamefont {Zhang},
  \citenamefont {Zhang},\ and\ \citenamefont {L.}}]{PhysRevLett.125.183001}%
  \BibitemOpen
  \bibfield  {author} {\bibinfo {author} {\bibfnamefont {L.}~\bibnamefont
  {Zhang}}, \bibinfo {author} {\bibfnamefont {L.}~\bibnamefont {Zhang}}, \ and\
  \bibinfo {author} {\bibfnamefont {X.-J.}\ \bibnamefont {L.}},\ }\bibfield
  {title} {\enquote {\bibinfo {title} {Unified theory to characterize floquet
  topological phases by quench dynamics},}\ }\href {\doibase
  10.1103/PhysRevLett.125.183001} {\bibfield  {journal} {\bibinfo  {journal}
  {Phys. Rev. Lett.}\ }\textbf {\bibinfo {volume} {125}},\ \bibinfo {pages}
  {183001} (\bibinfo {year} {2020})}\BibitemShut {NoStop}%
\bibitem [{\citenamefont {Lu}\ \emph {et~al.}(2020)\citenamefont {Lu},
  \citenamefont {Wang},\ and\ \citenamefont {Liu}}]{LU20202080}%
  \BibitemOpen
  \bibfield  {author} {\bibinfo {author} {\bibfnamefont {Yue-Hui}\ \bibnamefont
  {Lu}}, \bibinfo {author} {\bibfnamefont {Bao-Zong}\ \bibnamefont {Wang}}, \
  and\ \bibinfo {author} {\bibfnamefont {Xiong-Jun}\ \bibnamefont {Liu}},\
  }\bibfield  {title} {\enquote {\bibinfo {title} {Ideal weyl semimetal with 3d
  spin-orbit coupled ultracold quantum gas},}\ }\href@noop {} {\bibfield
  {journal} {\bibinfo  {journal} {Science Bulletin}\ }\textbf {\bibinfo
  {volume} {65}},\ \bibinfo {pages} {2080--2085} (\bibinfo {year}
  {2020})}\BibitemShut {NoStop}%
\bibitem [{\citenamefont {Zhang}\ \emph {et~al.}(2021)\citenamefont {Zhang},
  \citenamefont {Zhang},\ and\ \citenamefont {Liu}}]{PhysRevResearch.3.013229}%
  \BibitemOpen
  \bibfield  {author} {\bibinfo {author} {\bibfnamefont {L.}~\bibnamefont
  {Zhang}}, \bibinfo {author} {\bibfnamefont {L.}~\bibnamefont {Zhang}}, \ and\
  \bibinfo {author} {\bibfnamefont {X.-J.}\ \bibnamefont {Liu}},\ }\bibfield
  {title} {\enquote {\bibinfo {title} {Quench-induced dynamical topology under
  dynamical noise},}\ }\href {\doibase 10.1103/PhysRevResearch.3.013229}
  {\bibfield  {journal} {\bibinfo  {journal} {Phys. Rev. Research}\ }\textbf
  {\bibinfo {volume} {3}},\ \bibinfo {pages} {013229} (\bibinfo {year}
  {2021})}\BibitemShut {NoStop}%
\bibitem [{\citenamefont {Jia}\ \emph {et~al.}(2022)\citenamefont {Jia},
  \citenamefont {Zhou}, \citenamefont {Zhang}, \citenamefont {Zhang},\ and\
  \citenamefont {Liu}}]{2209.10394}%
  \BibitemOpen
  \bibfield  {author} {\bibinfo {author} {\bibfnamefont {W.}~\bibnamefont
  {Jia}}, \bibinfo {author} {\bibfnamefont {X.-C.}\ \bibnamefont {Zhou}},
  \bibinfo {author} {\bibfnamefont {L.}~\bibnamefont {Zhang}}, \bibinfo
  {author} {\bibfnamefont {L.}~\bibnamefont {Zhang}}, \ and\ \bibinfo {author}
  {\bibfnamefont {X.-J.}\ \bibnamefont {Liu}},\ }\href@noop {} {\enquote
  {\bibinfo {title} {Unified characterization for higher-order topological
  phase transitions},}\ } (\bibinfo {year} {2022}),\ \Eprint
  {http://arxiv.org/abs/arXiv:2209.10394} {arXiv:2209.10394} \BibitemShut
  {NoStop}%
\bibitem [{\citenamefont {Niu}\ \emph {et~al.}(2021)\citenamefont {Niu},
  \citenamefont {Yan}, \citenamefont {Zhou}, \citenamefont {Tao}, \citenamefont
  {Li}, \citenamefont {Liu}, \citenamefont {Zhang}, \citenamefont {Jia},
  \citenamefont {Liu}, \citenamefont {Yan} \emph {et~al.}}]{niu2021simulation}%
  \BibitemOpen
  \bibfield  {author} {\bibinfo {author} {\bibfnamefont {Jingjing}\
  \bibnamefont {Niu}}, \bibinfo {author} {\bibfnamefont {Tongxing}\
  \bibnamefont {Yan}}, \bibinfo {author} {\bibfnamefont {Yuxuan}\ \bibnamefont
  {Zhou}}, \bibinfo {author} {\bibfnamefont {Ziyu}\ \bibnamefont {Tao}},
  \bibinfo {author} {\bibfnamefont {Xiaole}\ \bibnamefont {Li}}, \bibinfo
  {author} {\bibfnamefont {Weiyang}\ \bibnamefont {Liu}}, \bibinfo {author}
  {\bibfnamefont {Libo}\ \bibnamefont {Zhang}}, \bibinfo {author}
  {\bibfnamefont {Hao}\ \bibnamefont {Jia}}, \bibinfo {author} {\bibfnamefont
  {Song}\ \bibnamefont {Liu}}, \bibinfo {author} {\bibfnamefont {Zhongbo}\
  \bibnamefont {Yan}},  \emph {et~al.},\ }\bibfield  {title} {\enquote
  {\bibinfo {title} {Simulation of higher-order topological phases and related
  topological phase transitions in a superconducting qubit},}\ }\href@noop {}
  {\bibfield  {journal} {\bibinfo  {journal} {Science Bulletin}\ }\textbf
  {\bibinfo {volume} {66}},\ \bibinfo {pages} {1168--1175} (\bibinfo {year}
  {2021})}\BibitemShut {NoStop}%
\bibitem [{\citenamefont {Yi}\ \emph {et~al.}(2019)\citenamefont {Yi},
  \citenamefont {Zhang}, \citenamefont {Zhang}, \citenamefont {Jiao},
  \citenamefont {Cheng}, \citenamefont {Wang}, \citenamefont {Xu},
  \citenamefont {Sun}, \citenamefont {Liu}, \citenamefont {Chen} \emph
  {et~al.}}]{yi2019observing}%
  \BibitemOpen
  \bibfield  {author} {\bibinfo {author} {\bibfnamefont {Chang-Rui}\
  \bibnamefont {Yi}}, \bibinfo {author} {\bibfnamefont {Long}\ \bibnamefont
  {Zhang}}, \bibinfo {author} {\bibfnamefont {Lin}\ \bibnamefont {Zhang}},
  \bibinfo {author} {\bibfnamefont {Rui-Heng}\ \bibnamefont {Jiao}}, \bibinfo
  {author} {\bibfnamefont {Xiang-Can}\ \bibnamefont {Cheng}}, \bibinfo {author}
  {\bibfnamefont {Zong-Yao}\ \bibnamefont {Wang}}, \bibinfo {author}
  {\bibfnamefont {Xiao-Tian}\ \bibnamefont {Xu}}, \bibinfo {author}
  {\bibfnamefont {Wei}\ \bibnamefont {Sun}}, \bibinfo {author} {\bibfnamefont
  {Xiong-Jun}\ \bibnamefont {Liu}}, \bibinfo {author} {\bibfnamefont {Shuai}\
  \bibnamefont {Chen}},  \emph {et~al.},\ }\bibfield  {title} {\enquote
  {\bibinfo {title} {Observing topological charges and dynamical bulk-surface
  correspondence with ultracold atoms},}\ }\href@noop {} {\bibfield  {journal}
  {\bibinfo  {journal} {Physical review letters}\ }\textbf {\bibinfo {volume}
  {123}},\ \bibinfo {pages} {190603} (\bibinfo {year} {2019})}\BibitemShut
  {NoStop}%
\bibitem [{\citenamefont {Wang}\ \emph
  {et~al.}(2021{\natexlab{b}})\citenamefont {Wang}, \citenamefont {Cheng},
  \citenamefont {Wang}, \citenamefont {Zhang}, \citenamefont {Lu},
  \citenamefont {Yi}, \citenamefont {Niu}, \citenamefont {Deng}, \citenamefont
  {Liu}, \citenamefont {Chen},\ and\ \citenamefont {Pan}}]{Wang271}%
  \BibitemOpen
  \bibfield  {author} {\bibinfo {author} {\bibfnamefont {Z.-Y.}\ \bibnamefont
  {Wang}}, \bibinfo {author} {\bibfnamefont {X.-C.}\ \bibnamefont {Cheng}},
  \bibinfo {author} {\bibfnamefont {B.-Z.}\ \bibnamefont {Wang}}, \bibinfo
  {author} {\bibfnamefont {J.-Y.}\ \bibnamefont {Zhang}}, \bibinfo {author}
  {\bibfnamefont {Y.-H.}\ \bibnamefont {Lu}}, \bibinfo {author} {\bibfnamefont
  {C.-R.}\ \bibnamefont {Yi}}, \bibinfo {author} {\bibfnamefont
  {S.}~\bibnamefont {Niu}}, \bibinfo {author} {\bibfnamefont {Y.}~\bibnamefont
  {Deng}}, \bibinfo {author} {\bibfnamefont {X.-J.}\ \bibnamefont {Liu}},
  \bibinfo {author} {\bibfnamefont {S.}~\bibnamefont {Chen}}, \ and\ \bibinfo
  {author} {\bibfnamefont {J.-W.}\ \bibnamefont {Pan}},\ }\bibfield  {title}
  {\enquote {\bibinfo {title} {Realization of an ideal weyl semimetal band in a
  quantum gas with 3d spin-orbit coupling},}\ }\href {\doibase
  10.1126/science.abc0105} {\bibfield  {journal} {\bibinfo  {journal}
  {Science}\ }\textbf {\bibinfo {volume} {372}},\ \bibinfo {pages} {271--276}
  (\bibinfo {year} {2021}{\natexlab{b}})}\BibitemShut {NoStop}%
\bibitem [{\citenamefont {Yan}\ \emph {et~al.}(2018)\citenamefont {Yan},
  \citenamefont {Song},\ and\ \citenamefont {Wang}}]{PhysRevLett.121.096803}%
  \BibitemOpen
  \bibfield  {author} {\bibinfo {author} {\bibfnamefont {Z.}~\bibnamefont
  {Yan}}, \bibinfo {author} {\bibfnamefont {F.}~\bibnamefont {Song}}, \ and\
  \bibinfo {author} {\bibfnamefont {Z.}~\bibnamefont {Wang}},\ }\bibfield
  {title} {\enquote {\bibinfo {title} {Majorana corner modes in a
  high-temperature platform},}\ }\href {\doibase
  10.1103/PhysRevLett.121.096803} {\bibfield  {journal} {\bibinfo  {journal}
  {Phys. Rev. Lett.}\ }\textbf {\bibinfo {volume} {121}},\ \bibinfo {pages}
  {096803} (\bibinfo {year} {2018})}\BibitemShut {NoStop}%
\bibitem [{\citenamefont {Wang}\ \emph {et~al.}(2018)\citenamefont {Wang},
  \citenamefont {Liu}, \citenamefont {Lu},\ and\ \citenamefont
  {Zhang}}]{PhysRevLett.121.186801}%
  \BibitemOpen
  \bibfield  {author} {\bibinfo {author} {\bibfnamefont {Q.}~\bibnamefont
  {Wang}}, \bibinfo {author} {\bibfnamefont {C.-C.}\ \bibnamefont {Liu}},
  \bibinfo {author} {\bibfnamefont {Y.-M.}\ \bibnamefont {Lu}}, \ and\ \bibinfo
  {author} {\bibfnamefont {F.}~\bibnamefont {Zhang}},\ }\bibfield  {title}
  {\enquote {\bibinfo {title} {High-temperature majorana corner states},}\
  }\href {\doibase 10.1103/PhysRevLett.121.186801} {\bibfield  {journal}
  {\bibinfo  {journal} {Phys. Rev. Lett.}\ }\textbf {\bibinfo {volume} {121}},\
  \bibinfo {pages} {186801} (\bibinfo {year} {2018})}\BibitemShut {NoStop}%
\bibitem [{\citenamefont {Lin}\ and\ \citenamefont
  {Hughes}(2018)}]{PhysRevB.98.241103}%
  \BibitemOpen
  \bibfield  {author} {\bibinfo {author} {\bibfnamefont {M.}~\bibnamefont
  {Lin}}\ and\ \bibinfo {author} {\bibfnamefont {T.~L.}\ \bibnamefont
  {Hughes}},\ }\bibfield  {title} {\enquote {\bibinfo {title} {Topological
  quadrupolar semimetals},}\ }\href {\doibase 10.1103/PhysRevB.98.241103}
  {\bibfield  {journal} {\bibinfo  {journal} {Phys. Rev. B}\ }\textbf {\bibinfo
  {volume} {98}},\ \bibinfo {pages} {241103} (\bibinfo {year}
  {2018})}\BibitemShut {NoStop}%
\bibitem [{\citenamefont {Wieder}\ \emph {et~al.}(2020)\citenamefont {Wieder},
  \citenamefont {Wang}, \citenamefont {Cano}, \citenamefont {Dai},
  \citenamefont {Schoop}, \citenamefont {Bradlyn},\ and\ \citenamefont
  {Bernevig}}]{Wieder2020}%
  \BibitemOpen
  \bibfield  {author} {\bibinfo {author} {\bibfnamefont {B.~J.}\ \bibnamefont
  {Wieder}}, \bibinfo {author} {\bibfnamefont {Z.}~\bibnamefont {Wang}},
  \bibinfo {author} {\bibfnamefont {J.}~\bibnamefont {Cano}}, \bibinfo {author}
  {\bibfnamefont {X.}~\bibnamefont {Dai}}, \bibinfo {author} {\bibfnamefont
  {L.~M.}\ \bibnamefont {Schoop}}, \bibinfo {author} {\bibfnamefont
  {B.}~\bibnamefont {Bradlyn}}, \ and\ \bibinfo {author} {\bibfnamefont
  {B.~A.}\ \bibnamefont {Bernevig}},\ }\bibfield  {title} {\enquote {\bibinfo
  {title} {Strong and fragile topological dirac semimetals with higher-order
  fermi arcs},}\ }\href {\doibase 10.1038/s41467-020-14443-5} {\bibfield
  {journal} {\bibinfo  {journal} {Nat. Commun.}\ }\textbf {\bibinfo {volume}
  {11}},\ \bibinfo {pages} {627} (\bibinfo {year} {2020})}\BibitemShut
  {NoStop}%
\bibitem [{\citenamefont {Ghorashi}\ \emph {et~al.}(2019)\citenamefont
  {Ghorashi}, \citenamefont {Hu}, \citenamefont {Hughes},\ and\ \citenamefont
  {Rossi}}]{PhysRevB.100.020509}%
  \BibitemOpen
  \bibfield  {author} {\bibinfo {author} {\bibfnamefont {S.~A.~A.}\
  \bibnamefont {Ghorashi}}, \bibinfo {author} {\bibfnamefont {X.}~\bibnamefont
  {Hu}}, \bibinfo {author} {\bibfnamefont {T.~L.}\ \bibnamefont {Hughes}}, \
  and\ \bibinfo {author} {\bibfnamefont {E.}~\bibnamefont {Rossi}},\ }\bibfield
   {title} {\enquote {\bibinfo {title} {Second-order dirac superconductors and
  magnetic field induced majorana hinge modes},}\ }\href {\doibase
  10.1103/PhysRevB.100.020509} {\bibfield  {journal} {\bibinfo  {journal}
  {Phys. Rev. B}\ }\textbf {\bibinfo {volume} {100}},\ \bibinfo {pages}
  {020509} (\bibinfo {year} {2019})}\BibitemShut {NoStop}%
\bibitem [{\citenamefont {Ghorashi}\ \emph {et~al.}(2020)\citenamefont
  {Ghorashi}, \citenamefont {Li},\ and\ \citenamefont
  {Hughes}}]{PhysRevLett.125.266804}%
  \BibitemOpen
  \bibfield  {author} {\bibinfo {author} {\bibfnamefont {S.~A.~A.}\
  \bibnamefont {Ghorashi}}, \bibinfo {author} {\bibfnamefont {T.}~\bibnamefont
  {Li}}, \ and\ \bibinfo {author} {\bibfnamefont {T.~L.}\ \bibnamefont
  {Hughes}},\ }\bibfield  {title} {\enquote {\bibinfo {title} {Higher-order
  weyl semimetals},}\ }\href {\doibase 10.1103/PhysRevLett.125.266804}
  {\bibfield  {journal} {\bibinfo  {journal} {Phys. Rev. Lett.}\ }\textbf
  {\bibinfo {volume} {125}},\ \bibinfo {pages} {266804} (\bibinfo {year}
  {2020})}\BibitemShut {NoStop}%
\bibitem [{\citenamefont {Wu}\ \emph {et~al.}(2020{\natexlab{b}})\citenamefont
  {Wu}, \citenamefont {Yu}, \citenamefont {Zhou}, \citenamefont {Zhao},\ and\
  \citenamefont {Yang}}]{PhysRevB.101.205134}%
  \BibitemOpen
  \bibfield  {author} {\bibinfo {author} {\bibfnamefont {W.}~\bibnamefont
  {Wu}}, \bibinfo {author} {\bibfnamefont {Z.-M.}\ \bibnamefont {Yu}}, \bibinfo
  {author} {\bibfnamefont {X.}~\bibnamefont {Zhou}}, \bibinfo {author}
  {\bibfnamefont {Y.~X.}\ \bibnamefont {Zhao}}, \ and\ \bibinfo {author}
  {\bibfnamefont {S.~A.}\ \bibnamefont {Yang}},\ }\bibfield  {title} {\enquote
  {\bibinfo {title} {Higher-order dirac fermions in three dimensions},}\ }\href
  {\doibase 10.1103/PhysRevB.101.205134} {\bibfield  {journal} {\bibinfo
  {journal} {Phys. Rev. B}\ }\textbf {\bibinfo {volume} {101}},\ \bibinfo
  {pages} {205134} (\bibinfo {year} {2020}{\natexlab{b}})}\BibitemShut
  {NoStop}%
\bibitem [{\citenamefont {Roy}(2020)}]{PhysRevB.101.220506}%
  \BibitemOpen
  \bibfield  {author} {\bibinfo {author} {\bibfnamefont {B.}~\bibnamefont
  {Roy}},\ }\bibfield  {title} {\enquote {\bibinfo {title} {Higher-order
  topological superconductors in $\mathcal{P}$-, $\mathcal{T}$-odd quadrupolar
  dirac materials},}\ }\href {\doibase 10.1103/PhysRevB.101.220506} {\bibfield
  {journal} {\bibinfo  {journal} {Phys. Rev. B}\ }\textbf {\bibinfo {volume}
  {101}},\ \bibinfo {pages} {220506} (\bibinfo {year} {2020})}\BibitemShut
  {NoStop}%
\bibitem [{\citenamefont {Wang}\ \emph {et~al.}(2020)\citenamefont {Wang},
  \citenamefont {Lin}, \citenamefont {Jiang}, \citenamefont {Guo},\ and\
  \citenamefont {Jiang}}]{PhysRevLett.125.146401}%
  \BibitemOpen
  \bibfield  {author} {\bibinfo {author} {\bibfnamefont {H.-X.}\ \bibnamefont
  {Wang}}, \bibinfo {author} {\bibfnamefont {Z.-K.}\ \bibnamefont {Lin}},
  \bibinfo {author} {\bibfnamefont {B.}~\bibnamefont {Jiang}}, \bibinfo
  {author} {\bibfnamefont {G.-Y.}\ \bibnamefont {Guo}}, \ and\ \bibinfo
  {author} {\bibfnamefont {J.-H.}\ \bibnamefont {Jiang}},\ }\bibfield  {title}
  {\enquote {\bibinfo {title} {Higher-order weyl semimetals},}\ }\href
  {\doibase 10.1103/PhysRevLett.125.146401} {\bibfield  {journal} {\bibinfo
  {journal} {Phys. Rev. Lett.}\ }\textbf {\bibinfo {volume} {125}},\ \bibinfo
  {pages} {146401} (\bibinfo {year} {2020})}\BibitemShut {NoStop}%
\bibitem [{\citenamefont {Chen}\ \emph {et~al.}(2021)\citenamefont {Chen},
  \citenamefont {Liu}, \citenamefont {Wang}, \citenamefont {Lu},\ and\
  \citenamefont {Xie}}]{PhysRevLett.127.066801}%
  \BibitemOpen
  \bibfield  {author} {\bibinfo {author} {\bibfnamefont {R.}~\bibnamefont
  {Chen}}, \bibinfo {author} {\bibfnamefont {T.}~\bibnamefont {Liu}}, \bibinfo
  {author} {\bibfnamefont {C.~M.}\ \bibnamefont {Wang}}, \bibinfo {author}
  {\bibfnamefont {H.-Z.}\ \bibnamefont {Lu}}, \ and\ \bibinfo {author}
  {\bibfnamefont {X.~C.}\ \bibnamefont {Xie}},\ }\bibfield  {title} {\enquote
  {\bibinfo {title} {Field-tunable one-sided higher-order topological hinge
  states in dirac semimetals},}\ }\href {\doibase
  10.1103/PhysRevLett.127.066801} {\bibfield  {journal} {\bibinfo  {journal}
  {Phys. Rev. Lett.}\ }\textbf {\bibinfo {volume} {127}},\ \bibinfo {pages}
  {066801} (\bibinfo {year} {2021})}\BibitemShut {NoStop}%
\bibitem [{\citenamefont {You}\ \emph {et~al.}(2018)\citenamefont {You},
  \citenamefont {Devakul}, \citenamefont {Burnell},\ and\ \citenamefont
  {Neupert}}]{PhysRevB.98.235102}%
  \BibitemOpen
  \bibfield  {author} {\bibinfo {author} {\bibfnamefont {Y.}~\bibnamefont
  {You}}, \bibinfo {author} {\bibfnamefont {T.}~\bibnamefont {Devakul}},
  \bibinfo {author} {\bibfnamefont {F.~J.}\ \bibnamefont {Burnell}}, \ and\
  \bibinfo {author} {\bibfnamefont {T.}~\bibnamefont {Neupert}},\ }\bibfield
  {title} {\enquote {\bibinfo {title} {Higher-order symmetry-protected
  topological states for interacting bosons and fermions},}\ }\href {\doibase
  10.1103/PhysRevB.98.235102} {\bibfield  {journal} {\bibinfo  {journal} {Phys.
  Rev. B}\ }\textbf {\bibinfo {volume} {98}},\ \bibinfo {pages} {235102}
  (\bibinfo {year} {2018})}\BibitemShut {NoStop}%
\bibitem [{\citenamefont {Zhao}\ \emph {et~al.}(2021)\citenamefont {Zhao},
  \citenamefont {Qiang}, \citenamefont {Lu},\ and\ \citenamefont
  {Xie}}]{PhysRevLett.127.176601}%
  \BibitemOpen
  \bibfield  {author} {\bibinfo {author} {\bibfnamefont {P.-L.}\ \bibnamefont
  {Zhao}}, \bibinfo {author} {\bibfnamefont {X.-B.}\ \bibnamefont {Qiang}},
  \bibinfo {author} {\bibfnamefont {H.-Z.}\ \bibnamefont {Lu}}, \ and\ \bibinfo
  {author} {\bibfnamefont {X.~C.}\ \bibnamefont {Xie}},\ }\bibfield  {title}
  {\enquote {\bibinfo {title} {Coulomb instabilities of a three-dimensional
  higher-order topological insulator},}\ }\href {\doibase
  10.1103/PhysRevLett.127.176601} {\bibfield  {journal} {\bibinfo  {journal}
  {Phys. Rev. Lett.}\ }\textbf {\bibinfo {volume} {127}},\ \bibinfo {pages}
  {176601} (\bibinfo {year} {2021})}\BibitemShut {NoStop}%
\bibitem [{\citenamefont {Wang}\ \emph {et~al.}(2012)\citenamefont {Wang},
  \citenamefont {Qi},\ and\ \citenamefont {Zhang}}]{PhysRevB.85.165126}%
  \BibitemOpen
  \bibfield  {author} {\bibinfo {author} {\bibfnamefont {Z.}~\bibnamefont
  {Wang}}, \bibinfo {author} {\bibfnamefont {X.-L.}\ \bibnamefont {Qi}}, \ and\
  \bibinfo {author} {\bibfnamefont {S.-C.}\ \bibnamefont {Zhang}},\ }\bibfield
  {title} {\enquote {\bibinfo {title} {Topological invariants for interacting
  topological insulators with inversion symmetry},}\ }\href {\doibase
  10.1103/PhysRevB.85.165126} {\bibfield  {journal} {\bibinfo  {journal} {Phys.
  Rev. B}\ }\textbf {\bibinfo {volume} {85}},\ \bibinfo {pages} {165126}
  (\bibinfo {year} {2012})}\BibitemShut {NoStop}%
\end{thebibliography}

\end{document}